\documentclass[12pt]{uwthesis17}

\usepackage{amsmath,amsfonts,amsthm,amssymb}
\usepackage[toc,page]{appendix}
\usepackage{fancyhdr}
\usepackage{color}
\usepackage{boxedminipage}
\usepackage{changepage}
\usepackage{graphicx}
\usepackage{subcaption}
\usepackage{array}

\newcolumntype{C}{>{\centering\arraybackslash}p{1em}}
\usepackage{booktabs}
\usepackage{multirow}
\usepackage{makecell}
\usepackage{caption}
\title{Generating spiky solutions of Einstein field equations with the Stephani transformation}
\author{Muhammad Zubair Ali Moughal}

\def\lb{\label}

\def\Sm{\Sigma_-}

\def\Sp{\Sigma_+}

\def\sc{\sigma_\times}

\def\be{\begin{equation}}
\def\ee{\end{equation}}

\def\Sp{\Sigma_+}
\def\Spo{\Sigma_{+0}}
\def\Sno{\Sigma_{-0}}

\begin{document}
\pagenumbering{roman}
\maketitle
\setcounter{page}{2}
\begin{abstract}
The Geroch/Stephani transformation is a solution-generating transformation, and may generate spiky solutions. The spikes in solutions generated so far are either early-time permanent spikes or transient spikes. We want to generate a solution with a late-time permanent spike. We achieve this by applying the Stephani transformation with the rotational Killing vector field of the locally rotationally symmetric Jacobs solution. The late-time permanent spike occurs along the cylindrical axis. The generated solution also features a rich variety of transient structures. We introduce a new technique to analyse these structures. Our findings lead us to discover a transient behaviour, which we call the overshoot transition.
\end{abstract} 

\begin{acknowledgements}
In the name of Allah, The Most Gracious and The Most Merciful, who provided me the opportunity to unveil the concealed realities in the world of Mathematics.

In the first place, I owe my deepest gratitude to my supervisor, Dr. Woei Chet Lim.  It would have been next to impossible to write this dissertation without the guidance and valuable input/feedback of my respected supervisor. He is a source of true inspiration  as he has always supported and guided me during my stay  in New Zealand -- especially his financial and personal support for providing a grant to present my work at the 22nd International Conference on General Relativity and Gravitation (GR22) and the 10th Australasian Conference on General Relativity and Gravitation (ACGRG10).

I deeply acknowledge the role of the Department of Mathematics at University of Waikato which facilitated and provided learning and a healthy environment to complete my thesis. I am also thankful to the other PhD lab students Ejaz, Liam, Fahim, Hamish, Chris and Nick for a healthy working environment.

My acknowledgment will never be complete without the special mention of my friends: Bilal, Hassam, Humair, Irfan, Atta, Irfan Habib, Shaheer, Mairaj and many more (it is not feasible to name all here). Thanks for being around and sharing several good times together during my stay in the University and New Zealand.

I owe a special debt to my affectionate and lovely parents. The credit, for my enjoying this status in my life, goes to my parents. I donot have words to convey my cordial regards and thanks to my mother for her utmost efforts, sacrifices, and prayers. I am also grateful to my father for all that he did for my bright future. Besides I cannot dare to forget to mention the co-operation of my sweet sister, brothers and their naughty kids. I appreciate the moral support of my friends in Pakistan as well, specially Khalil, Kamran, Hassan, Faraz and Waqas.

Lastly, I would like to thank the Higher Education Commission (H.E.C) of the government of Pakistan that provided me with the  scholarship to do this PhD.
\end{acknowledgements}
\tableofcontents
\listoffigures
\pagenumbering{arabic}
\setcounter{page}{1}

\chapter{Introduction}
\section{What is spike}
The big bang theory postulates that the universe has always been expand-
ing. Extrapolating this into the past, there was a time when the universe
was incredibly dense and hot, such that even the laws of Einstein's general
relativity, which predicted the big bang, fail. Such time or place is called
a singularity. Similarly, after depleting their nuclear fuel, massive stars can
collapse under their own weight and become black holes. Inside a black hole,
the collapse also inevitably leads to an incredibly dense and hot state where
the laws of general relativity fail again. In the final stage before
the laws fail, spacetime undergoes a kind of chaotic dynamics called Mixmaster or BKL dy-
namics ~\cite{art:BKL1970, art:BKL1982, art:LK63, art:Misner1969b}. Under the influence of Mixmaster dynamics, a collapsing object
is crushed or stretched at predictable and alternating speeds along its three
dimensions. But what is really being crushed and stretched is space itself,
regardless of the presence of any object there. Spikes form when adjacent parts of space experience vastly different rates of crushing and stretching. It is important to understand spikes
because it occurs during a regime that transitions into the quantum regime,
and quantum gravity researchers need to understand what happens during
the transition. In another context, spike can also help explain the formation of large scale structure at late times.

On approach to the singularity a generic solution is approximated by a
sequence of Kasner states, described by the Kasner solution of Einstein
field equations (EFEs). The transition between Kasner states is approximated by
another exact solution, the Taub solution. Furthermore, this transition can
be achieved in two ways, as reflected by the sign of a curvature variable. In a
spatially inhomogeneous model, this curvature variable may change sign from
one place to another. As a result, transition fails to occur in a normal way
at places where this curvature variable is zero. A different, inhomogeneous,
dynamics occur instead, and this is the spike \cite{thesis:Lim2004}.

What is the mechanism behind spike formation? Briefly, when a solution
is close to an unstable background solution that is represented by a saddle
point in the state space, it generally becomes unstable. If the initial condition
is such that the solution straddles the separatrix of the saddle point, then
as the solution becomes unstable, some region of spacetime evolves one way,
the other region evolves in a different way. The boundary between these two
parts evolves in a way that is different from both regions (because on this
boundary, its state lies exactly on the separatrix), and this creates a spiky
inhomogeneity in the neighborhood of this boundary \cite{thesis:Lim2004, art:ColeyLim2014}.

\section{The journey from numerical to exact solution}
It is a general feature of solutions of partial differential equations that spikes occur ~\cite{inbook:Wei2008}. As Einstein field equations (EFEs) of general relativity are a set of partial differential equations, spikes can arise
in the solutions of these equations  ~\cite{thesis:Lim2004}. 
Spikes were first discovered in numerical simulations by Berger and Moncrief in 1993 ~\cite{art:BergerMoncrief1993}. In their numerical study on the so-called $T^3$
Gowdy model\footnote{ $T^3$ Gowdy models are orthogonally transitive $G_2$ models with toroidal spatial topology. See Appendix \ref{App:KVFsgroup_actions}.} \cite{art:Gowdy1971}, where the BKL dynamics terminates at a final Kasner state,  they observed the development of
large spatial derivatives near the singularity, which they
termed as spiky features.  These spikes are permanent spikes as their amplitude does not tend to zero towards singularity \cite{art:BergerLiving2002}.
Furthermore, based on the work
by Grubi$\check{\text{s}}$i$\acute{\text{c}}$ and Moncrief of the same year \cite{art:GrubisicMoncrief1993}, these structures where found to occur in the neighbourhood of  isolated spatial surfaces. Toward the end of the 1990s, Berger, Moncrief and
co-workers had found further numerical evidence that the
BKL picture seemed to be correct generically but there were difficulties in simulating these spikes  ~\cite{art:Bergeretall1998, art:BergerMoncrief1998, art:BergerGarfinkle1998, art:Bergeretal2001}. Also, Hern in his PhD thesis \cite{thesis:Hern1999} resolved individual spatially spiky features to high
numerical accuracy but for a short time interval.
The observed inhomogeneity of the curvature invariants makes it clear that the spikes are physical features of the spacetime  not effects of the coordinate system.

In 2001 Rendall and Weaver \cite{art:RendallWeaver2001} made a significant analytic step toward the understanding of
spikes. They discovered  a composition of two transformations that can map a spike-free solution to a  solution with spike. They applied the solution-generating transformation and Fuchsian methods\footnote{The theory of Fuchsian equations has been applied to analyse singularities in	a variety of classes of spacetimes in general relativity. In \cite{art:KichenassamyRendall1998, art:AnderssonRendall2001},  Fuchsian algorithm is applied to Einstein’s equations to establish the existence of a family of solutions. Recent work can be found in \cite{art:RodnianskiSpeck2018}, \cite{art:RodnianskiSpeck20182}.} in \cite{art:KichenassamyRendall1998, art:Rendall2000},
to produce asymptotic expansions for spikes. In their numerical study they find false and true(real) spikes. False spikes are an effect of parameterisation of the metric, not a geometrical one, while true spikes are the geometric change, which they check by observing highly non-uniform behaviour in curvature invariants. The work on spiky
features in Gowdy spacetimes by Rendall and Weaver was
followed up by Garfinkle and Weaver in 2003, who used two
different complementary numerical techniques \cite{art:GarfinkleWeaver2003}. In particular they studied the so-called (transient and recurring) high-velocity spikes and
found that they eventually evolve into permanent low-velocity
spikes. Also, Lim in his PhD thesis \cite{thesis:Lim2004} applied the Rendall-Weaver transformation on the Wainwright-Marshman solution \cite{art:WainwrightMarshman1979}. He obtained a new explicit vacuum OT $G_2$ (Appendix \ref{App:KVFsgroup_actions}) solution that develops a permanent spike.

After the 1993 discovery of spikes many researchers tried to understand the behaviour and dynamics of these structure. They put a lot of effort to solve it through numerical solutions and analytical approximation. In the beginning these analytical approximation were insufficient and the  numerical simulations lack resolution. In 2008 \cite{art:Lim2008}, Lim discovered the first exact spike solution. Iteratively, he applied the Rendall-Weaver transformation on Kasner seed solution. The generated spike solution admits two Killing vector fields (KVFs) and it is an orthogonally transitive (OT) $G_2$ solution (see Appendix \ref{App:KVFsgroup_actions}). This exact solution is the gateway to the understanding and analysis of the spikes. There was a question whether numerical solutions match the exact solution. The answer is yes -- in 2009 Lim $et$ $al.$ \cite{art:Limetal2009} using a new zooming technique, provided highly accurate numerical evidence. 

Nungesser and Lim \cite{art:NungesserLim2013} found the inhomogeneous electromagnetic spike solution. They use the existing relation between vacuum Gowdy spacetime and electromagnetic Gowdy spacetime to find this new explicit solution. Beyer and Hennig \cite{art:BeyerHennig2014} derived a family of Gowdy-symmetric generalized Taub--NUT solutions and found both false and true spikes.

Coley and Lim \cite{art:ColeyLim2012} discussed the influence of spikes on matter that leads to the formation of large scale structure at the early universe. They concentrated on how spikes generate matter overdensities in
a radiation fluid in a special class of inhomogeneous models. In 2014 Lim and Coley \cite{art:LimColey2014} examined the tilted fluid, whose tilt provided another mechanism in generating matter inhomogeneity through the divergence term.

Coley and Lim  \cite{art:ColeyLim2014} demonstrated the spike phenomenon by using the Lema\^{i}tre-Tolman-Bondi (LTB) model. The LTB model is an exact solution that makes it easier to construct the spike. In this paper, they showed that spikes do not form in the matter density directly, it forms in the curvature as in \cite{art:ColeyLim2012}. They also explained that spike can provide an alternate contribution to  the formation of large scale structures in the Universe.

The OT $G_2$ spike solution contains permanent spikes, and there is a debate whether in the non-OT $G_2$ solution these permanent spikes are unresolved spike transitions  or are really permanent. In other words, would the yet undiscovered non-OT $G_2$ spike solution contain permanent spikes? Numerical evidence suggests that the permanent spikes are unresolved spike transitions.
Heinzle $et$ $al.$ \cite{art:HeinzleUgglaLim2012} described how BKL and spike oscillations arise from concatenations of exact solutions, suggesting the existence of hidden symmetries and showing that the results of BKL are part of a greater picture. Woei Chet Lim used Geroch's transformation\footnote{In 1971-72, Geroch wrote two papers on generating  new exact solutions of Einsteins field equations by using a transformation \cite{art:Geroch1971,art:Geroch1972}. The transformation acts on a vacuum solution of Einstein field equations that possesses at least one KVF. This KVF remains preserved in the transformed solution. The stiff fluid version of Geroch transformation was given by Stephani in 1988 \cite{art:Stephani1988}.} to discover the non-OT $G_2$ spike solution in 2015. He applied the transformation to a Kasner seed solution, with a generic
linear combination of KVFs \cite{art:Lim2015}. He showed that non-OT $G_2$ spike solution always resolves its spike as opposed to previous OT $G_2$ solutions. This method shows a new way to generate various kinds of spikes.

The above advancements were made in the particular case where spikes vary in only one direction. Spikes that vary in two or three directions are much more complex; sheets of spikes can intersect each other and interact. Similarly, in non-vacuum models, sheets of overdensity in the fluid can intersect in filaments and points to make even more pronounced overdensity in the fluid -- a web of large scale structures form. The space between the sheets are filled with underdensed fluid, and the underdensity
becomes more pronounced -- voids form. The search for these complex structure led researchers to apply the Stephani transformation on various seed solutions. Coley and Lim generalised the non-OT $G_2$ vacuum spike solution by applying Stephani transformation on Jacobs solution \cite{art:ColeyLim2016}. 

Coley $et$ $al.$ \cite{art:Coleyetal2016, art:Gregorisetal2017} found the first exact spike solution in which two spikes intersect. They applied the Stephani transformation on a family of Bianchi type V solution. These are the first $G_1$ stiff fluid spike solutions. In the generated solution, they observed some interesting phenomena at early times. They discussed many cases and some of them have permanent spike. But the most interesting one is the intersecting spike. This is the first exact spike solution that has an intersecting spike. In this case the intersecting planes are $Y = 0$ and $Z = 0$. Intersecting spikes epitomise a prototypical intersecting wall. The density is higher on the walls but highest at the intersection. We note that the Universe is dominated by bubbles of large voids surrounded by denser walls \cite{art:ColeyWiltshire2017}. The existence of non-linear structures at early times in the universe may support the large scale observational anomaly \cite{art:Dolgov2016}. Another interesting result in one of the cases of this paper is two phenomena at the same time. i.e. it has spike crossing at early times with a close-to-FL background. 
These spikes form at early times.

\section{Goals}
With the exception of the LTB models \cite{art:ColeyLim2014}, the exact solutions which are discussed in the previous section have spikes at early times. LTB models are silent\footnote{ In silent model  there is no exchange of information between different fluid element either by sound waves (p=0) or gravitational waves ($H_{ab}=0$) \cite[Chapter 13]{book:WE}.}, so our main goal is to find spikes at late times in non-silent models. For this we will use seed solutions that have a rotational KVF. Before this no one used the rotational KVF in a transformation. i.e. all the KVFs used are translational. We will apply the Stephani transformation on the LRS Jacobs solution. Our second goal is to develop a new technique to carry out intermediate time analysis of inhomogeneous structures.

\section{Overview}
The thesis consists of two main results. The first is the generation of new spiky solutions and their properties using existing method of analysis (Chapter 3-6). The second is the development of a new method of analysis, and its application to the spiky solution (Chapter 7).

In Chapter 2, we review some background material. We write the general metric in Iwasawa frame variables. We describe the Geroch/Stephani transformation with a KVF  adopted to the Iwasawa frame, and give the formulas for the transformation of the Iwasawa frame variables. 

In Chapter 3, we choose a seed solution, the LRS Jacobs solution and set it up for the Geroch/Stephani transformation. The linear combination of KVFs introduces a parameter $k$. The solution is cast in cylindrical coordinates, and we take note of its false spikes.

In Chapters 4 and 5, we apply the Stephani transformation to the seed solution for the cases $k=0$ and $k\neq 0$ respectively. We analyse the dynamics of the generated solution at early and late times\footnote{Early time means  the time towards Big bang and late times means the time away from Big bang.}. The case $k=0$ has a late-time permanent spike forming along the rotational axis. The case $k\neq0$ has  both true and false spikes.

We develop a heuristic for permanent spike in Chapter 6, for an arbitrary metric. We define a general way of finding a permanent spike and compare it to previous results.

In Chapter 7 we develop a new technique to explore the dynamics of the $k \neq 0$ case and revise the description of transient spikes. We also discover and describe the overshoot transition.

In the concluding Chapter 8 we summarise the new results in this thesis,
and remark on future research.

\chapter{Background material}
Einstein's field equations (EFEs) are 16 coupled nonlinear partial differential
equations relating a set of symmetric tensors that explain the gravitational
effects. In general relativity these gravitational effects are produced by a
given mass distribution. These field equations were presented by Einstein in 1915.
Mathematically they are written as
\begin{equation}
\label{a3a}
R_{a b}-\frac{1}{2}g_{	a	b}R+g_{a b}\Lambda=\frac{8\pi G}{c^{4}}T_{a b}. 
\end{equation}
where $g_{a b}$\ is the metric tensor, $R_{a b}$ is the Ricci curvature
tensor, $R$ is the Ricci curvature scalar, $\Lambda$ is the
cosmological constant, and $T_{a b}$ is the
stress--energy tensor. Because of the symmetry of $T_{a b}$ and $R_{a b}$, the genuine number of
equations decreases to $10$, while there are Bianchi identities (four
differential identities) satisfied by $R_{a b}$ that are one for each
coordinate, so it reduces the number of independent equation's to $6$.  Einstein felt cosmological constant desirable at that time but Hubble's
observation of the expansion of the universe made him reject the cosmological constant. But,
recent astronomical observations suggest it strongly and consider that it is
small but not zero but we shall set $\Lambda=0$. Also, in
gravitational units we take $8\pi G=c=1$. So equation $\left(  \ref{a3a}%
\right)  $ becomes%

\begin{equation}
R_{a b}-\frac{1}{2}g_{
	a
	b}R=T_{a b}. \label{A5}
\end{equation}
In $\left(  \ref{a3a}\right)  ,$ the Ricci curvature tensor is obtained by
contracting the Riemann curvature tensor. So first we have to write Riemann curvature tensor, which is
\begin{equation}
R_{abk}^{l}=\frac{\partial}{\partial x^{b}}\Gamma_{ak}^{l}-\frac{\partial
}{\partial x^{k}}\Gamma_{ab}^{l}+\Gamma_{bs}^{l}\Gamma_{ak}^{s}-\Gamma
_{ks}^{l}\Gamma_{ab}^{s}, \label{A6}
\end{equation}
where
\begin{equation}
\Gamma_{bs}^{l}=\frac{1}{2}g^{lo}\left(  \frac{\partial}{\partial x^{s}}
g_{ob}+\frac{\partial}{\partial x^{b}}g_{os}-\frac{\partial}{\partial x^{o}
}g_{bs}\right)  . \label{A7}
\end{equation}
are the Christoffel symbols.\newline So Ricci curvature tensor is
\begin{equation}
R_{ab}=R_{alb}^{l}=\frac{\partial}{\partial x^{l}}\Gamma_{ab}^{l}
-\frac{\partial}{\partial x^{b}}\Gamma_{al}^{l}+\Gamma_{ab}^{l}\Gamma_{ls}
^{s}-\Gamma_{al}^{s}\Gamma_{bs}^{l}, \label{A8}
\end{equation}
and Ricci curvature scalar is
\begin{equation}
R=g^{ib}R_{ib}. \label{A9}
\end{equation}
Equation $\left(  \ref{A5}\right)  $ is also written as
\begin{equation}
G_{a b}=8\pi T_{a b}. \label{A11}
\end{equation}
where
\begin{equation}
G_{a b}=R_{a \nu}-\frac{1}{2}g_{	a	b}R \label{A10}
\end{equation}
is the Einstein tensor.
The stress--energy tensor $T_{a b}$ for a perfect fluid  with respect to timelike vector field $\mathbf{u}$
\begin{equation}
\label{a3EMT}
T_{ab} = \rho u_au_b + p(g_{ab} + u_au_b)
\end{equation}
with energy density $ \rho > 0$, pressure $p$ and (unit timelike) fluid 4-vector $\mathbf{u}$. We assume that the equation of state of the perfect fluid is of the form
$p = (\gamma -1)\rho$, where $0 \leq \gamma \leq 2$ is constant. The cases $\gamma = 1$ (dust) and $\gamma =4/3$ (radiation) are of primary physical interest. The cosmological constant $\Lambda$ in
the EFEs can be treated as a perfect fluid with $\rho=\Lambda$ and $p = -\Lambda$, i.e $\gamma=0$; and the value $\gamma=2$ corresponds to a stiff fluid.

A cosmological model $(\mathcal{M}, \mathbf{g},\mathbf{u})$ is determined by a Lorentzian metric $\mathbf{g}$ defined on a manifold $\mathcal{M}$, and a family of fundamental observers, whose congruence of worldlines is represented by the unit timelike vector field $\mathbf{u}$, which we often identify with the matter 4-velocity. The dynamics of the model is governed by EFEs (\ref{a3a}) with suitable matter content (\ref{a3EMT}). It is helpful to classify cosmological solutions of the EFEs using the dimension of orbits of the symmetry group admitted by the metric (see Appendix \ref{App:KVFsgroup_actions}). This classification scheme forms a hierarchy of cosmological models of increasing complexity (that can be found in Section 1.2.2 of \cite{book:WE}). In this thesis, we are interested in Bianchi cosmologies. 
\section{Bianchi cosmologies}
A Bianchi cosmological model $(\mathcal{M}, \mathbf{g},\mathbf{u})$ is a model whose metric admit a three-- dimensional group of isometries acting simply transitively on spacelike hypersurfaces, which are hypersurfaces of homogeneity in  spacetime. A Bianchi cosmology thus admits a Lie algebra of KVFs with basis $\pmb{\xi }_a$, $\alpha=1 ,2 ,3,$ and structure constants $C^\mu_{\alpha \beta}$: 
\begin{equation}
[\pmb\xi_\alpha,\pmb\xi_\beta]=C^\mu_{\alpha \beta}\pmb\xi_\mu
\end{equation}
The $\pmb{\xi }_a$ are tangent to the group orbits, which are called the hypersurface of homogeneity. 
The Bianchi cosmology can be classified by classifying the Lie algebras of KVFs, and hence the associated isometry\footnote{An isometry of a manifold $(\mathcal{M}, \mathbf{g})$ is a mapping of $\mathcal{M}$ into itself that leaves the metric $\mathbf{g}$ invariant.} of the group $G_3$. Bianchi cosmologies are classified \cite{art:EllisMacCallum1969} in Table \ref{tab:tabel1}.
\begin{table}[htb]
	\centering    
	\caption{Classification of Bianchi cosmologies.}
	\begin{tabular}{*{5}{c}}
		\toprule
		\multicolumn{3}{c}{Group type} & \multirow{1}{*}{Eigenvalues}\\
		Class A & Class B & \ \ \ \ \ \ &   of $n_{\alpha \beta}$ & \\
		\midrule
		IX & & \ \ \ \ \ \ + & + & +\\
		VIII & & \ \ \ \ \ \ + & + & $-$ \\
		VII$_0$ & VII$_h$ &  \ \ \ \ \ \ + & $-$ & 0 \\
		VI$_0$ & VI$_h$ & \ \ \  \ \ \ + & + & 0 \\
		II & IV & \ \ \  \ \ \ + & 0 & 0 \\
		I & V & \ \ \ \ \ \  0 & 0 & 0 \\
		\bottomrule
	\end{tabular}
	\label{tab:tabel1}
\end{table}
\section{1+3 Orthonormal frame formalism}
In the orthonormal frame approach one does not use the metric $g$ directly
(as done in the metric approach), but chooses at each point of the spacetime
manifold $(\mathcal{M}, \mathbf{g})$ a set of four linearly independent 1-forms $\{\pmb\omega^a\}$ such that
the line element can be locally expressed as 
\begin{equation}
ds^2 = \eta_{ab}\pmb\omega^a\pmb\omega^b,
\end{equation}
where $\eta_{ab} =diag(-1, 1, 1, 1)$. The corresponding vector fields {$\mathbf{e}_a$} are then mutually
orthogonal and of unit length -- they form an orthonormal basis, with $\mathbf{e}_0$
being timelike (and thus defining a timelike congruence).
The gravitational field is described by the commutation functions $\gamma^c_{ab}$ of the orthonormal frame, defined by
\begin{equation}
[\mathbf{e}_a, \mathbf{e}_b] = \gamma^c_{ab}\mathbf{e}_c.
\end{equation}
The first step is to perform a 1+3 decomposition of the commutation functions as follows:
\begin{align}
\label{1+3d1}
[\mathbf{e}_a, \mathbf{e}_b] = \dot{u}_\alpha \mathbf{e}_0 -[H \delta_\alpha {}^\beta+\sigma_\alpha {}^\beta-\epsilon_\alpha {}^{\beta \gamma}(\omega_\gamma-\Omega_\gamma)]\mathbf{e}_\beta,
\end{align}
\begin{align}
\label{1+3d2}
[\mathbf{e}_a, \mathbf{e}_b] = -2\epsilon_{\alpha\beta} {}^\mu \omega_\mu \mathbf{e}_0+[\epsilon_{\alpha\beta \nu}\eta^{\mu \nu}+a_\alpha \delta_\beta{}^\mu-a_\beta \delta_\alpha{}^\mu]\mathbf{e}_\mu 
\end{align}
The variables in (\ref{1+3d1}) and (\ref{1+3d2}) have physical or geometrical meanings, as follows. The variable $H$ is the Hubble scalar, $\sigma_{\alpha\beta}$ the rate of shear tensor, $\dot{u}_\alpha$ the acceleration vector, and $\omega_\alpha$ the rate of vorticity vector of the timelike congruence defined by $\mathbf{e}_0$, while $\Omega_\alpha$ is the angular velocity of the spatial frame ${\mathbf{e}_\alpha}$ with respect to a nonrotating frame ($\Omega_\alpha= 0$). 
We shall thus refer to $n_{\alpha\beta}$ and  $a_{\alpha}$ as the spatial curvature variables. Collectively, the variables above describe the
gravitational field. We shall refer to them as the gravitational field variables,
and denote them by the state vector
\begin{align}
\label{1+3d3}
\mathbf{X}_{grav} = (H,\sigma_{\alpha\beta}, \dot{u}_\alpha,\Omega_\alpha,n_{\alpha\beta}, a_{\alpha}),
\end{align}
The matter content of a cosmological model is described by the stress energy tensor $T_{ab}$, which is decomposed into irreducible parts with respect
to $\mathbf{e}_0$ in the following way (let $\textbf{e}_0 = \textbf{u}$ below):
\begin{equation}
\label{1+3d4}
T_{ab} = \rho u_au_b + 2q_{(a}u_{b)}+ph_{ab} + \pi_{ab},
\end{equation}
where
\begin{equation}
\label{1+3d5}
q_{a}u^{b} =0, \ \ \ \pi_{ab}u^b=0,\ \ \ \pi_{a}^{a}=0,\ \ \ \pi_{ab}=\pi_{ba},
\end{equation}
and $h_{ab} = g_{ab} + u_au_b$ is the projection tensor which locally projects into the
3-space orthogonal to $\textbf{u}$. Since we are using an orthonormal frame, we have
$g_{ab}=\eta_{ab}$, $u^a = (1, 0, 0, 0)$, and $q_0 = 0, \pi_{0a}=0$. The variables $(\rho, p, q_\alpha, \pi_{\alpha \beta})$ have physical meanings: $\rho$ is the energy density, $p$ is the (isotropic) pressure,
$q_\alpha$ is the energy flux density and $\pi_{\alpha \beta}$ is the anisotropic pressure (see, for
example, van Elst  Uggla 1997 [73, page 2677]). We shall refer to these
variables as the matter variables, and denote them by the state vector
\begin{align}
\label{1+3d6}
\mathbf{X}_{matter} = (\rho, q_\alpha, p, \pi_{\alpha \beta})
\end{align}
The dynamics of the variables in  (\ref{1+3d3}) and (\ref{1+3d6}) is described by the EFEs,
the Jacobi identities (using $\textbf{e}_a$) and the contracted Bianchi identities
respectively. The evolution of $p$ and $\pi_{\alpha \beta}$ has to be specified by giving an
equation of state for the matter content (e.g. perfect fluid). The variables $\dot{u}_\alpha$
and $\Omega_\alpha$ correspond to the temporal and spatial gauge freedom respectively. 
The variables in (\ref{1+3d3}) and (\ref{1+3d6}) are scale-dependent and dimensional,
and are unsuitable for describing the asymptotic behaviour of cosmological models near the initial singularity, since they typically diverge. It is
thus essential to introduce scale-invariant (dimensionless) variables, which one hopes will be bounded as the initial singularity is approached. So we use  the Hubble-normalised gravitational and matter variables respectively as follows:
\begin{align}
\label{1+3d7}
(\Sigma_{\alpha\beta}, \dot{U}_\alpha,R_\alpha,N_{\alpha\beta}, A_{\alpha}) &= (\sigma_{\alpha\beta}, \dot{u}_\alpha,\Omega_\alpha,n_{\alpha\beta}, a_{\alpha})/H,\\
(\Omega, Q_\alpha, P, \Pi_{\alpha \beta},\Omega_{\Lambda})&= (\rho, q_\alpha, p, \pi_{\alpha \beta},\Lambda)/(3H^2).
\end{align}
\section{The Iwasawa frame}
Assume zero vorticity (zero shift). The spatial metric components are given by the formula $g_{\alpha \beta} = e_a{}^i e_b{}^j \delta_{ij}$, where $\alpha$, $\beta$, $i$, $j$ = $1,2,3$.
The Iwasawa frame is a choice of orthonormal frame that makes $e^\alpha{}_i$ (and equivalently $e_\alpha{}^i$) upper triangular, as follows.
The frame coefficients $e^\alpha{}_i$ simplify from 9 components to 6 components, represented by $b^1$, $b^2$, $b^3$, $n_1$, $n_2$ and $n_3$.

\begin{align}
e^\alpha{}_i = \left( \begin{array}{ccc}
e^1{}_1 & e^1{}_2 & e^1{}_3 \\	
e^2{}_1	& e^2{}_2 & e^2{}_3 \\
e^3{}_1	& e^3{}_2 & e^3{}_3
\end{array} \right)
&= \left( \begin{array}{ccc}
e^{-b^1} & 0 & 0 \\
0 & e^{-b^2} & 0 \\
0 & 0 & e^{-b^3}
\end{array} \right)
\left( \begin{array}{ccc}
1 & n_1 & n_2 \\
0 & 1   & n_3 \\
0 & 0   & 1
\end{array} \right)
\notag\\
&= \left( \begin{array}{ccc}
e^{-b^1} & e^{-b^1} n_1 & e^{-b^1} n_2 \\
0 & e^{-b^2} & e^{-b^2} n_3 \\
0 & 0 & e^{-b^3}
\end{array} \right)
\end{align}
\begin{align}
e_\alpha{}^i = \left( \begin{array}{ccc}
e_1{}^1 & e_2{}^1 & e_3{}^1 \\
e_1{}^2 & e_2{}^2 & e_3{}^2 \\
e_1{}^3 & e_2{}^3 & e_3{}^3   
\end{array} \right)
&= \left( \begin{array}{ccc} 
1 & -n_1	& n_1 n_3 - n_2 \\
0 & 1   & -n_3 \\
0 & 0  	& 1
\end{array} \right)
\left( \begin{array}{ccc}
e^{b^1} & 0 & 0 \\
0 & e^{b^2} &	0 \\
0 & 0 &	e^{b^3}
\end{array} \right)
\notag\\
&= \left( \begin{array}{ccc}
e^{b^1} & -e^{b^2} n_1 & e^{b^3} (n_1 n_3 - n_2) \\
0 & e^{b^2} & -e^{b^3} n_3 \\
0 & 0 & e^{b^3}
\end{array} \right)
\end{align}

The frame derivative operators $\mathbf{e}_0 = N^{-1} \partial_0$, $\mathbf{e}_\alpha = e_\alpha{}^i \partial_i$ in the Iwasawa frame are \cite{art:HeinzleUgglaRohr2009}
\begin{align}
\mathbf{e}_0 &= \frac{1}{N} \partial_0
\\
\mathbf{e}_1 &= e^{b^1} \partial_1
\\
\mathbf{e}_2 &= e^{b^2} [ -n_1 \partial_1 + \partial_2 ]
\\
\mathbf{e}_3 &= e^{b^3} [ (n_1 n_3 - n_2) \partial_1 - n_3 \partial_2 + \partial_3 ].
\end{align}

\section{The metric}

In the Iwasawa frame, the metric components in terms of the $b$'s and $n$'s are given by
\begin{align}
g_{00} &= -N^2
\\
g_{11} &= e^{-2b^1},\quad g_{12} = e^{-2b^1} n_1,\quad g_{13} = e^{-2b^1} n_2
\\
g_{22} &= e^{-2b^2} + e^{-2b^1} n_1^2,\quad g_{23} =  e^{-2b^1} n_1 n_2 + e^{-2b^2} n_3
\\
g_{33} &= e^{-2b^3} + e^{-2b^1} n_2^2 + e^{-2b^2} n_3^2.
\end{align}

If the metric is given, we can compute the $b$'s and $n$'s as follows.
\begin{align}
b^1 &= -\tfrac12 \ln g_{11}
\\
n_1 &= \frac{g_{12}}{g_{11}}
\\
n_2 &= \frac{g_{13}}{g_{11}}
\\
b^2 &= -\tfrac12 \ln ( g_{22} - g_{12} n_1 )
\\
n_3 &= (g_{23}-g_{12} n_2) e^{2b^2}
\\
b^3 &= -\tfrac12 \ln ( g_{33} - g_{13} n_2 - e^{-2b^2} n_3^2 ).
\end{align}
The determinant $g$ of the  metric satisfies
\be
\sqrt{-g} = N e^{-b^1-b^2-b^3}.
\ee

\section{The Geroch/Stephani transformation}

Consider a solution $g_{ab}$ of the vacuum Einstein's field equations with a KVF $\xi^a$. 
The Geroch transformation~\cite{art:Geroch1971,art:Geroch1972} (see also \cite[Section 10.3]{book:Exactsol2002}) is an algorithm for generating new solutions, by exploiting the KVF $\xi^a$.
The algorithm involves solving the following PDEs
\begin{gather}
\label{omega_PDE}
\nabla_a \omega =\varepsilon_{abcd}\xi ^b\nabla^c \xi^d,
\\
\label{alpha_PDE}
\nabla_{[a}\alpha_{b]} =\frac{1}{2}\varepsilon_{abcd} \nabla^c \xi^d,\quad
\xi^a \alpha_a =\omega,
\\
\label{beta_PDE}
\nabla_{[a}\beta_{b]}=2\lambda \nabla_a \xi_b + \omega \varepsilon_{abcd}  \nabla^c \xi^d,
\quad
\xi^a \beta_a =\lambda^2 +\omega^2 -1
\end{gather}
for $\omega$, $\alpha_a$ and $\beta_a$, where $\lambda=\xi^a \xi_a$, $\nabla_{a}$ is the covariant derivative and $\varepsilon_{abcd}$  is the totally antisymmetric permutation tensor, with 
$\varepsilon^{0123}=\frac{1}{\sqrt{-g}}$ \cite{book:WE}.

Next, define $\tilde{\lambda}$ and $\eta_a$ as
\begin{align}
\tilde{\lambda}&=\lambda \Big[(\cos\theta-\omega\sin\theta)^2 +\lambda^2 \sin^2\theta\Big]^{-1},
\\
\label{eta}
\eta_a &=\tilde{\lambda}^{-1} \xi_a +2 \alpha_a \cos\theta\sin\theta-\beta_a \sin^2\theta,
\end{align} 
for any constant $\theta$. 
Then the new metric is given by
\be
\tilde{g}_{a b}=\frac{\lambda}{\tilde{\lambda}}(g_{a b}-\lambda^{-1} \xi_a\xi_b)+\tilde{\lambda} \eta_a \eta_b.
\ee
This new metric is again a solution of the vacuum Einstein's field equations with the same KVF.
$\theta=0$ gives the trivial transformation $\bar{g}_{a b} = g_{ab}$.

Notice from (\ref{eta}) that $\alpha_a$ appears in the new metric only through $\eta_a$, and if $\theta$ is chosen to be $\pi/2$ then $\alpha_a$ does not appear at all. 
We shall exploit this simplification. In this case the new metric simplifies to
\be
\tilde{g}_{a b}= (\lambda^2+\omega^2) g_{a b} + \frac{\lambda}{\lambda^2+\omega^2}\beta_a \beta_b - \xi_a \beta_b - \beta_a \xi_b.
\ee
Stephani~\cite{art:Stephani1988} generalised the Geroch transformation to the case of comoving stiff fluid
if the KVF is spacelike (and to the case of perfect fluid with equation of state $p=-\rho/3$ if the KVF is timelike, which we do not study here). The algorithm is the same as before, with the new stiff fluid density given by
\be
\tilde{\rho} = \frac{\rho}{\lambda^2+\omega^2}.
\ee

\section{Geroch/Stephani transformation adapted to the Iwasawa frame}

Before applying the Geroch transformation or the Stephani transformation, we set up the coordinates such that the KVF to be used has the form
\begin{equation}
\xi^a = (0,\ 1,\ 0,\ 0),
\end{equation}
to adapt to the Iwasawa frame for simplicity.
If the seed metric in these coordinates has the general form
\begin{equation}
g_{ab} = \begin{bmatrix}
-N^2	&	0	&	0	&	0	\\
0	&	g_{11}	&	g_{12}	&	g_{13}	\\
0	&	g_{12}	&	g_{22}	&	g_{23}	\\
0	&	g_{13}	&	g_{23}	&	g_{33}
\end{bmatrix},
\end{equation}
then the generated metric has the form
\begin{equation}
\lb{g_tilde}
\tilde{g}_{ab} = \begin{bmatrix}
-FN^2    &       0       &       0       &       0       \\
0       &       \tilde{\lambda}  &       g_{12}-\beta_2 \tilde{\lambda}  &       g_{13}-\beta_3 \tilde{\lambda}  \\
0       &       \tilde{g}_{12}  &       Fg_{22}-2g_{12}\beta_2+\beta_2^2\tilde{\lambda}  &       Fg_{23}-g_{12}\beta_3-g_{13}\beta_2+\beta_2\beta_3\tilde{\lambda}  \\
0       &       \tilde{g}_{13}  &       \tilde{g}_{23}  &       Fg_{33}-2g_{13}\beta_3+\beta_3^2\tilde{\lambda}  
\end{bmatrix},
\end{equation}
where
\begin{equation}
\tilde{\lambda} = \frac{\lambda}{F},\quad	F = \omega^2 + \lambda^2,\quad \lambda =\xi^a\xi_a = g_{11} = e^{-2b^1}.
\end{equation}

The twist of the KVF, $\omega$ has gradient
\be
\omega_{a}=(\omega_0,\omega_1,\omega_2,\omega_3),
\ee
whose components satisfy
\begin{align}
\omega_0&=-N e^{b^2+b^3}\lambda^{3/2}\partial_3 n_{1}\\
\omega_1&=0\\
\omega_2&=-N^{-1}e^{-b^2+b^3}\lambda^{3/2}(n_3\partial_0 n_{1}-\partial_0 n_{2})\\
\omega_3&=-N^{-1}e^{b^{2}+b^{3}}\lambda^{3/2}(e^{-2b^3}\partial_0 n_{1} + e^{-2b^2}n_3^2\partial_0 n_{1}-e^{-2b^2}n_3\partial_0 n_{2}).
\end{align}
The covector
\be
\beta_a = (0,F-1,\beta_2,\beta_3)
\ee
satisfies the following partial differential equations:
\begin{align}
\partial_0\beta_{2}&=2\lambda n_{1}\partial_0\lambda +2\lambda^2\partial_0 n_{1} +2\omega \lambda^{-1}e^{2b^3}\sqrt{-g}\partial_3\lambda-2\omega N^2 \lambda^2 (\sqrt{-g})^{-1}n_1\partial_3 n_1\\
\partial_0\beta_{3}&=2\lambda n_{2}\partial_0\lambda +2\lambda^2\partial_0n_{2}+2\omega \lambda^{-1}e^{2b^3}\sqrt{-g}n_3\partial_3\lambda-2\omega N^2 \lambda^2 (\sqrt{-g})^{-1}n_2\partial_3n_1\\
\partial_3\beta_{2}-\partial_2\beta_{3}&=2\lambda n_{1}\partial_3\lambda +2\lambda^2\partial_3 n_{1}+2\omega \lambda^{-1}N^{-2}\sqrt{-g}\partial_0\lambda-2\omega e^{-2b^3} \lambda^2 (\sqrt{-g})^{-1}n_1\partial_0n_1 \notag\\
&\quad+N^{-1}e^{-b^2+b^3}\omega\lambda^{3/2}[-n_1n_3^2\partial_0 n_{1}+n_2n_3\partial_0 n_{1}-n_1n_3\partial_0 n_{2}
+n_2\partial_0 n_{2}].
\end{align}

Expressing the metric $\tilde{g}_{ab}$ in (\ref{g_tilde}) in $\tilde{N}$, $\tilde{b}$'s and $\tilde{n}$'s gives
\begin{align}
\label{Geroch_bn_first}
\tilde{N}&=N\sqrt{F}\\
\tilde{b}^1&=b^1 + \frac12\ln F \\
\tilde{b}^2&=b^2-\frac12\ln F \\
\tilde{b}^3&=b^3-\frac12\ln F \\
\tilde{n}_{1}&=n_{1}F-\beta_{2}\\
\tilde{n}_{2}&=n_{2}F-\beta_{3}\\
\label{Geroch_bn_last}
\tilde{n}_{3}&=n_{3}.
\end{align}


In simpler cases, if the seed metric has the form

\begin{equation}
\label{gmetric1}
g_{ab} = \begin{bmatrix}
-N^2    &       0       &       0       &       0       \\
0       &       g_{11}  &       g_{12}  &       0  \\
0       &       g_{12}  &       g_{22}  &       0  \\
0       &       0	&       0	&       g_{33}  
\end{bmatrix},
\end{equation}
i.e. if $n_2=0=n_3$,
then the generated metric has the form
\begin{equation}
\label{gmetric2}
\tilde{g}_{ab} = \begin{bmatrix}
-FN^2    &       0       	&       0       &       0       \\
0       &       \tilde{\lambda}  &       g_{12}-\beta_2 \tilde{\lambda}  &       0  \\
0       &       \tilde{g}_{12}  &       Fg_{22}-2g_{12}\beta_2+\beta_2^2\tilde{\lambda}  &       0  \\
0       &        0		&       0  &       Fg_{33} 
\end{bmatrix}.
\end{equation}
The twist of the KVF, $\omega$ has gradient
\begin{align}
\omega_{a}=(-N e^{b^{2}+b^{3}}\lambda^{3/2}{\partial_3 n_{1}},0,0,-N^{-1} e^{b^2-b^3}\lambda^{3/2}{\partial_0 n_{1}}), 
\end{align}
The covector
\be
\beta_a = (0,F-1,\beta_2,0)
\ee
satisfies the following partial differential equations:
\begin{align}
\partial_0\beta_{2}&=2\lambda n_{1}\partial_0\lambda +2\lambda^2\partial_0 n_{1} +2\omega \lambda^{-1}e^{2b^3}\sqrt{-g}\partial_3\lambda-2\omega N^2 \lambda^2 (\sqrt{-g})^{-1}n_1\partial_3 n_1\\
\partial_{3}\beta_{2}&=2\lambda n_{1}\partial_{3}\lambda +2\lambda^2\partial_{3}n_{1}+2\omega \lambda^{-1}N^{-2}\sqrt{-g}\partial_{0}\lambda-2\omega e^{-2b^3} \lambda^2 (\sqrt{-g})^{-1}n_1\partial_{0}n_1,
\end{align}

Expressing the metric $\tilde{g}_{ab}$ in (\ref{g_tilde}) in $b$'s and $n$'s gives
\begin{align}
\label{Geroch_bn_kzero_first}
\tilde{N}&=N\sqrt{F}\\
\tilde{b}^1&=b^1+\frac12\ln F \\
\tilde{b}^2&=b^2-\frac12\ln F \\
\tilde{b}^3&=b^3-\frac12\ln F \\
\tilde{n}_{1}&=n_{1}F-\beta_{2}\\
\tilde{n}_{2}&=0\\
\label{Geroch_bn_kzero_last}
\tilde{n}_{3}&=0.
\end{align}
\chapter{The seed solution}
As discussed in the introduction, past applications of Geroch/Stephani transformation used translational KVFs\footnote{We had Cartesian coordinates mostly and the KVFs are translational.}. In this thesis, we will use rotational KVFs\footnote{A rotational KVF is a KVF that is present in axis-symmetric solutions. The length of a rotational KVF vanishes at the axis of rotation.}. Stephani transformation requires the matter to be a stiff fluid, so we start by looking at locally rotationally symmetric (LRS) solutions \cite[page 22]{book:WainwrightEllis1997} with a stiff fluid. The simplest such solution is the flat FLRW solutions \cite[page 53]{book:WainwrightEllis1997}, but it does not generate as much structure as the next simplest solution, the LRS Jacobs (Bianchi type I) solution, which we shall use as the seed solution.

The Jacobs solution \cite[page 189]{book:WainwrightEllis1997} is given by the line element 
\be
ds^2=-dt^2+t^{2p_{1}}dx^2+t^{2p_{2}}dy^2+t^{2p_{3}}dz^2,
\ee
where the coordinates are $(t,x,y,z)$,
and
\begin{align}
p_1 &= \frac13(1+\Spo + \sqrt3\Sno),\\
p_2 &= \frac13(1+\Spo - \sqrt3\Sno),\\
p_3 &= \frac13(1-2\Spo).
\end{align}
The non-zero Hubble-normalised shear components are $\Spo$ and $\Sno$, and they are constant, with $\Spo^2+\Sno^2\leq1$.
The stiff fluid has pressure $p$ and density $\rho$ given by
\be
p = \rho = \frac{1-\Spo^2-\Sno^2}{3t^2}.
\ee

To impose the LRS condition; it is simplest to set $\Sno=0$, so the parameter $\Spo$ takes values from $-1$ to $1$.
$\Spo=-1$ gives the LRS Kasner solution \cite[page 132]{book:WainwrightEllis1997} with $(p_1,p_2,p_3)=(0,0,1)$ (also known as the Taub form of flat spacetime);
$\Spo=1$ gives the LRS Kasner solution with $(p_1,p_2,p_3)=(\frac23,\frac23,-\frac13)$;
$\Spo=0$ gives the flat FLRW solution with stiff fluid.

The LRS Jacobs solution admits four KVFs, namely
\be
\partial_x,\quad \partial_y,\quad \partial_z,\quad -y \partial_x + x \partial_y,
\ee
where the fourth one is rotational.\\
We intend to apply the Stephani transformation with the general linear combination of the KVFs:
\be
c_1 \partial_x + c_2 \partial_y + c_3 \partial_z + c_4(-y\partial_x + x\partial_y)
= (c_1-c_4 y) \partial_x + (c_2 + c_4 x)\partial_y + c_3 \partial_z
\ee

Observe that $c_1$ and $c_2$ can be eliminated without loss of generality by a translation in $x$ and $y$ directions. We set $c_4=1$ and $c_3=k$, so the KVF reads
\be
-y \partial_x + x \partial_y + k \partial_z.
\ee
This KVF forms an Abelian OT $G_2$ group with exactly one other KVF (namely a linear combination of $\partial_z$ and $-y \partial_x + x \partial_y$). By Geroch's theorem~\cite[Appendix B]{art:Geroch1972}, the generated metric will admit an Abelian OT $G_2$ group.

There is a rotational symmetry about the $z$-axis, so we adopt cylindrical coordinates $(r,\psi,z)$, but we want to arrange the coordinates in the following order: $(t,\psi,z,r)$, due to the way we adapt the orthonormal frame to the coordinates.
In these coordinates, the KVF reads
\be
\partial_\psi + k \partial_z.
\ee
We want to simplify the KVF to just $\partial_\psi$ for the application of the Stephani transformation, so we make a further change of coordinates,
by introducing
\be
Z = z - k\psi.
\ee
Then, in the coordinates $(t,\psi,Z,r)$, the KVF is simply $\partial_\psi$, but
the line element now reads
\begin{align}
\label{seed1}
ds^2 &= -dt^2+(k^2t^{2p_3}+r^2t^{2p_1})d\psi^2+2kt^{2p_3}d\psi dZ
+t^{2p_3}dZ^2+t^{2p_1}dr^2.
\end{align}
This shall be the seed solution to which we apply the Stephani transformation. It has the simple form (\ref{gmetric1}).
For later convenience we list the $b$'s and $n$'s of this line element and define the  squared norm of the KVF below.
\be
\lambda  = k^2t^{2p_3}+r^2t^{2p_1},
\ee
\begin{align}
N &= 1 \\
b^1 &= -\frac12\ln\lambda\\
b^2 &=-\frac12 \ln \dfrac{r^2 t^{2p_1+2p_3}}{\lambda} \\
b^3 &= -\frac12\ln (t^{2p_1})\\
n_1 &= \frac{kt^{2p_3}}{\lambda} \\
n_2 &= 0 = n_3.
\end{align}
Observe that
\be
b^1+b^2+b^3 = -\ln (rt).
\ee
The $\psi$-$Z$ area element
\be
\mathcal{A}=e^{-b^1-b^2}=rt^{p_1+p_3}
\ee
and volume element
\be
\mathcal{V}=e^{-b^1-b^2-b^3}=rt
\ee
are always expanding.

\begin{figure}
	\begin{center}
		\includegraphics[width=8cm]{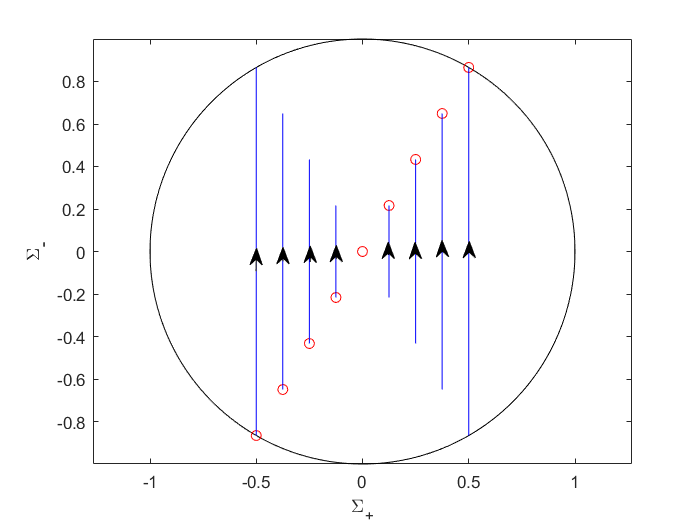}	
	\end{center}
	\caption{State space orbits projected on the $(\Sp,\Sm)$ plane for various values of $\Spo$. A circle represents the orbit along $r=0$, which is a fixed point. At $t$ increases, $r\neq0$ orbits move away from these fixed points for $\Spo >0$, and towards these fixed points $\Spo <0$.}
	\label{fig_spsmbt}
\end{figure}

Here we list the dynamical variables of the seed solution using the formulas in Appendix~\ref{App:formulas}.
To write the expressions in a more compact form, we list several intermediate expressions, particularly the partial derivatives of the essential variable.
\begin{align}
\partial_0 \lambda &= 2p_3 k^2t^{2p_3-1}+ 2p_1 r^2t^{2p_1-1}\\
\partial_3 \lambda &= 2rt^{2p_1}
\end{align}
So the expressions are
\begin{align}
H &= \frac1{3 t} \\
\Theta_{11} &= \frac12 \left( \frac{\partial_0 \lambda}{\lambda} \right)\\
\Theta_{22} &=\frac12 \left( \frac{2p_1+2p_3}{t}-\frac{\partial_0 \lambda}{\lambda}\right)\\
\Theta_{33} &=\frac12 \left(\frac{2p_1}{t}\right)\\
\Theta_{12} &=\frac12\dfrac{kt^{-p_1+p_3}}{r} \left( -\frac{\partial_0 \lambda}{\lambda}+2\frac{p_3}{t}\right)\\
\dot{u}_3 &=0\\ 
n_{11} &=\dfrac{kt^{-2p_1+p_3}}{r} \left(-\frac{\partial_3 \lambda}{\lambda}\right) \\
n_{12} &= \frac12 t^{-p_1} \left(\frac{1}{r} -\frac{\partial_3 \lambda}{\lambda}\right) \\
a_3 &= -\frac12\dfrac{t^{-p_1}}{r}\\
\Sigma_{11}&= \frac32 \left( \frac{t\partial_0 \lambda}{\lambda} \right)-1\\
\Sigma_{22}&=\frac32 \left( {2p_1+2p_3}-\frac{t\partial_0 \lambda}{\lambda}\right)-1\\
\Sigma_{33}&=3p_1-1\\
\Sigma_{12}&=\frac32\dfrac{kt^{1-p_1+p_3}}{r} \left( -\frac{\partial_0\lambda}{\lambda}+2\frac{p_3}{t}\right).
\end{align}
Define Hubble-normalised expansion shear components $\Sp$ and $\Sm$ as
\begin{align}
\label{ESC1}
\Sp &=-\frac{1}{2}\Sigma_{33}\\
\label{ESC2}
\Sm &=\frac{\Sigma_{11}-\Sigma_{22}}{2\sqrt3},
\end{align}
which gives
\begin{align}
\Sp&=-\frac12 \Spo\\
\Sm&=\frac{\sqrt3}{2}\left(t(\ln\lambda)_t-\frac{2-\Spo}{3}\right).
\end{align}
Figure \ref{fig_spsmbt} shows that state space orbits projected on the $(\Sp,\Sm)$ plane for various values of $\Spo$. The $r=0$ orbits are fixed points.  $r\neq0$ orbits move away from these fixed points as $t$ increases but only in $\Sm$ direction.
The solution is undefined at $r=0$ if $k=0$ (coordinate singularity).
It is straightforward to analyse $l=t(\ln\lambda)_t$.
If $k=0$ then $l=2p_1$.
If $k\neq0$, then $l=2p_3$ at $r=0$.
For $r\neq0$ write
\be
\label{3.50}
l=p_3(1-\tau)+p_1(1+\tau), \quad\tau=\tanh(\Spo(\ln t)+\ln|k|-\ln r).
\ee
For $\Spo>0$, $l$ goes from $2p_3$ to $2p_1$ as $t$ goes from $0$ (early times) to $\infty$ (late times). For $\Spo<0$, $l$ goes from $2 p_1$ to $2p_3$. For $\Spo=0$, $l=\frac23$. So $l$ has a simple sigmoid transitional dynamics. It has a discontinuous limit along $r=0$ (at late times for $\Spo>0$, at early times for $\Spo<0$). We take this opportunity to introduce the transition time, which roughly means when the transition occurs. For $l$ we define the transition time to be $\tau=0$, that is when
\be
\Spo \ln t +\ln |k|-\ln r=0,
\ee
or equivalently when the two competing terms in $\lambda$ are equal. Solving for $t$ gives the transition time
\be
t=\left(\frac{r}{|k|}\right)^{-\frac1\Spo}.
\ee
From (\ref{3.50}), observe that $\Spo$ appears as the coefficient of $\ln t$ in the $\tanh$ function.
So smaller $|\Spo|$ leads to milder slope/longer transition duration.\\
To summarise, the seed solution (\ref{seed1}) has permanent false spikes in the case $k\neq0$:\\
\textbf{1}. At $r=0$ at late times for $0<\Spo\leq1$.\\
\textbf{2}. At $r=0$ at early times for $-1\leq\Spo<0$.\\
If $k=0$, the seed solution has no false spikes.

In the next two chapters we apply the Stephani transformation on this seed solution.

\chapter{Generated Solution, $k=0$} 
In this chapter, we apply Stephani transformation on the seed solution (\ref{seed1}) for the case $k=0$ (that is, using only the rotational KVF) and discuss the dynamics of the generated solution. Computation is carried out using Maple.
\section{Applying the Stephani transformation}

We now carry out the Stephani transformation with the rotational KVF $\partial_\psi$.
First we find the squared norm of the KVF,
\be
\lambda  = r^2t^{2p_1},
\ee
and the twist of the KVF,
\begin{align}
\omega &= \omega_{0}.
\end{align}
The combination $\lambda^2+\omega^2$ will appear frequently, so for brevity we introduce
\be
F = \lambda^2+\omega^2.
\ee

The next step is to find a particular solution $\beta_{a}$ for the constrained system
\be
\nabla_{[a}\beta_{b]} = 2 \lambda\nabla_{a}\xi_b + \omega\epsilon_{abcd}\nabla^c \xi^d,\quad{ \xi^{a}\beta_{a} =F-1}.
\ee
We obtain
\be
\beta_{a}=(0, F-1,\beta_2, 0),
\ee
where
\begin{align}
\beta_2 = 2p_1\omega_0r^2 +\frac{4\omega_0}{1+p_3}t^{1+p_3}
\end{align}
is found by integration.

The generated metric is then given through $b$'s and $n$'s by the formulas (\ref{Geroch_bn_kzero_first})--(\ref{Geroch_bn_kzero_last}), dropping tildes for brevity.
\begin{align}
N&=F^{1/2} \\
b^1&= -\frac12\ln \frac{\lambda}{F}\\
b^2&= -\frac12 \ln \dfrac{F r^2 t^{2p_1+2p_3}}{\lambda} \\
b^3&= -\frac12\ln (Ft^{2p_1})\\
n_1&= -\beta_2 \\
n_2&=0\\
n_3&=0.
\end{align}
Observe that
\be
b^1+b^2+b^3 = -\frac12\ln (F r^2 t^{4p_1+2p_3}) = -\frac12\ln (F r^2 t^2).
\ee

\section{The dynamical variables}

Here we list the dynamical variables of the generated solution using the formulas
in Appendix~\ref{App:formulas}.
To write the expressions in a more compact form, we list several intermediate expressions, particularly the partial derivatives of the essential variables.
\be
\partial_0 \lambda = 2p_1 r^2t^{2p_1-1}
\ee
\be
\partial_3 \lambda = 2rt^{2p_1}
\ee
\be
\partial_0 \omega =0
\ee
\be
\partial_3 \omega = 0
\ee
\be
\partial_0 F = 2\lambda \partial_0 \lambda + 2\omega\partial_0\omega
\ee
\be
\partial_3 F = 2\lambda \partial_3 \lambda + 2\omega\partial_3\omega
\ee
\be
\partial_0 \beta_2 = 4\omega t^{p_3}
\ee
\be
\partial_3 \beta_2 = 4{p_1}\omega r.
\ee
Now using the above convention, we get
\begin{align}
H &= \frac16 F^{-1/2}\left(\frac{\partial_0 F}{F} + \frac{2}{t}\right)\\
\Theta_{11} &= \frac12 F^{-1/2} \left( \frac{\partial_0 \lambda}{\lambda} - \frac{\partial_0 F}{F}\right)\\
\Theta_{22} &=\frac12 F^{-1/2} \left(\frac{\partial_0 F}{F} -\frac{\partial_0 \lambda}{\lambda} + \frac{2p_1+2p_3}{t}\right)\\
\Theta_{33} &=\frac12 F^{-1/2} \left(\frac{\partial_0 F}{F} +  \frac{2p_1}{t}\right)\\
\Theta_{12} &= \dfrac{\lambda t^{-p_1-p_3}}{2rF^{3/2}}{\partial_0 \beta_2}\\
\dot{u}_3 &=-\frac12 F^{-1/2}t^{-p_1} \left(\frac{\partial_3 F}{F}\right)\\
n_{11} &=- \dfrac{\lambda t^{-2p_1-p_3}}{rF^{3/2}}{\partial_3 \beta_2}\\
n_{12} &= \frac12{F^{-1/2}t^{-p_1}} \left( \dfrac{\partial_3 F}{F}-\frac{\partial_3 \lambda}{\lambda}+\frac{1}{r}\right) \\
a_3 &= -\frac12\dfrac{F^{-1/2}t^{-p_1}}{r}.
\end{align}
\section{Dynamics of the solution}
The state space orbits of a solution, projected onto the $(\Sp,\Sm)$ plane, can provide some insight into the dynamics of the solution. Recall that $(\Sp,\Sm)$  are defined in terms of the diagonal components of the Hubble-normalised expansion shear as
\begin{align}
\label{ESC1}
\Sp &=-\frac{1}{2}\Sigma_{33},\\
\label{ESC2}
\Sm &=\frac{\Sigma_{11}-\Sigma_{22}}{2\sqrt3}.
\end{align}
This reduces to
\be
\label{SpsmkzeroG}
\Sp=-\frac{\Spo+f}{2+f},\quad\Sm=\frac{\sqrt{3}(\Spo-f)}{2+f},
\ee
where 
\be
\label{fkzeroG}
f=t(\ln F)_t=\frac{4p_1r^4t^{4p_1}}{r^4t^{4p_1}+\omega_{0}^2},\ \ \  p_1=\frac13(1+\Spo).
\ee

We carry out the asymptotic analysis of $f$, similar to what we did for $l$ in Chapter 3.
\subsection{Case $\omega_{0}\neq 0$}
Along $r=0$, we have $f=0$ and
\begin{align}
(\Sp,\Sm)=(-\frac12\Spo,\frac{\sqrt{3}}{2}\Spo).
\end{align}
Along $r\neq0$,  provided that $p_1\neq0$, we have
\begin{align}
f\rightarrow 
\begin{cases}
0  &\text{as} \ \ \ t\rightarrow 0\\
4p_1  &\text{as} \ \ \ t\rightarrow \infty,
\end{cases}
\end{align}
and
\begin{align}
(\Sp,\Sm)\rightarrow
\begin{cases}
(-\frac{1}{2}\Spo,\frac{\sqrt{3}}{2}\Spo)&\text{as} \ \ \ t\rightarrow 0\\
(-\frac{(7\Spo+4)}{2(2\Spo+5)},-\frac{\sqrt3(4+\Spo)}{2(2\Spo+5)})&\text{as} \ \ \ t\rightarrow \infty.
\end{cases}
\end{align}
There is a permanent spike at $r=0$ at late times, for $p_1\neq0$ $\Leftrightarrow$ $\Spo>-1$.
\begin{figure}
	\begin{center}
		\includegraphics[width=8cm]{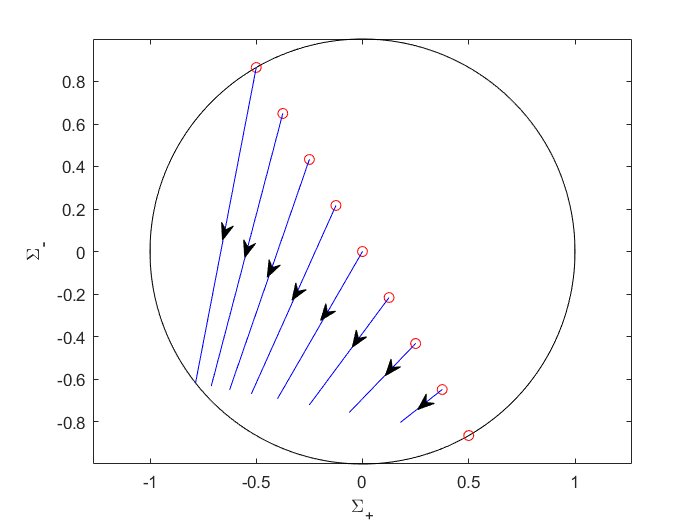}
	\end{center}
	\caption{State space orbits projected on the $(\Sp,\Sm)$ plane for various values of $\Spo$ for $\omega_0\neq0$. A circle represents the orbit along $r=0$, which is a fixed point. $r\neq0$ orbits move away from these fixed points as $t$ increases, mimicking the orbits of Taub solutions.}	
	\label{fig_spsmkn0}
\end{figure}
Figure \ref{fig_spsmkn0} shows the state space orbits projected on the $(\Sp,\Sm)$ plane for various values of $\Spo$ for $\omega_{0}\neq0$. The $r=0$ orbits are fixed points, while the $r\neq0$ orbits move away from these fixed points as $t$ increases, mimicking the orbits of Taub solutions \cite[page 136]{book:WainwrightEllis1997}.

The exceptional case is when $\Spo=-1$, where $(\Sp,\Sm)=(\frac12,-\frac{\sqrt{3}}{2})$ for all $t$ and $r$.
But this is not a Kasner solution. 
Recall that the timelike KVF is preserved. The Weyl scalars (see Appendix \ref{App:Weyl_S}) are time-independent :
\begin{align}
CC&=-\frac{(192(-r^{12}+15\omega_0^2r^8-15\omega_0^4r^4+\omega_0^6))}{(r^4+\omega_0^2)^6}\\
CCs&=-\frac{384\omega_0r^2(-r^4+3\omega_0^2)(-3r^4+\omega_0^2)}{(r^4+\omega_0^2)^6}\\
CCC&=-\frac{768r^2(-r^4+3\omega_0^2)(-r^{12}+33\omega_0^2r^8-27\omega_0^4r^4+3\omega_0^6)}{(r^4+\omega_0^2)^9}\\
CCCs&=\frac{768\omega_0(-3r^4+\omega_0^2)(-3r^{12}+27\omega_0^2r^8-33\omega_0^4r^4+\omega_0^6)}{(r^4+\omega_0^2)^9}.
\end{align}

\begin{figure}
	\begin{center}
		\begin{subfigure}[t]{7.5cm}
			\includegraphics[width=8.5cm]{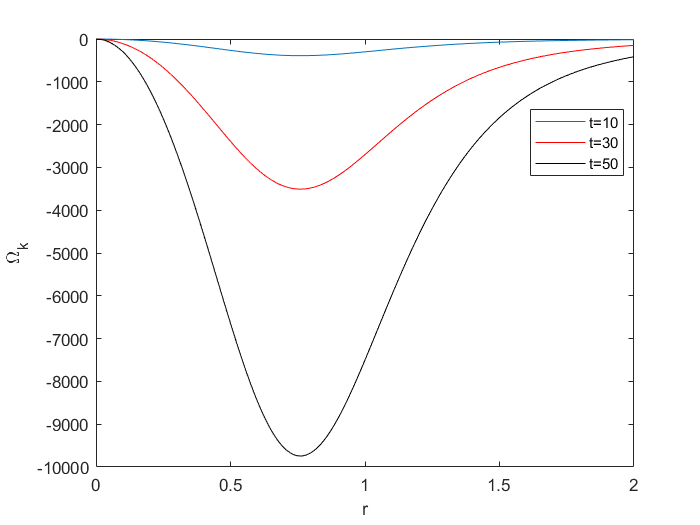}
			\caption{$\Omega_k$}
		\end{subfigure}
		\begin{subfigure}[t]{7.5cm}
			\includegraphics[width=8.5cm]{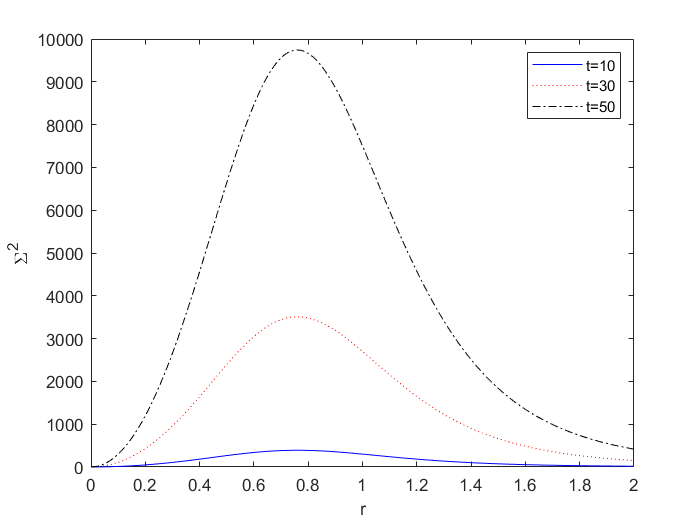}
			\caption{$\Sigma^2$}
		\end{subfigure}
	\end{center}
	\caption{Snapshots of $\Omega_k$ and $\Sigma^{2}$ for $\omega_0=1$ and $\Spo=-1$, showing that the inhomogeneities do not become narrow.}
	\label{Spo-1variables1}
\end{figure}

The inhomogeneities do not become narrow, so there are no spikes. Figure \ref{Spo-1variables1} shows the snapshots of $\Omega_{k}$ and $\Sigma^2$, whose amplitude increases indefinitely, but whose width is constant.
\subsection{Case $\omega_0=0$}
In this case (\ref{fkzeroG}) reduces to $f=4p_1$, and (\ref{SpsmkzeroG}) simplifies to 
\begin{align}
\Sp &=- \frac{7\Spo+4}{2(2\Spo+5)}, & \Sm=-\frac{\sqrt3(\Spo+4)}{2(2\Spo+5)}.
\end{align}
From the fluid density
\be
\rho =\frac{1-\Spo^2}{3r^4t^{2+4p_1}}
\ee
we conclude that for $-1<\Spo<1$ a physical singularity exists at $r=0$.

For the vacuum cases $\Spo=\pm1$, we can reach the same conclusion by examining the Weyl scalars. For $\Spo=-1$ we have \footnote{This solution is equivalent to a Kasner solution (in the broader sense), or a Levi-Civita solution \cite[pages 197 and 343]{book:Exactsol2002}.}
\begin{align}
CC&=\frac{192}{r^{12}}\\
CCs&=0\\
CCC&=-\frac{768}{r^{18}}\\
CCCs&=0.
\end{align}
And for $\Spo=1$ we have
\begin{align}
CC&=\frac{64(7r^4-18r^2t^{2/3}+27t^{4/3})}{9r^{12}t^{28/3}}\\
CCs&=0\\
CCC&=\frac{256(r^6t^{2/3}-5r^4t^{4/3}+9r^2t^2-9t^{8/3})}{3r^{18}t^{44/3}}\\
CCCs&=0.
\end{align}
Because there is a physical singularity at $r=0$, we should not say that a spike occurs there.
\section{Radius of spike}
Following $\cite{art:Lim2008}$, we define the coordinate radius of spike (or inhomogeneities) to be the value of $r$ where
\be
\label{radius1}
\lambda=|\omega|,
\ee
which gives
\be
\text{coordinate radius}=\sqrt{|\omega_0|}t^{-\frac13(1+\Spo)}.
\ee
For $\Spo=-1$ the coordinate radius is constant (so the inhomogeneities cannot be called as spike).

For $\Spo>-1$, the coordinate radius tends to zero as $t\rightarrow\infty$, indicating the formation of permanent spike along $r=0$ at late times.

We must measure the radius of the spike with respect to some important scale like horizon -- the particle or event horizon. From (\ref{gmetric2}) and (\ref{seed1}), we obtain
\begin{align}
\text{coordinate radius of particle horizon}&=\int_{0}^{t} \frac{N}{\sqrt{g_{33}}}\text{ }dt\\
&=\int_{0}^{t} t^{-\frac13(1+\Spo)}\text{ }dt\\
&=\frac{3}{2-\Spo}t^{\frac13(2-\Spo)}
\end{align}
Their ratio is
\be
\frac{\text{coordinate radius of spike}}{\text{coordinate radius of particle horizon}}=\left(\frac{2-\Spo}{3}\right)\sqrt{|\omega_0|}t^{-1}.
\ee

The physical radius of the spike is
\be
s=\int_{0}^{\text{coordinate radius of spike}}\sqrt{g_{rr}}\text{ }dr.
\ee
where
\be
\sqrt{g_{rr}} =\sqrt{Ft^{2p_1}}.\\
\ee
\text{Let} 
\begin{align} 
u &=\frac{rt^{\frac13+\frac13\Spo}}{|\omega_0|^\frac12} &\text{  and } &
& du =\frac{t^{\frac13+\frac13\Spo}}{|\omega_0|^\frac12}\text{ }dr
\end{align}
and the limits are $u=0$ and $1$.
Then 
\be
s =|\omega_0|^{3/2}\int_{0}^{{1}}\sqrt{1+u^4}\text{ } du.
\ee
i.e. the physical radius of the spike is constant. As the universe expands, in comparison  the spike becomes narrower and narrower.
\section{Discussion}
We set out to generate a solution with a permanent spike that forms at late times. We have achieved this by applying the Stephani transformation on the LRS Jacobs solution, using the rotational KVF. The generated solution is cylindrically symmetric, and has a spike along its rotational axis for the case $\Spo>-1$, $\omega_{0}\neq0$. This is the first generated solution with a late-time permanent spike, and the first generated solution with a spike along a line.\footnote{Such features can also be achieved through silent LTB and Szekeres models \cite{art:ColeyLim2014} without using solution-generating transformations.}
The spike produces an overdensity along the axis at late times, which is conducive to large scale structure formation. Thus the generated solution can serve as a prototypical model for formation of galactic filaments along web-like strings.

\chapter{Generated solutions, $k\neq0$}

In the previous chapter we have discussed the dynamics of generated solution for the case $k=0$. We have seen the spike at late times.\\
In this chapter we are going to generalise the result to the case $k\neq0$ (that is, using the general linear combination of KVFs).
\section{Applying the Stephani transformation}

We now carry out the Stephani transformation with the general KVF $\partial_\psi$.
First we find the squared norm of the KVF,
\be
\lambda = k^2t^{2p_3}+r^2t^{2p_1},
\ee
and the twist of the KVF,
\begin{align}
\omega= \dfrac{2k}{1+p_3} t^{1+p_3}+k\Spo r^2+\omega_{0}.
\end{align}
The combination $\lambda^2+\omega^2$ will appear frequently, so for brevity we introduce
\be
F = \lambda^2+\omega^2.
\ee

The next step is to find a particular solution $\beta_{a}$ for the constrained system
\be
\nabla_{[a}\beta_{b]} = 2 \lambda\nabla_{a}\xi_b + \omega\epsilon_{abcd}\nabla^c \xi^d,\quad{ \xi^{a}\beta_{a} =F-1}.
\ee
We obtain
\be
\beta_{a}=(0, F-1,\beta_2, 0),
\ee
where
\begin{align}
\beta_2 
= 2p_1\omega_0r^2+ p_1\Spo kr^4 + \left(\frac{4\omega+2k(1-p_3)r^2}{1+p_3} \right)t^{1+p_3}+k^3 t^{4p_3} + \frac{4k}{(1+p_3)^2} t^{2+2p_3} 
\end{align}
The generated metric is then given through $b$'s and $n$'s by the formulas (\ref{Geroch_bn_kzero_first})--(\ref{Geroch_bn_kzero_last}), dropping tildes for brevity.
\begin{align}
N&=F^{1/2} \\
b^1&= -\frac12\ln \frac{\lambda}{F}\\
b^2&= -\frac12 \ln \dfrac{F r^2 t^{2p_1+2p_3}}{\lambda} \\
b^3&= -\frac12\ln (Ft^{2p_1})\\
n_1&= \frac{Fkt^{2p_3}}{\lambda} - \beta_2 \\
n_2&=0\\
n_3&=0.
\end{align}
Its $\psi$-$Z$ area element 
\be
\mathcal{A}=rt^{p_1+p_3}
\ee
is the same as the seed solution's, and is always expanding. Its volume element 
\be
\mathcal{V}=rt \sqrt{F}
\ee
is different from the seed solution's and is not always expanding. This means the Hubble scalar $H$ can become negative for some parameter values, and Hubble-normalised variables would blow up. In this case we use $\beta$-normalisation, which is based on the ever-expanding area element \cite{art:vEUW2002}.
\section{The dynamical variables}
Here we list the dynamical variables of the generated solution using the formulas
in Appendix~\ref{App:formulas}.
To write the expressions in a more compact form, we list several intermediate expressions, particularly the partial derivatives of the essential variables.
\begin{align}
\partial_0 \lambda &= 2p_3 k^2t^{2p_3-1}+ 2p_1 r^2t^{2p_1-1}\\
\partial_3 \lambda &= 2rt^{2p_1}\\
\partial_0 \omega &= 2k t^{p_3}\\
\partial_3 \omega &= 2k(3p_1-1)r\\
\partial_0 F &= 2\lambda \partial_0 \lambda + 2\omega\partial_0\omega\\
\partial_3 F &= 2\lambda \partial_3 \lambda + 2\omega\partial_3\omega\\
\partial_0 \beta_2& = 4k p_3\lambda t^{2p_3-1} + 4\omega t^{p_3}\\
\partial_3 \beta_2 &= 4{p_1}\omega r.
\end{align}
By using the above convention, we get this list of dynamical variables.
\begin{align}
H &= \frac16 F^{-1/2}\left( \frac{\partial_0 F}{F} + \frac{2}{t}\right) \\
\Theta_{11} &= \frac12 F^{-1/2} \left( \frac{\partial_0 \lambda}{\lambda} - \frac{\partial_0 F}{F}\right)\\
\Theta_{22} &=\frac12 F^{-1/2} \left(\frac{\partial_0 F}{F} -\frac{\partial_0 \lambda}{\lambda} + \frac{2p_1+2p_3}{t}\right)\\
\Theta_{33} &=\frac12 F^{-1/2} \left(\frac{\partial_0 F}{F} +  \frac{2p_1}{t}\right)\\
\Theta_{12} &=\frac12\dfrac{kF^{-1/2}t^{-p_1+p_3}}{r} \left(  \dfrac{\partial_0 F}{F}-\frac{\partial_0 \lambda}{\lambda}+2\frac{p_3}{t}\right)- \dfrac{\lambda t^{-p_1-p_3}}{2rF^{3/2}}{\partial_0 \beta_2}\\
\dot{u}_3 &=-\frac12 F^{-1/2}t^{-p_1} \left(\frac{\partial_3 F}{F}\right)\\
n_{11} &=\dfrac{kF^{-1/2}t^{-2p_1+p_3}}{r} \left( \dfrac{\partial_3 F}{F}-\frac{\partial_3 \lambda}{\lambda}\right)- \dfrac{\lambda t^{-2p_1-p_3}}{rF^{3/2}}{\partial_3 \beta_2}\\
n_{12} &= \frac12{F^{-1/2}t^{-p_1}} \left( \dfrac{\partial_3 F}{F}-\frac{\partial_3 \lambda}{\lambda}+\frac{1}{r}\right) \\
a_3 &= -\frac12\dfrac{F^{-1/2}t^{-p_1}}{r}.
\end{align}
\section{Dynamics of the solution}
\label{Dynamics of the solution}
Now we find $\Sp$ and $\Sm$ by using the equations (\ref{ESC1}), (\ref{ESC2}).  We have
\be
\label{GSigma}
\Sp=-\frac{\Spo+f}{2+f},\\
\quad\Sm=\frac{\sqrt{3}(l-f-\frac13(2-\Spo))}{2+f}
\ee
where $f=t(\ln F)_t$, $l=t(\ln \lambda)_t$.
The $\beta$-normalised $(\Sp,\Sm)$ are
\be
\Sp=-\frac{\Spo+f}{2-\Spo}\\
\quad\Sm=\frac{\sqrt{3}(l-f)}{2-\Spo}-\frac{1}{\sqrt{3}}.
\ee
We have analysed the asymptotic dynamics of $l$ in Chapter 3. Recall that $l$ has a false spike at $r=0$ at late times for $0<\Spo\leq1$, and at early times for $-1\leq\Spo<0$. The asymptotic dynamics of $f$ can be analysed as follows. $f$ consists of terms involving $\lambda$ and terms involving $\omega$. $\lambda$ contains two different power terms, $t^{2p_1}$ and $t^{2p_3}$; while $\omega$ contains $t^{1+p_3}$ and constant. At late times, the term with the biggest exponent dominates; at early times the term with the smallest exponent dominates.
\begin{figure}
	\begin{center}
		\begin{subfigure}[t]{6.5cm}
			\includegraphics[width=6.5cm]{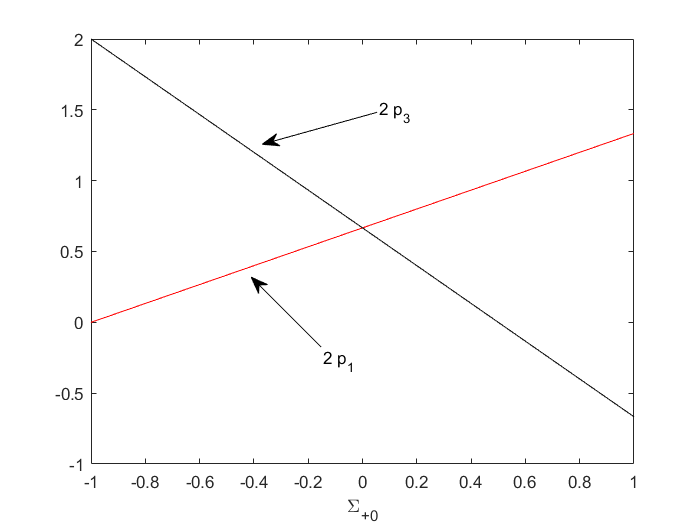}
			\caption{$\lambda$ exponents}
		\end{subfigure}
		\begin{subfigure}[t]{6.5cm}
			\includegraphics[width=6.5cm]{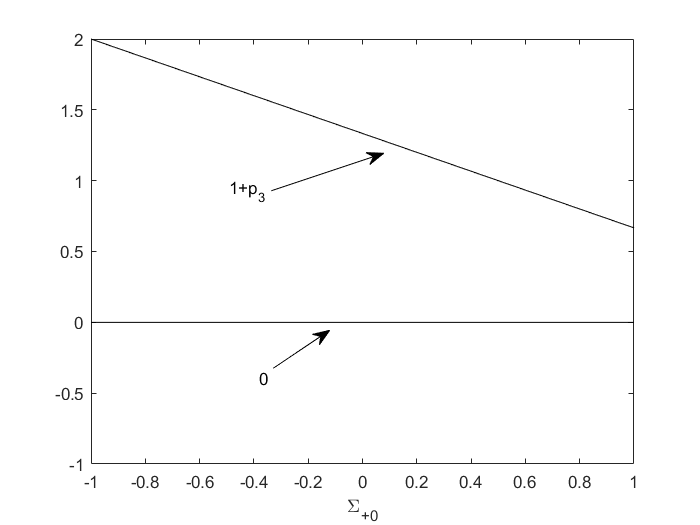}
			\caption{$\omega$ exponents }
		\end{subfigure}
		\begin{subfigure}[t]{6.5cm}
			\includegraphics[width=6.5cm]{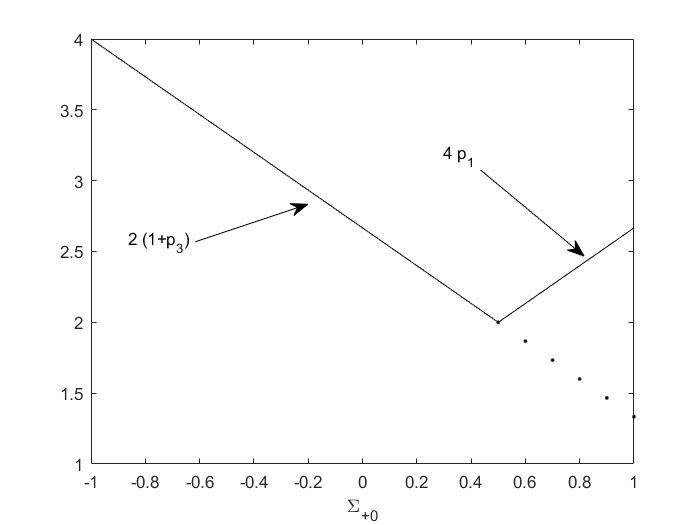}
			\caption{Dominant exponents in $f$ at late times}
		\end{subfigure}
		\begin{subfigure}[t]{6.5cm}
			\includegraphics[width=6.5cm]{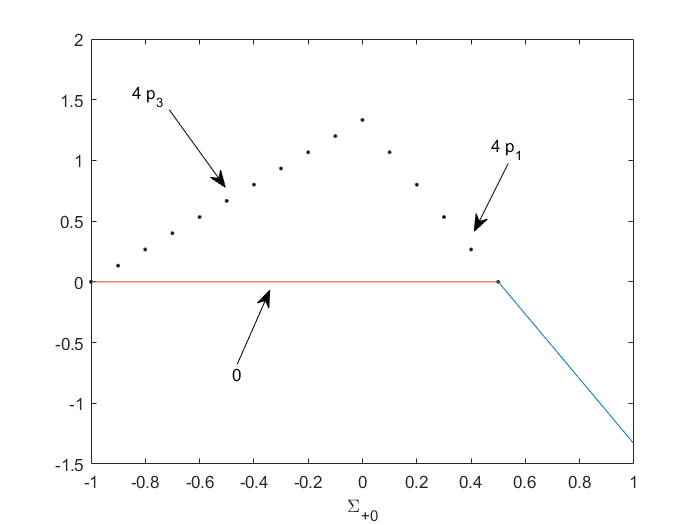}
			\caption{Dominant exponents in $f$ at early times}
		\end{subfigure}
	\end{center}
	\caption{Exponents in $\lambda$, $\omega$ and $f$. Dotted lines indicate the next dominant exponent.}
	\label{fig_exp1}
\end{figure}
Because $f$ has the form
\be
\label{Gf}
f=\frac{t(\lambda^2)_t+t(\omega^2)_t}{\lambda^2+\omega^2},
\ee 
and $\lambda^2$ and $\omega^2$ are sums of power of $t$, the asymptotic limits of $f$ are determined by coefficients of the dominant terms in the general case. When there are two equally dominant terms, care is taken to include all equally dominant terms. It turns out that
\begin{align}
\lim\limits_{t\rightarrow\infty}f=&
\begin{cases} 
4p_1 & \text{for } \ \ \ \frac12\leq\Spo\leq1 \\
2(1+p_3) & \text{for } \ \ \ -1\leq\Spo\leq\frac12
\end{cases}
\end{align}
and
\begin{align}
\lim\limits_{t\rightarrow0}f=&
\begin{cases} 
4p_3 & \text{for} \ \ \ \frac12\leq\Spo\leq1 \\
0 & \text{for} \ \ \  -1\leq\Spo\leq\frac12.
\end{cases}
\end{align}
In some special cases, the coefficient of a dominant term can become zero. Then we find the next dominant term. These special cases are: $r=0$, which kills the $t^{2p_1}$ term in $\lambda$, $r=\sqrt{\frac{-\omega_0}{k\Spo}}$, which kills the constant term in $\omega$, and $k=0$, which kills the $t^{2p_3}$ term. The case $k=0$ has been dealt with in Chapter 4. Figure \ref{fig_exp1} shows the exponents in $\lambda$, $\omega$ and $f$.\\
Along $r=0$,
\begin{align}
\lim\limits_{t\rightarrow\infty}f=2(1+p_3) &\ \ \text{for all} \ \ \  -1\leq\Spo\leq1.
\end{align}
Along $r=\sqrt{\frac{-\omega_0}{k\Spo}}$,
\begin{align}
\lim\limits_{t\rightarrow0}f=&
\begin{cases} 
4p_3 & \text{for} \ \ \ 0\leq\Spo\leq1 \\
4p_1 & \text{for} \ \ -1\leq\Spo\leq0
\end{cases}
\end{align}
In the very special case $\omega_0=0=r$, $f$ is equal to 4.\\
In summary, $f$ has a spike at $r=0$ at late times for $\frac12<\Spo\leq1$, a spike at $r=0$ at early times for $\Spo=-1$, $\omega_{0}=0$, and a spike at $r=\sqrt{\frac{-\omega_0}{k\Spo}}\neq0$ at early times
for $-1<\Spo<0$, $\omega_0k>0$, or $0<\Spo<\frac12$, $\omega_0k<0.$\\
Combining the effects of $f$ and $l$, we see that $(\Sp,\Sm)$ has a spike at $r=0$ at late times for $0<\Spo\leq1$, a spike at $r=0$ at early times for $-1<\Spo<0$, and a spike at $r=\sqrt{\frac{-\omega_0}{k\Spo}}\neq0$ at early times for $-1<\Spo<0$, $\omega_0k>0$, or   $0<\Spo<\frac12$, $\omega_0k<0$. Later we will see that the spike caused by $l$ is not real.
\subsection{\textbf{At late times} $(t\rightarrow \infty)$}
\begin{figure}
	\begin{center}
		\begin{subfigure}[t]{5.5cm}
			\includegraphics[width=5.5cm]{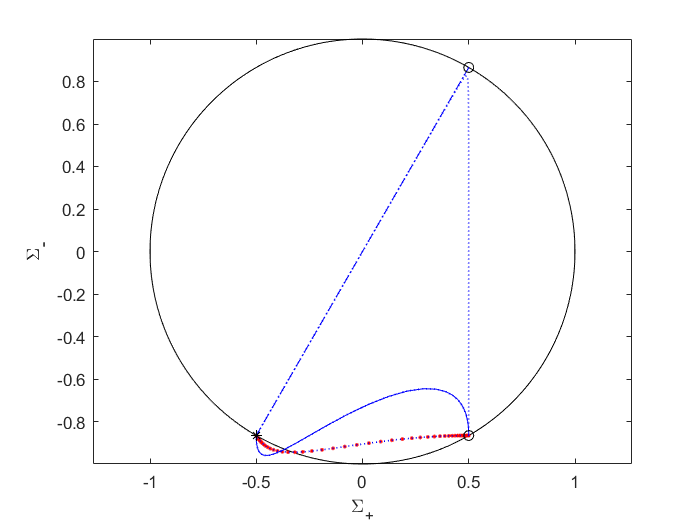}
			\caption{$\Spo=-1$ }
		\end{subfigure}
		\begin{subfigure}[t]{5.5cm}
			\includegraphics[width=5.5cm]{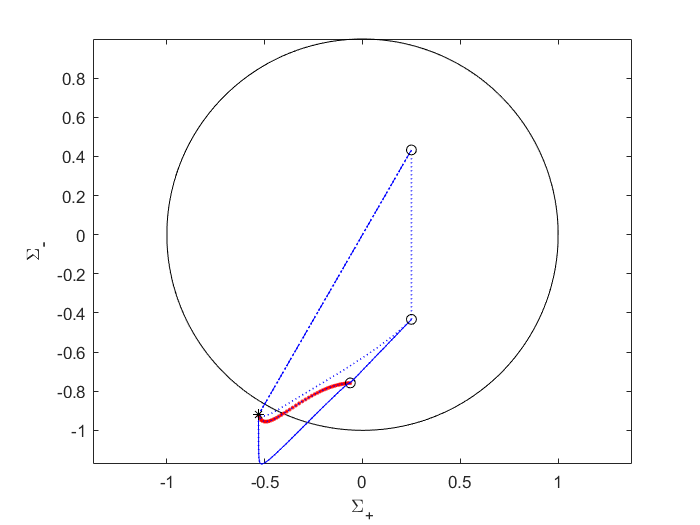}
			\caption{$\Spo=-0.5$ }
		\end{subfigure}
		\begin{subfigure}[t]{5.5cm}
			\includegraphics[width=5.5cm]{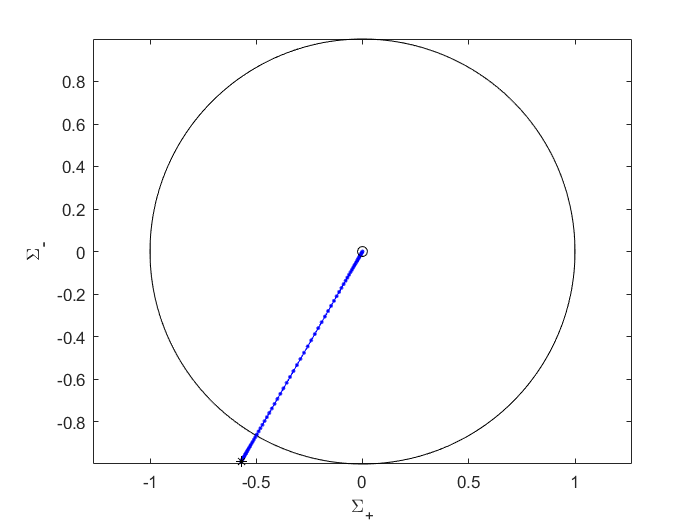}
			\caption{\textbf{$\Spo=0$ }}
		\end{subfigure}
	\end{center}
	\caption{State space orbits projected on the $(\Sp,\Sm)$ plane for $k=1$, $\Spo=-1,-0.5, 0$ and $\omega_0=1$. All orbits end up at the same point indicated by an asterisk.}
	\label{fig5.2}
\end{figure}

\begin{figure}
	\begin{center}
		\begin{subfigure}[t]{5.5cm}
			\includegraphics[width=5.5cm]{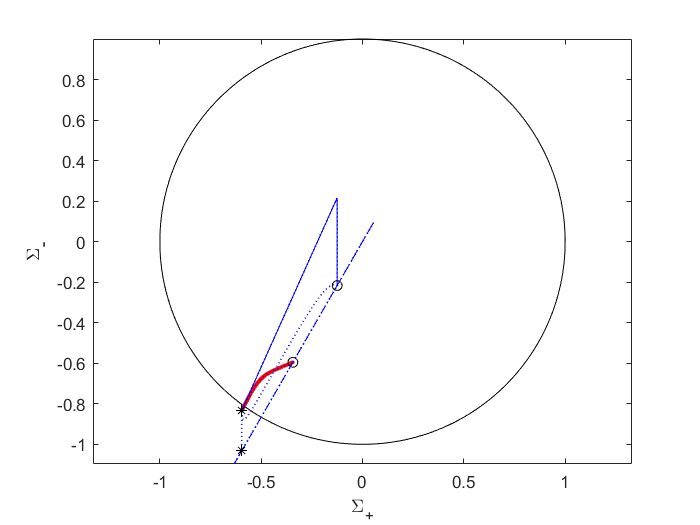}
			\caption{Combine for all values of $r$}
		\end{subfigure}
		\begin{subfigure}[t]{5.5cm}
			\includegraphics[width=5.5cm]{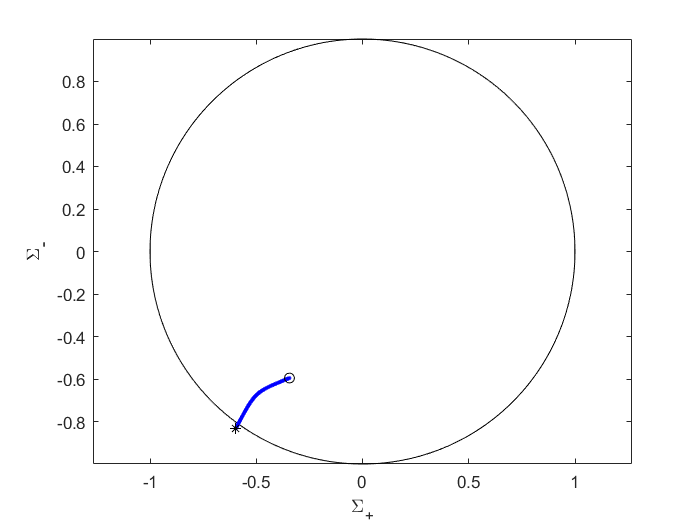}
			\caption{For $r=\sqrt{-\frac{\omega_0}{k\Spo}}$ }
		\end{subfigure}
		\begin{subfigure}[t]{5.5cm}
			\includegraphics[width=5.5cm]{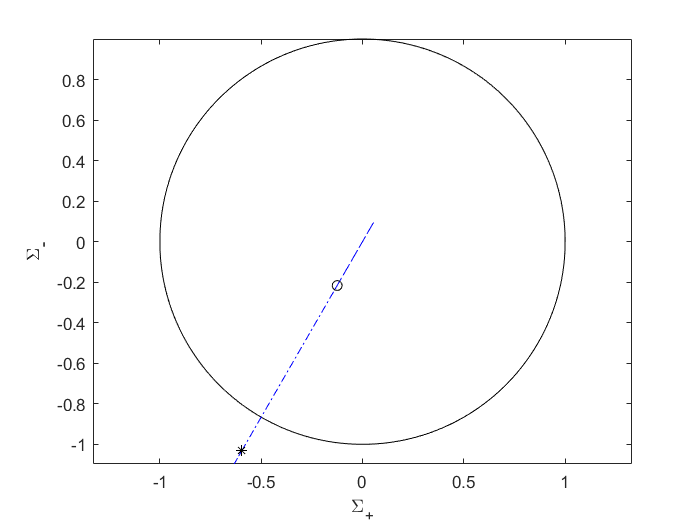}
			\caption{For $r=0$}
		\end{subfigure}
		\begin{subfigure}[t]{5.5cm}
			\includegraphics[width=5.5cm]{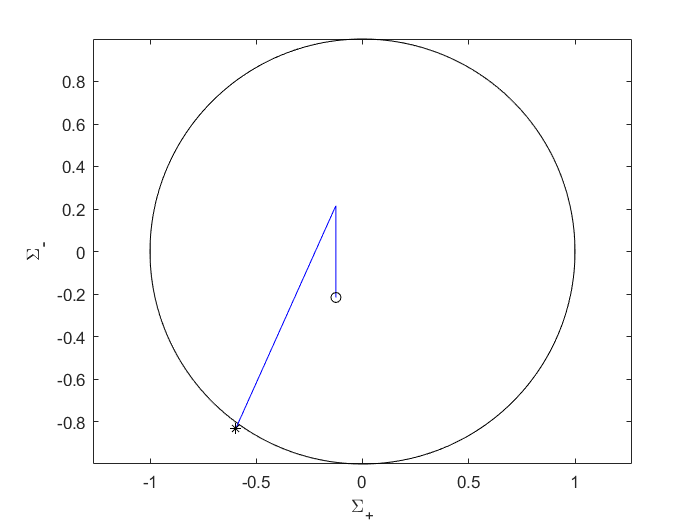}
			\caption{For a large value of $r$}
		\end{subfigure}
	\end{center}
	\caption{State space orbits for $k=1$, $\Spo=0.25$ and $\omega_0=-1$.}
	\label{fig5.3}
\end{figure}

\begin{figure}
	\begin{center}
		\begin{subfigure}[t]{5.5cm}
			\includegraphics[width=5.5cm]{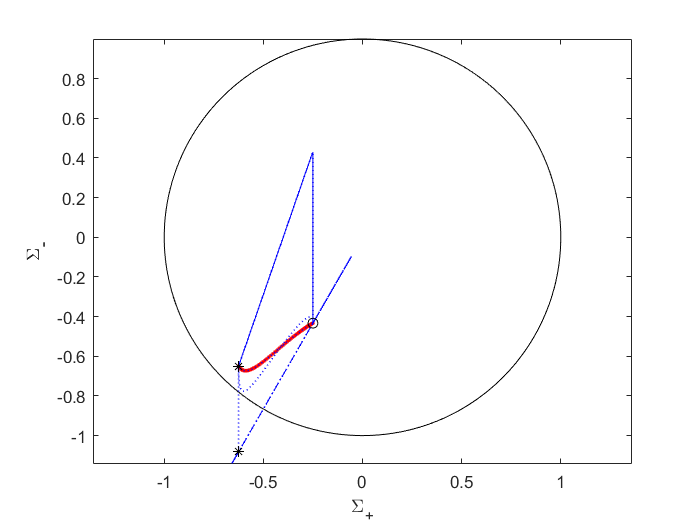}
			\caption{Combine for all values of $r$}
		\end{subfigure}
		\begin{subfigure}[t]{5.5cm}
			\includegraphics[width=5.5cm]{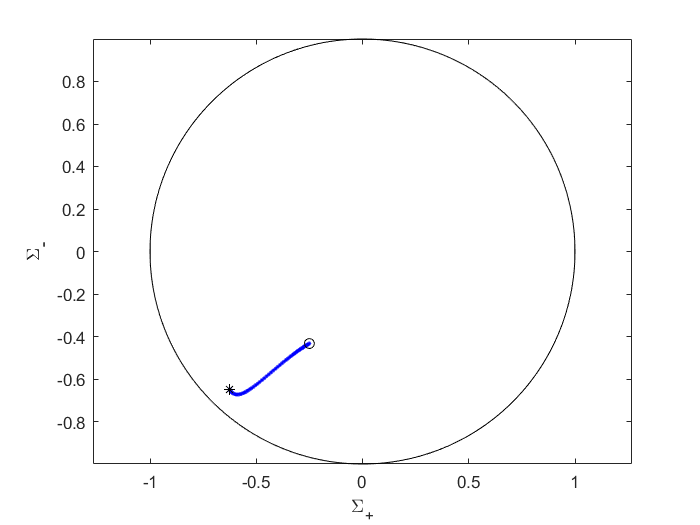}
			\caption{For $r=\sqrt{-\frac{\omega_0}{k\Spo}}$ }
		\end{subfigure}
		\begin{subfigure}[t]{5.5cm}
			\includegraphics[width=5.5cm]{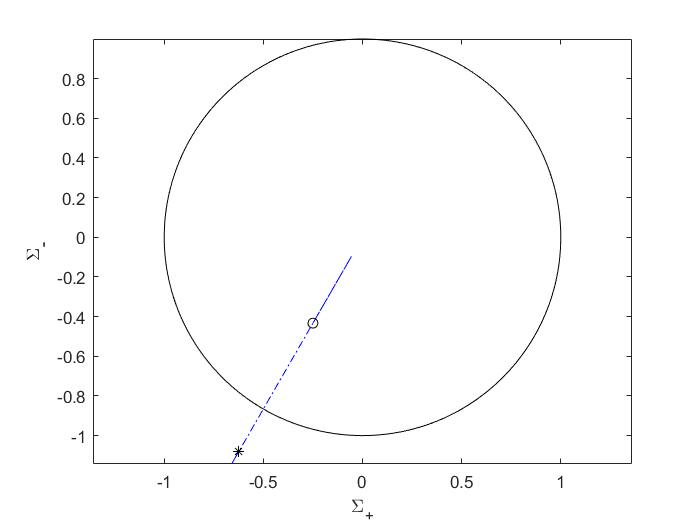}
			\caption{For $r=0$}
		\end{subfigure}
		\begin{subfigure}[t]{5.5cm}
			\includegraphics[width=5.5cm]{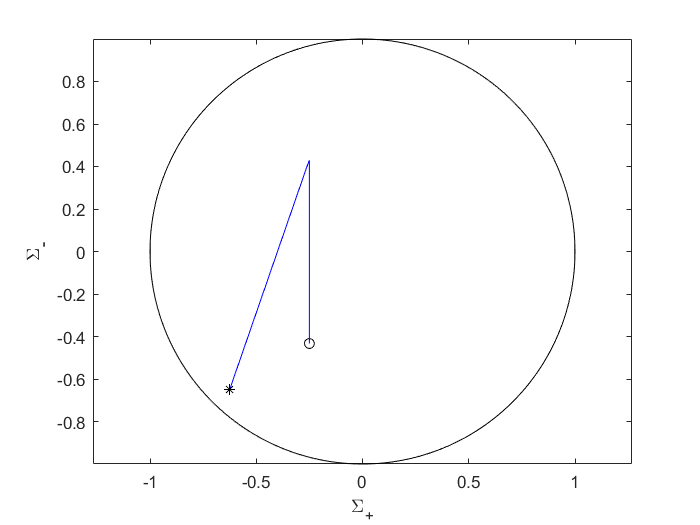}
			\caption{For a large value of $r$}
		\end{subfigure}
	\end{center}
	\caption{State space orbits for $k=1$, $\Spo=0.5$ and $\omega_0=-1$.}
	\label{fig5.4}
\end{figure}

\begin{figure}
	\begin{center}
		\begin{subfigure}[t]{5.5cm}
			\includegraphics[width=5.5cm]{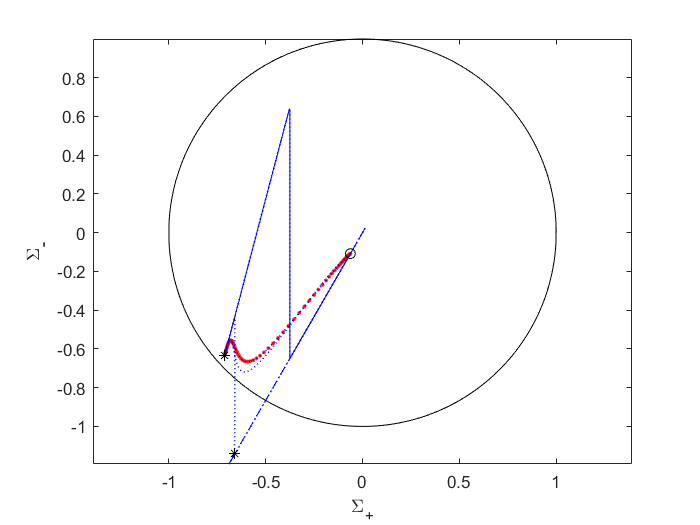}
			\caption{Combine for all values of $r$}
		\end{subfigure}
		\begin{subfigure}[t]{5.5cm}
			\includegraphics[width=5.5cm]{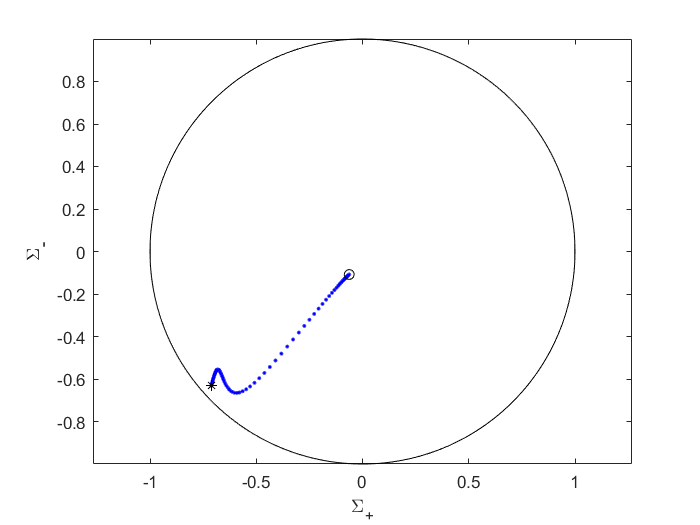}
			\caption{For $r=\sqrt{-\frac{\omega_0}{k\Spo}}$ }
		\end{subfigure}
		\begin{subfigure}[t]{5.5cm}
			\includegraphics[width=5.5cm]{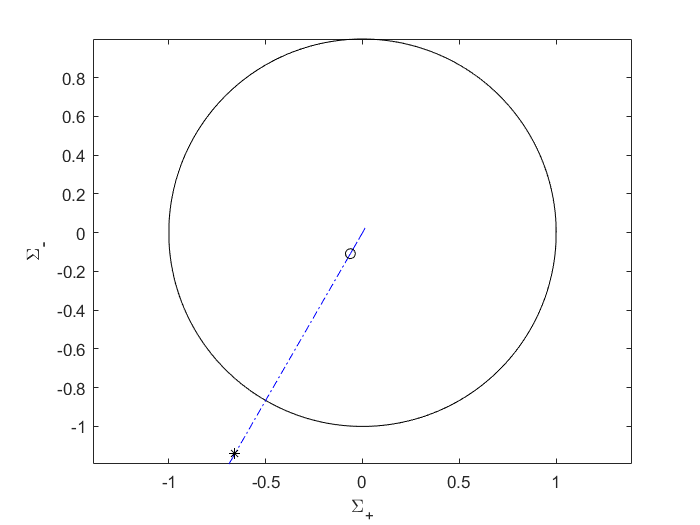}
			\caption{For $r=0$}
		\end{subfigure}
		\begin{subfigure}[t]{5.5cm}
			\includegraphics[width=5.5cm]{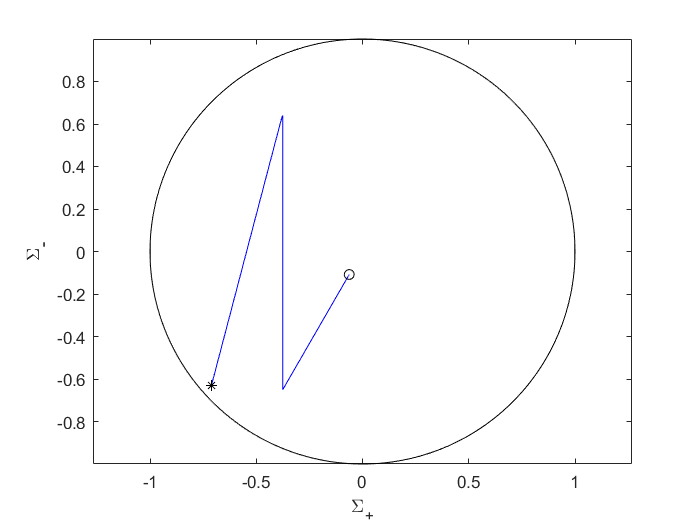}
			\caption{For a large value of $r$}
		\end{subfigure}
		
	\end{center}
	\caption{State space orbits for $k=1$, $\Spo=0.75$ and $\omega_0=-1$.}
	\label{fig5.5}
\end{figure}

\begin{figure}
	\begin{center}
		\begin{subfigure}[t]{6cm}
			\includegraphics[width=6cm]{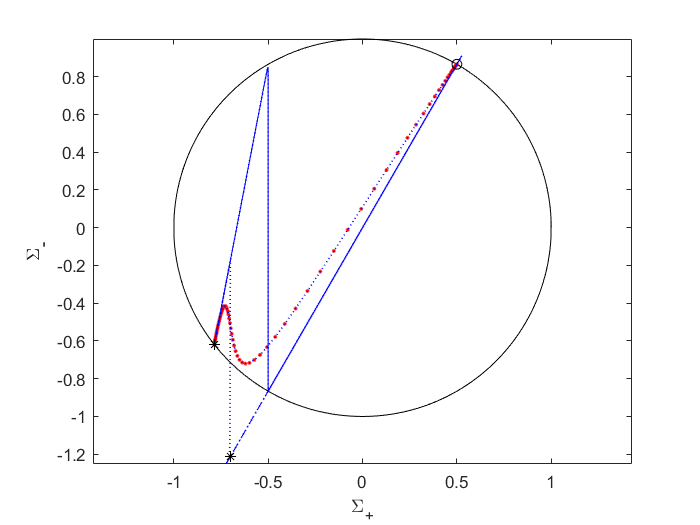}
			\caption{Combine for all values of $r$}
		\end{subfigure}
		\begin{subfigure}[t]{6cm}
			\includegraphics[width=6cm]{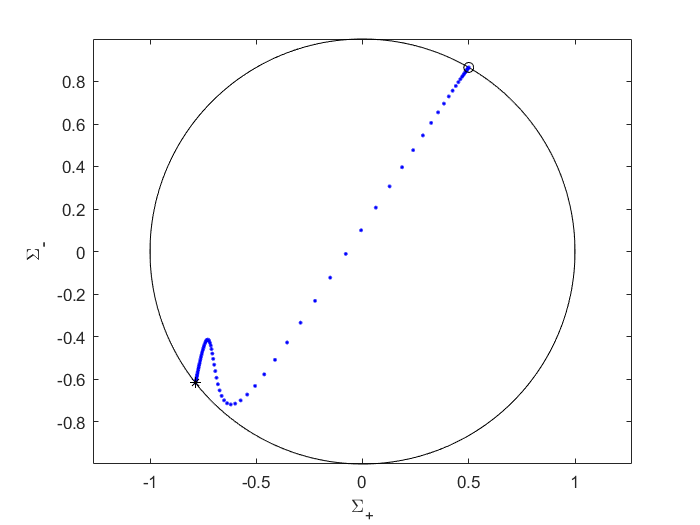}
			\caption{For $r=\sqrt{-\frac{\omega_0}{k\Spo}}$ }
		\end{subfigure}
		\begin{subfigure}[t]{6cm}
			\includegraphics[width=6cm]{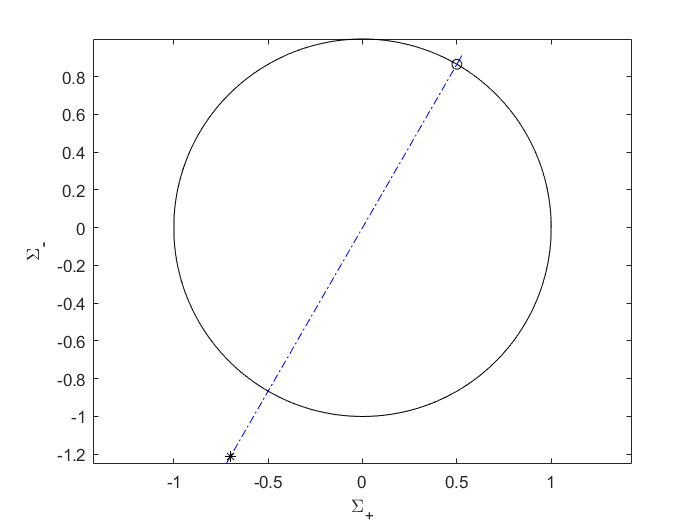}
			\caption{For $r=0$}
		\end{subfigure}
		\begin{subfigure}[t]{6cm}
			\includegraphics[width=6cm]{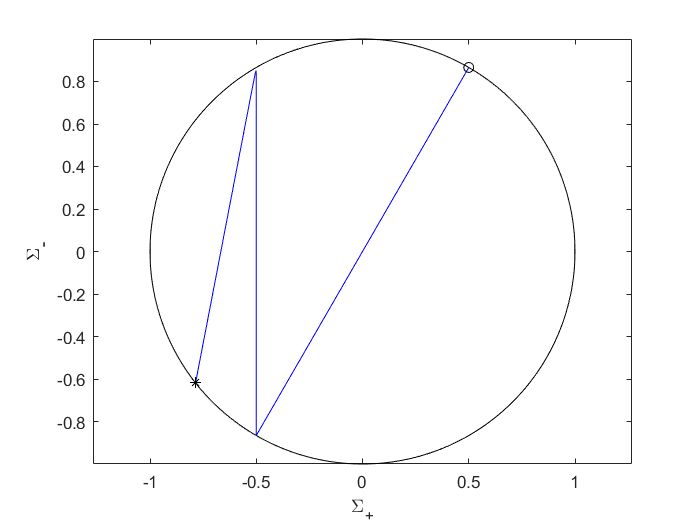}
			\caption{For a large value of $r$}
		\end{subfigure}
	\end{center}
	\caption{State space orbits for $k=1$, $\Spo=1$ and $\omega_0=-1$.}
	\label{fig5.6}
\end{figure}

For $-1<\Spo\leq0$, we do not have a spike.
\begin{align}
(\Sp,\Sm)\rightarrow\left(-\frac{8-\Spo}{2(7-2\Spo)},-\frac{\sqrt{3}(8-\Spo)}{2(7-2\Spo)}\right).
\end{align}
See Figure \ref{fig5.2}.
\newpage
For $0<\Spo<0.5$, we have a spike at $r=0$.
\begin{align}
(\Sp,\Sm)\rightarrow&
\begin{cases} 
\left(-\frac{8-\Spo}{2(7-2\Spo)},-\frac{\sqrt{3}(8-7\Spo)}{2(7-2\Spo)}\right) & \text{along } r\neq 0 \\
\left(-\frac{8-\Spo}{2(7-2\Spo)},-\frac{\sqrt{3}(8-\Spo)}{2(7-2\Spo)}\right) & \text{along }r=0.
\end{cases}
\end{align}
See Figure \ref{fig5.3}. Later we will see that this spike is not real.
\newpage
For $\Spo=0.5$, we have a spike at $r=0$.
\begin{align}
(\Sp,\Sm)\rightarrow&
\begin{cases} 
\left(-\frac58,-\frac38\sqrt{3}\right) & \text{along } r\neq 0 \\
\left(-\frac58,-\frac58\sqrt{3}\right) & \text{along }r=0.
\end{cases}
\end{align}
See Figure \ref{fig5.4}. Later we will see that this spike is not real.
\newpage
For $0.5<\Spo\leq1$, we have a spike at $r=0$
\begin{align}
(\Sp,\Sm)\rightarrow&
\begin{cases} 
(-\frac{4+7\Spo}{2(5+2\Spo)},-\frac{\sqrt{3}(4+\Spo)}{2(5+2\Spo)}) & \text{along } r\neq 0 \\
(-\frac{8-\Spo}{2(7-2\Spo)},-\frac{\sqrt{3}(8-\Spo)}{2(7-2\Spo)}) & \text{along }r=0.
\end{cases}
\end{align}
See Figures \ref{fig5.5} and \ref{fig5.6}.
\newpage
%
\subsection{\textbf{At early times} $(t\rightarrow 0)$}
\begin{figure}
	\begin{center}
		\begin{subfigure}[t]{5.6cm}
			\includegraphics[width=5.5cm]{./CGSp0minus1dr.png}
			\caption{Combine for all values of $r$}
		\end{subfigure}
		\begin{subfigure}[t]{5.5cm}
			\includegraphics[width=5.5cm]{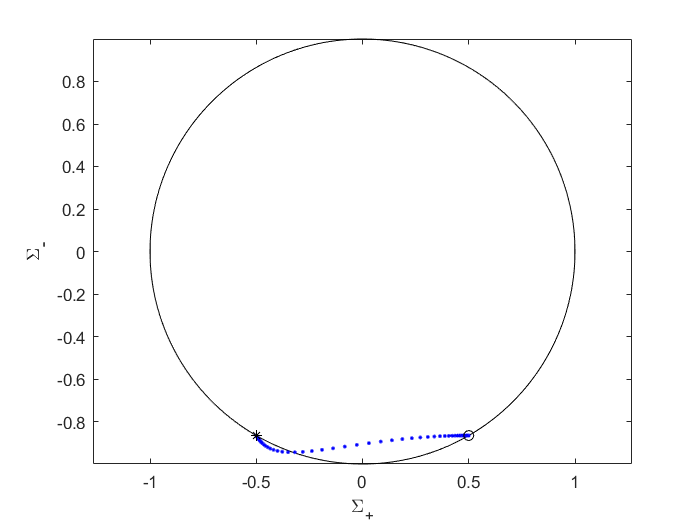}
			\caption{For $r=\sqrt{-\frac{\omega_0}{k\Spo}}$ }
		\end{subfigure}
		\begin{subfigure}[t]{5.5cm}
			\includegraphics[width=5.5cm]{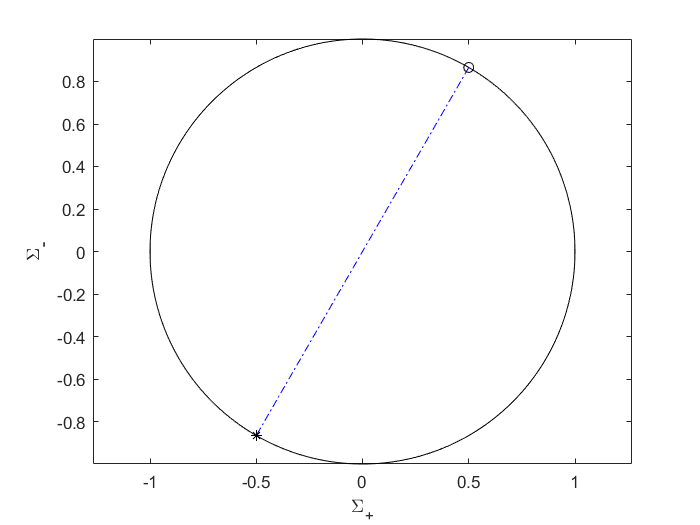}
			\caption{For $r=0$}
		\end{subfigure}
		\begin{subfigure}[t]{5.5cm}
			\includegraphics[width=5.5cm]{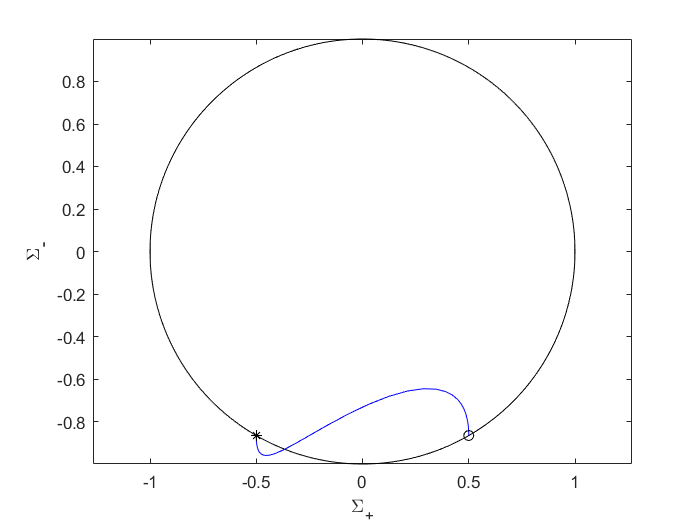}
			\caption{For a large value of $r$}
		\end{subfigure}
	\end{center}
	\caption{State space orbits for $k=1$, $\Spo=-1$ and $\omega_0=1$.}
	\label{fig5.7}
\end{figure}

\begin{figure}
	\begin{center}
		\begin{subfigure}[t]{5.5cm}
			\includegraphics[width=5.5cm]{./CGSp0minuspoint5dr.png}
			\caption{Combine for all values of $r$}
		\end{subfigure}
		\begin{subfigure}[t]{5.5cm}
			\includegraphics[width=5.5cm]{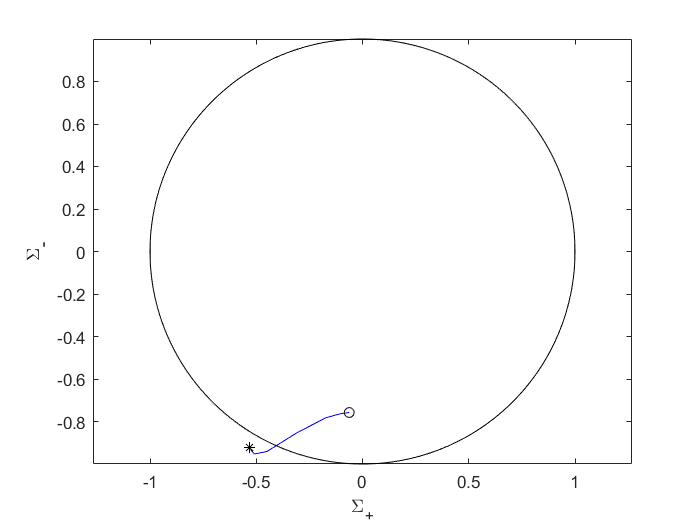}
			\caption{For $r=\sqrt{-\frac{\omega_0}{k\Spo}}$ }
		\end{subfigure}
		
		\begin{subfigure}[t]{5.5cm}
			\includegraphics[width=5.5cm]{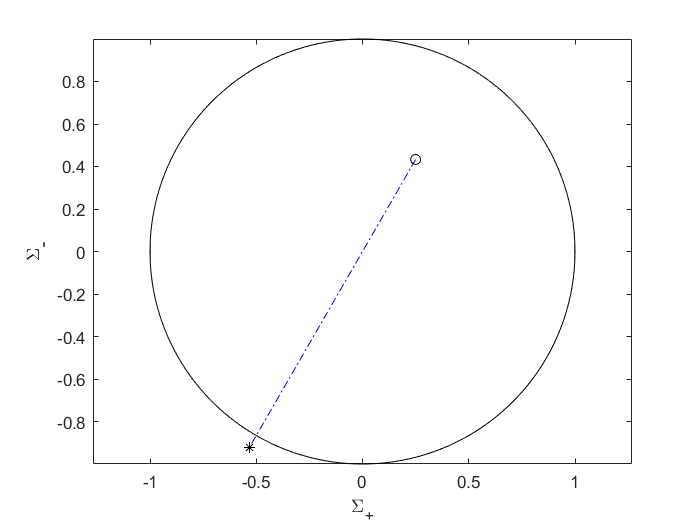}
			\caption{For $r=0$}
		\end{subfigure}
		\begin{subfigure}[t]{5.5cm}
			\includegraphics[width=5.5cm]{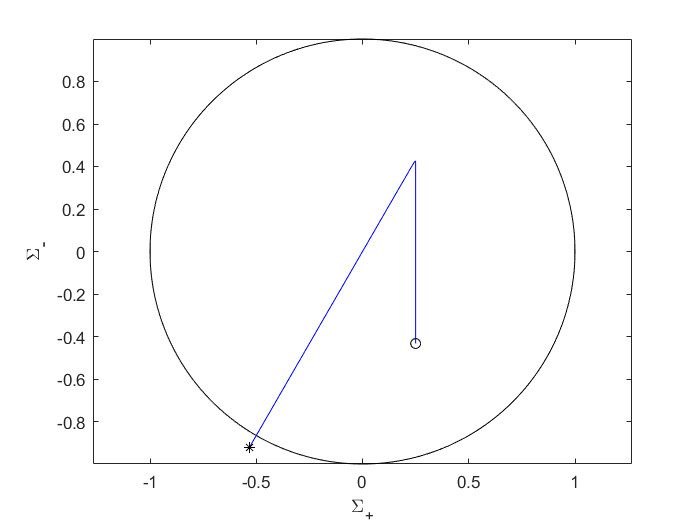}
			\caption{For a large value of $r$}
		\end{subfigure}
		
	\end{center}
	\caption{State space orbits for $k=1$, $\Spo=-0.5$ and $\omega_0=1$.}
	\label{fig5.8}
\end{figure}

For $\Spo=-1$, we have a spike along $r=0$.
\begin{align}
(\Sp,\Sm)\rightarrow&
\begin{cases}
(\frac{1}{2},\frac{\sqrt{3}}{2}) & \text{along } r=0 \\
(\frac{1}{2},-\frac{\sqrt{3}}{2}) & \text{along } r\neq0.
\end{cases}
\end{align}
See Figure \ref{fig5.7}. Later we will see that this spike is not real.
\newpage
For $-1<\Spo<0$, we have a spike along $r=0$, and if $\omega_0 k>0$, another spike along $r=\sqrt{-\frac{\omega_0}{k\Spo}}$.
\begin{align}
(\Sp,\Sm)\rightarrow&
\begin{cases}
\left(-\frac{\Spo}{2},-\frac{\sqrt{3}\Spo}{2}\right) & \text{along } r=0 \\
\left(-\frac{\Spo}{2},\frac{\sqrt{3}\Spo}{2}\right) & \text{along } r\neq 0,\sqrt{-\frac{\omega_0}{k\Spo}} \\
\left(-\frac{4+7\Spo}{2(5+2\Spo)},-\frac{\sqrt{3}(4+\Spo)}{2(5+2\Spo)}\right) & \text{along }r=\sqrt{-\frac{\omega_0}{k\Spo}}.
\end{cases}
\end{align} 
See Figure \ref{fig5.8}. Later we will see that the spike along $r=0$ is not real.

%
\newpage

\begin{figure}
	\begin{center}
		\begin{subfigure}[t]{5.5cm}
			\includegraphics[width=5.5cm]{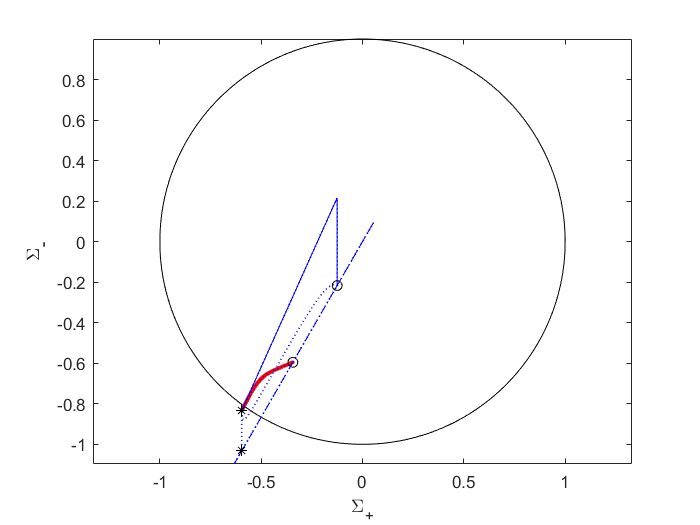}
			\caption{Combine for all values of $r$}
		\end{subfigure}
		\begin{subfigure}[t]{5.5cm}
			\includegraphics[width=5.5cm]{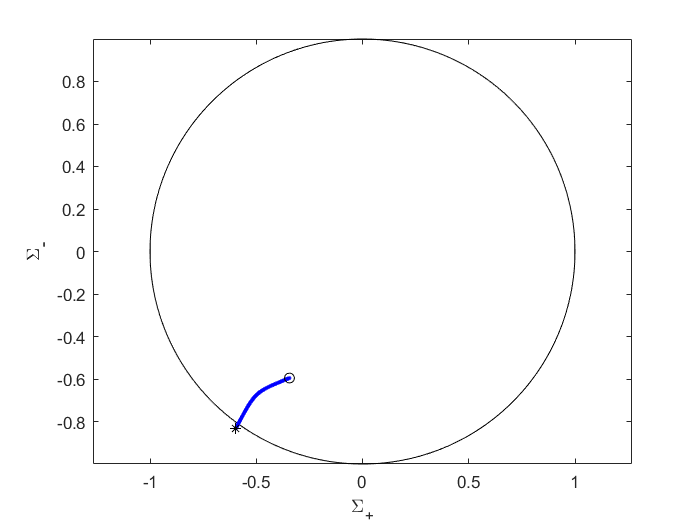}
			\caption{For $r=\sqrt{-\frac{\omega_0}{k\Spo}}$ }
		\end{subfigure}
		
		\begin{subfigure}[t]{5.5cm}
			\includegraphics[width=5.5cm]{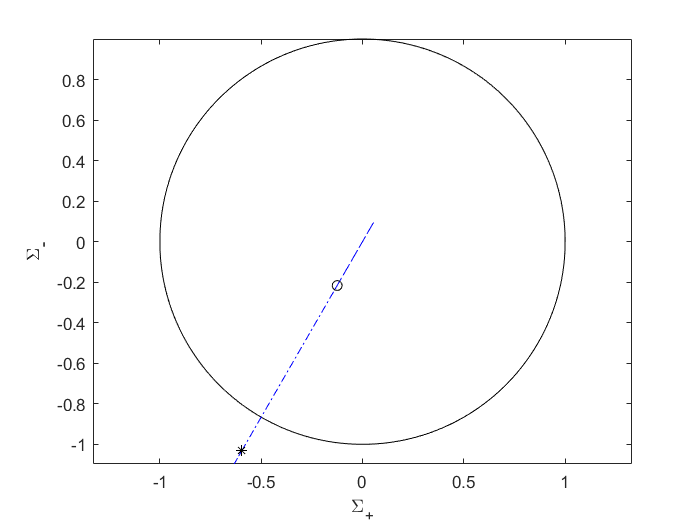}
			\caption{For $r=0$}
		\end{subfigure}
		\begin{subfigure}[t]{5.5cm}
			\includegraphics[width=5.5cm]{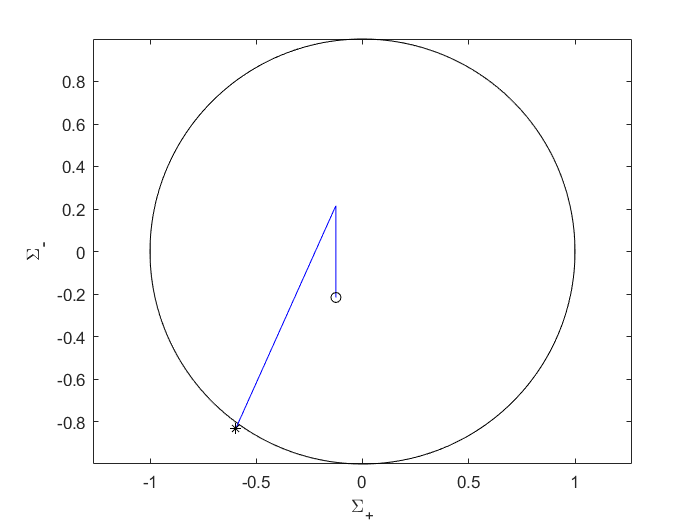}
			\caption{For a large value of $r$}
		\end{subfigure}
		
	\end{center}
	\caption{State space orbits for $k=1$, $\Spo=0.25$ and $\omega_0=-1$.}
	\label{fig5.9}
\end{figure}

\begin{figure}
	\begin{center}
		\begin{subfigure}[t]{5.5cm}
			\includegraphics[width=5.5cm]{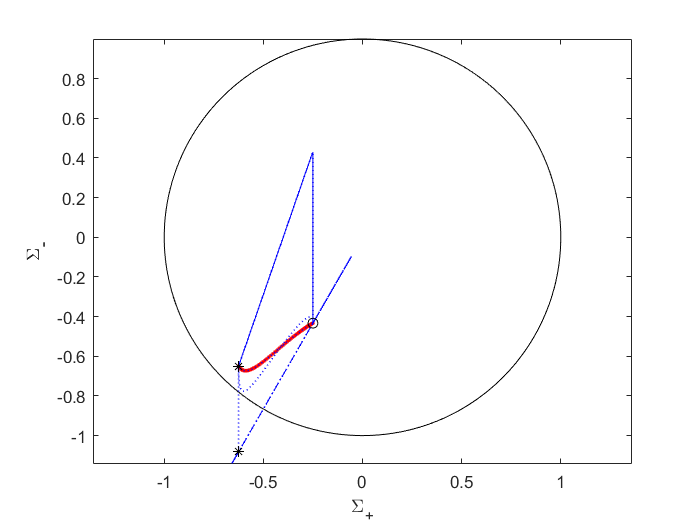}
			\caption{$\Spo=0.5$}
		\end{subfigure}
		\begin{subfigure}[t]{5.5cm}
			\includegraphics[width=5.5cm]{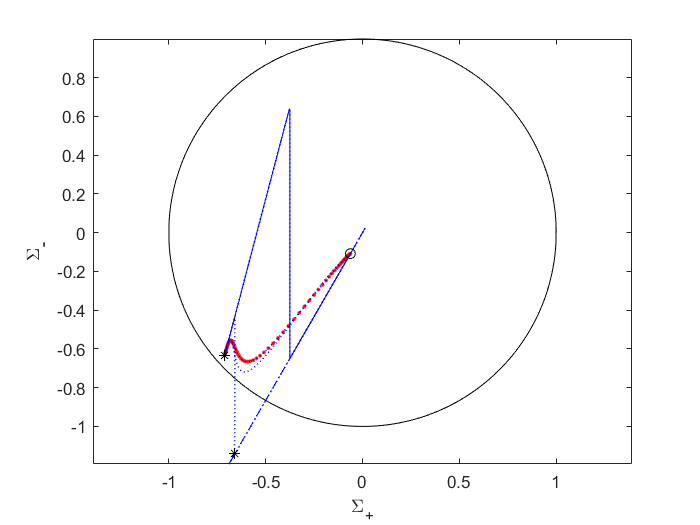}
			\caption{$\Spo=0.75$ }
		\end{subfigure}
		
		\begin{subfigure}[t]{5.5cm}
			\includegraphics[width=5.5cm]{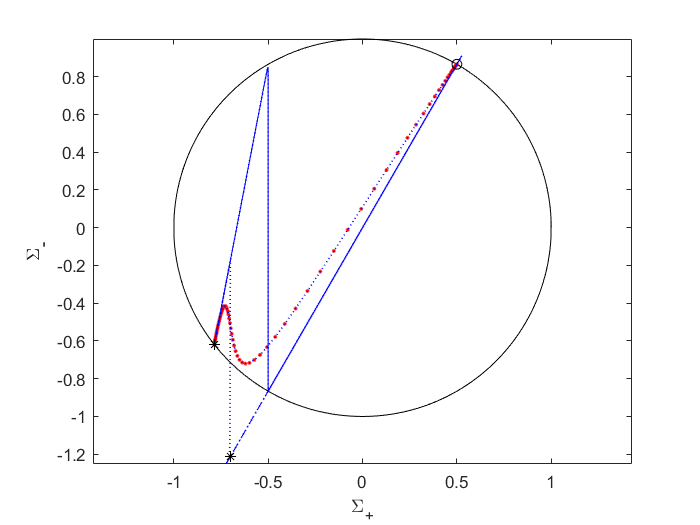}
			\caption{$\Spo=1$}
		\end{subfigure}
		
	\end{center}
	\caption{State space orbits for $k=1$ and $\omega_0=-1$.}
	\label{fig5.10}
\end{figure}

For $0<\Spo<0.5$, we have a spike at $r=\sqrt{-\frac{\omega_0}{k\Spo}}$ if $\omega_0 k<0$.
\begin{align}
(\Sp,\Sm)\rightarrow&
\begin{cases}
\left(-\frac{\Spo}{2},-\frac{\sqrt{3}\Spo}{2}\right) & \text{along } r\neq\sqrt{-\frac{\omega_0}{k\Spo}} \\
\left(-\frac{4-5\Spo}{2(5-4\Spo)},-\frac{\sqrt{3}(4-5\Spo)}{2(5-4\Spo)}\right) & \text{along }r=\sqrt{-\frac{\omega_0}{k\Spo}}.
\end{cases}
\end{align}
See Figure \ref{fig5.9}.
\newpage
For $0.5\leq\Spo\leq1$, there are no spikes.
\begin{align}
(\Sp,\Sm)\rightarrow \left(-\frac{4-5\Spo}{2(5-4\Spo)},-\frac{\sqrt{3}(4-5\Spo)}{2(5-4\Spo)}\right).
\end{align}
See Figure \ref{fig5.10}.
\section{When $\Spo=0$}


When $\Spo=0$ equation (\ref{GSigma}) simplifies to
\be
\label{Sigmaomegazero}
\Sp=-\frac{f}{2+f},\\
\quad\Sm=\frac{\sqrt{3}(l-f-\frac23)}{2+f}
\ee
where $f=t(\ln F)_t$, $l=t(\ln \lambda)_t$.

Now, for $\omega_0=0$,
\begin{align}
\lim\limits_{t\rightarrow 0}f=\frac43 &\ \ \text{for all} \ \ r.\\
\lim\limits_{t\rightarrow\infty}f=\frac83 &\ \ \text{for all} \ \ r.
\end{align}
For $\omega_0\neq0$,
\begin{align}
\lim\limits_{t\rightarrow 0}f= 0 &\ \ \text{for all} \ \ r.\\
\lim\limits_{t\rightarrow\infty}f=\frac83 &\ \ \text{for all} \ \ r.
\end{align}
For all values of $\omega_{0}$ and $t$, we have $l=\frac23$.

\begin{figure}
	\begin{center}
		\begin{subfigure}[t]{5.5cm}
			\includegraphics[width=5.5cm]{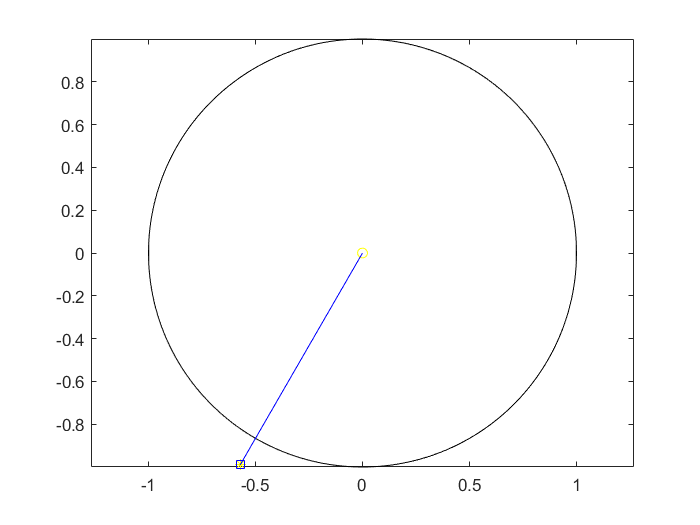}
			\caption{$\omega_0\neq0$}
		\end{subfigure}
		\begin{subfigure}[t]{5.5cm}
			\includegraphics[width=5.5cm]{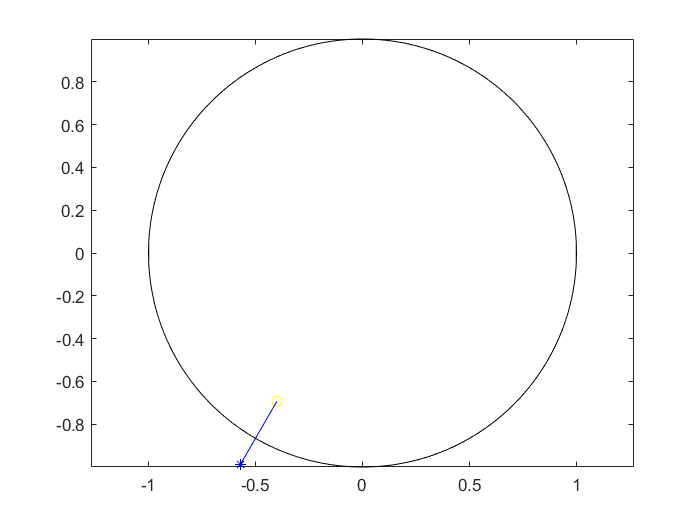}
			\caption{$\omega_0=0$ }
		\end{subfigure}
		
		\begin{subfigure}[t]{5.5cm}
			\includegraphics[width=5.5cm]{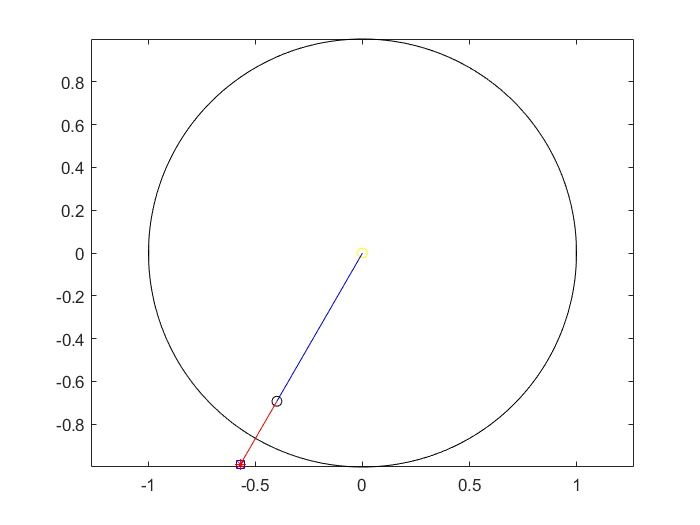}
			\caption{ for all $\omega_0$}
		\end{subfigure}
	\end{center}
	\caption{State space orbits for $k=1$ and $\Spo=0$.}
	\label{fig5.11}
\end{figure}

\begin{figure}
	\begin{center}
		\begin{subfigure}[t]{5.5cm}
			\includegraphics[width=5.5cm]{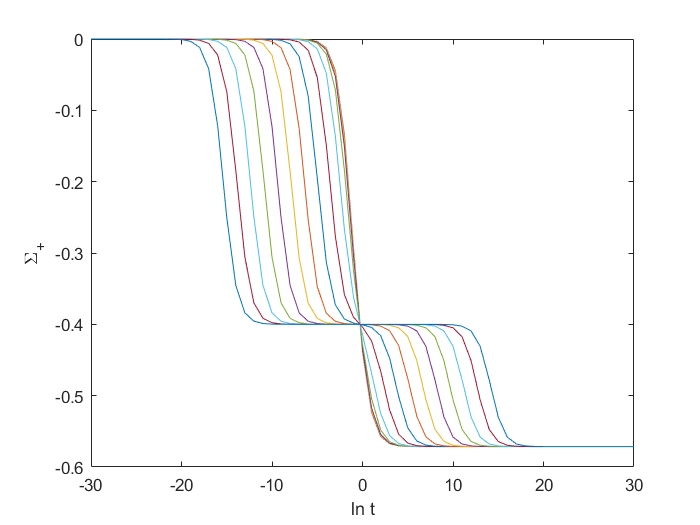}
			\caption{Evolution along different worldlines}
		\end{subfigure}
		\begin{subfigure}[t]{5.5cm}
			\includegraphics[width=5.5cm]{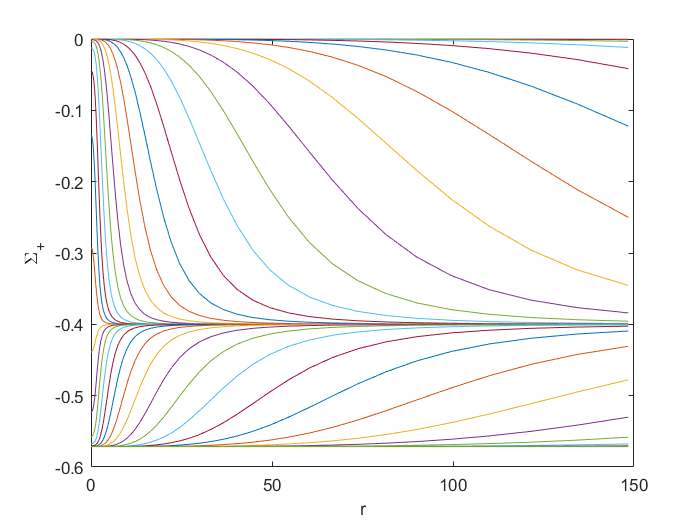}
			\caption{Snapshots at different times}
		\end{subfigure}
		\begin{subfigure}[t]{5.5cm}
			\includegraphics[width=5.5cm]{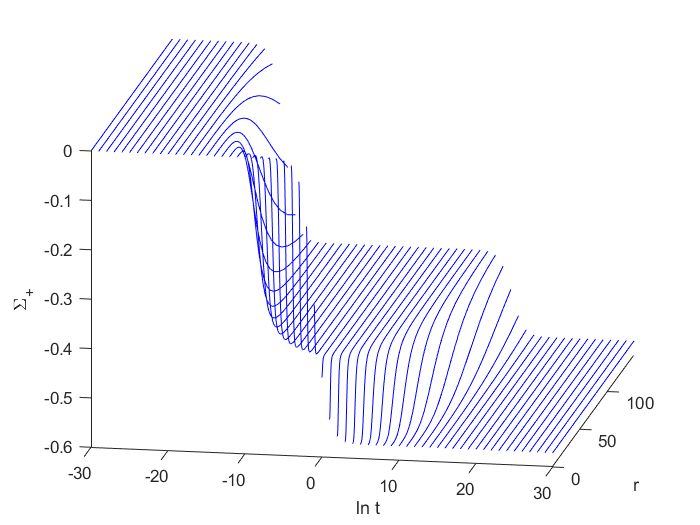}
			\caption{}
		\end{subfigure}
	\end{center}
	\caption{Plots of $\Sp$ for $k=1$, $\omega_{0}=1$ and $\Spo=0$, showing a transient spike along $r=0$.}
	\label{fig5.12}
\end{figure}

\begin{figure}
	\begin{center}
		\begin{subfigure}[t]{5.5cm}
			\includegraphics[width=5.5cm]{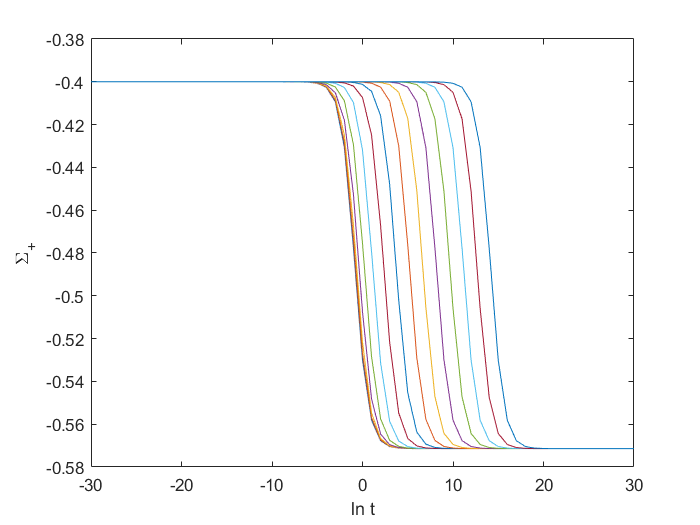}
			\caption{Evolution along different worldlines}
		\end{subfigure}
		\begin{subfigure}[t]{5.5cm}
			\includegraphics[width=5.5cm]{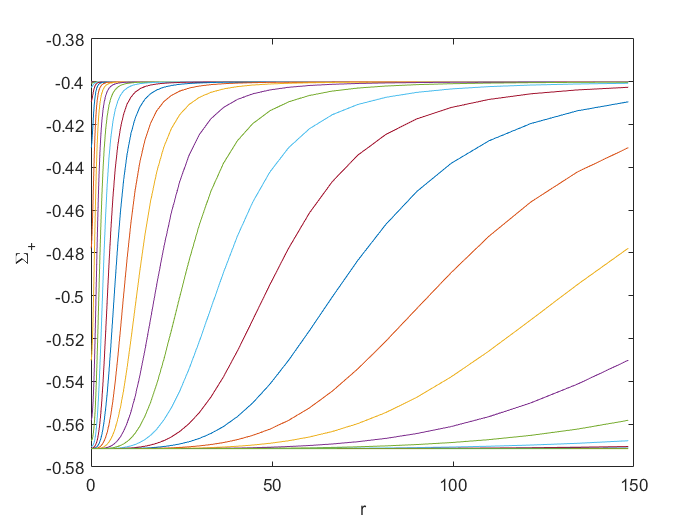}
			\caption{Snapshots at different times}
		\end{subfigure}
		\begin{subfigure}[t]{5.5cm}
			\includegraphics[width=5.5cm]{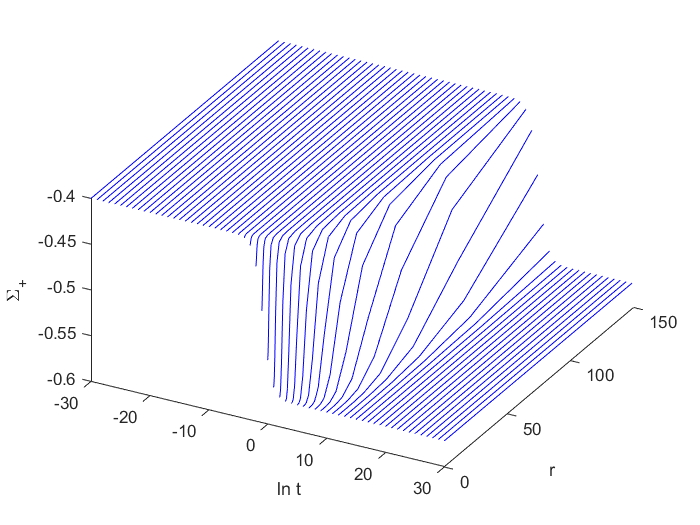}
			\caption{}
		\end{subfigure}
	\end{center}
	\caption{Plots of $\Sp$ for $k=1$, $\omega_{0}=0$ and $\Spo=0$, showing a transient spike along $r=0$.}
	\label{fig5.13}
\end{figure}

So after using these values, we get for $\omega_{0}=0$,
\begin{align}
\lim\limits_{t\rightarrow 0}(\Sp,\Sm)&=(-2/5,-2\sqrt{3}/5) \ \text{for all} \ \ r.\\
\lim\limits_{t\rightarrow\infty}(\Sp,\Sm)&=(-4/7,-4\sqrt{3}/7) \ \ \text{for all} \ \ r.
\end{align}
For $\omega_0\neq0$,
\begin{align}
\lim\limits_{t\rightarrow 0}(\Sp,\Sm)&=(0,0) \ \text{for all} \ \ r.\\
\lim\limits_{t\rightarrow\infty}(\Sp,\Sm)&=(-4/7,-4\sqrt{3}/7) \ \ \text{for all} \ \ r.
\end{align}
See Figure \ref{fig5.11}. While there are no permanent spikes, we see that in Figures \ref{fig5.12} and \ref{fig5.13} that there are transient spikes\footnote{Spike that form during transition for a short interval of time}. More analysis on transient spikes will be done in Chapter \ref{Chapter7}.

\newpage
\section{Radius of spike}
We define the coordinate radius of spike (or inhomogeneities) in (\ref{radius1}), which gives
\be
\text{coordinate radius }=\sqrt{\frac{(1-p_1)(k^2t^{1-p_1}-\omega_0)-kt^{2(1-p_1)}}{(1-p_1)(k\Spo-t^{2p_1})}}.
\ee
For $\Spo\geq0.5$, the coordinate radius tends to zero as $t\rightarrow\infty$, indicating the formation of permanent spike along $r=0$ at late times.


\section{Weyl scalars} 
To see whether the spikes are real or merely a coordinate effects at $r=0$ for the intervals $-1\leq \Spo< 0$ at early times and $0< \Spo\leq 0.5$ at late times, we are plotting the Weyl scalars. 

\begin{figure}
	\begin{center}
		\begin{subfigure}[t]{5.5cm}
			\includegraphics[width=5.5cm]{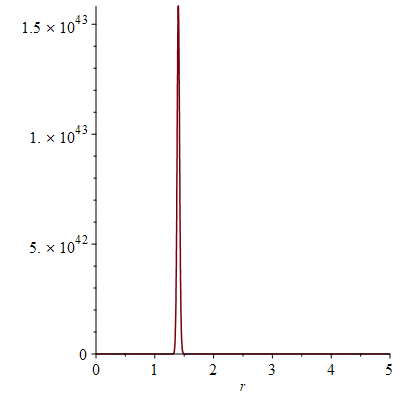}
			\caption{CC}
		\end{subfigure}
		\begin{subfigure}[t]{5.5cm}
			\includegraphics[width=5.5cm]{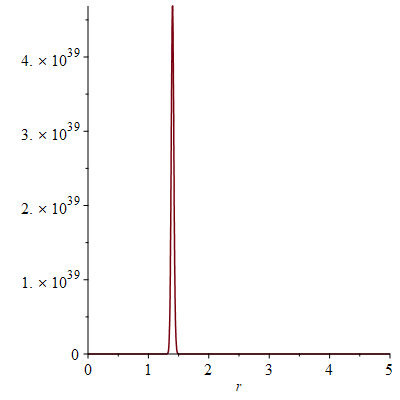}
			\caption{CCs}
		\end{subfigure}
		\begin{subfigure}[t]{5.5cm}
			\includegraphics[width=5.5cm]{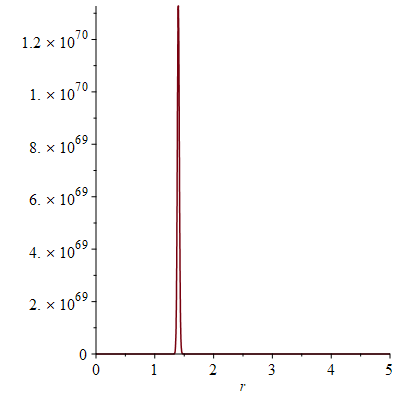}
			\caption{CCC}
		\end{subfigure}
		\begin{subfigure}[t]{5.5cm}
			\includegraphics[width=5.5cm]{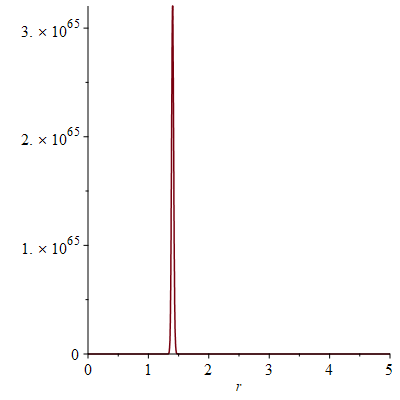}
			\caption{CCCs}
		\end{subfigure}
	\end{center}
	\caption{Weyl scalars when $t=0.001$, for $k=1$, $\omega_{0}=1$ and $\Spo=-\frac12$. Compare with Figure \ref{fig5.8}. This shows that the spike along $r=\sqrt{\frac{-\omega_{0}}{k\Spo}}$ is real, while the spike along $r=0$ is a coordinate effect  for the interval $-1\leq \Spo< 0$ at early times.}
	\label{fig5.14}
\end{figure}

First we plot Weyl scalars when $t=0.001$, for $k=1$, $\omega_{0}=1$ and $\Spo=-\frac12$ in Figure \ref{fig5.14}. Compare with Figure \ref{fig5.8}. This shows that the spike along $r=\sqrt{\frac{-\omega_{0}}{k\Spo}}$ is real, while the spike along $r=0$ is a coordinate effect. This strongly suggests that the spike along $r=0$ (caused by $l$) is a false spike for the intervals $-1\leq \Spo< 0$ at early times.

\begin{figure}
	\begin{center}
		\begin{subfigure}[t]{5.5cm}
			\includegraphics[width=5.5cm]{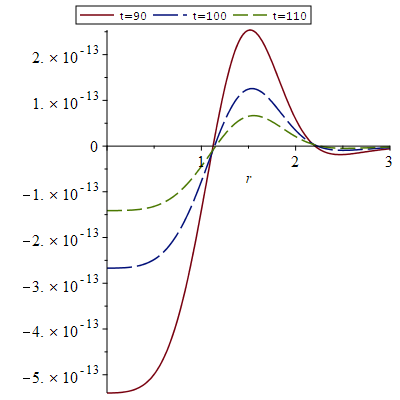}
			\caption{CC}
		\end{subfigure}
		\begin{subfigure}[t]{5.5cm}
			\includegraphics[width=5.5cm]{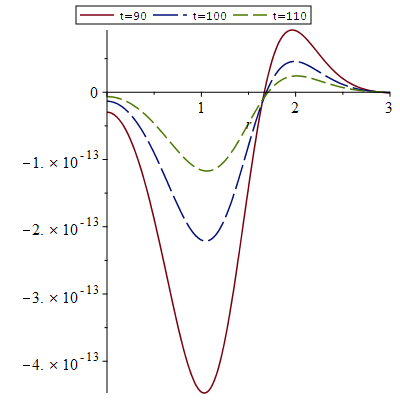}
			\caption{CCs}
		\end{subfigure}
		\begin{subfigure}[t]{5.5cm}
			\includegraphics[width=5.5cm]{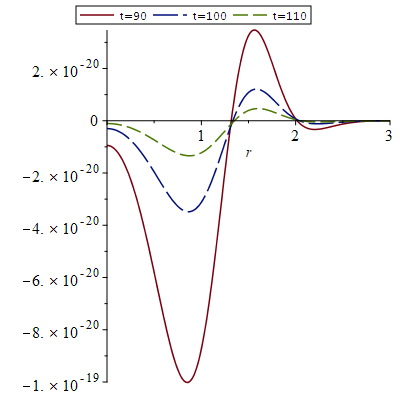}
			\caption{CCC}
		\end{subfigure}
		\begin{subfigure}[t]{5.5cm}
			\includegraphics[width=5.5cm]{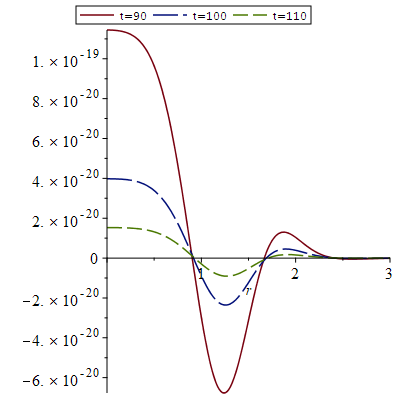}
			\caption{CCCs}
		\end{subfigure}
	\end{center}
	\caption{Weyl scalars at late times, for $k=1$, $\omega_{0}=-1$ and $\Spo=\frac13$. We see that curve is not narrowing as time increases. i.e it is not a spike. 
		Compare with Figures \ref{fig5.3}$-$\ref{fig5.4}. This shows that the spike along $r=0$ is not real for the interval $0< \Spo\leq 0.5$ at late times..}
	\label{fig5.15}
\end{figure}

Similarly, Figure \ref{fig5.15}, we plot Weyl scalars at late times, for $k=1$, $\omega_{0}=-1$ and $\Spo=\frac13$. We see that curve is not narrowing as time increases. i.e it is not a spike.
Compare with Figures \ref{fig5.3}$-$\ref{fig5.4}. This shows that the spike along $r=0$ is not real. This strongly suggests that the spike along $r=0$ (again caused by $l$) is a false spike for the interval $0< \Spo\leq 0.5$ at late times.

While we do not have a proof that the spike caused by $l$ is a false spikes, the Weyl scalar plots are strong evidence that the spike is a false spike.
\section{Discussion}
We have generated a solution that has permanent spikes that form at early and late times. We have done this by applying the Stephani transformation on the LRS Jacobs solution for the general case when $k\neq0$ by using the rotational KVF. The generated solution is cylindrically symmetric, and has a real spike along its rotational axis for the case $\frac12< \Spo\leq1$ at late times for $\omega_{0}\neq0$, and a real spike along the surface  $r=\sqrt{\frac{-\omega_{0}}{k\Spo}}$ at early times for the case $-1<\Spo<0$, $k\omega_{0}>0$ and the case $0<\Spo<\frac12$, $k\omega_0<0$.

We also see that transient spikes form in the case $\Spo=0$. In fact transient spikes also form in other cases. We will carry out intermediate times analysis in Chapter 7. 

Compared with the $k=0$ case, the $k\neq0$ case presents a variety of phenomena --- false spikes, transient spikes, and a second spike along $r=\sqrt{\frac{-\omega_{0}}{k\Spo}}$.

\chapter{A heuristic for permanent spikes}
Is there a quick way to determine whether a Geroch/Stephani transformation solution has permanent spike? If so, whether the permanent spike forms at early or late times? We shall develop a heuristic to answer these two questions.

For our experience in Chapter 4 and Chapter 5, we observed that the expression
\[
f=t(\ln F)_t=\frac{t(\lambda^2)_t+t(\omega^2)_t}{\lambda^2+\omega^2}
\]
has discontinuous limit if and only if a permanent spike forms.

In Chapter 4, we had, for $-1<\Spo\leq1$, discontinuous limit
\begin{align}
\label{6.1}
\lim\limits_{t\rightarrow\infty}f=
\begin{cases}
0  &\text{along} \ \ \ r=0\\
4p_1  &\text{along} \ \ r\neq0,
\end{cases}
\end{align}
which corresponds to a late-time permanent spike at the cylindrical axis, $r=0$,

Similarly, in Chapter 5, we had, for $\frac12\leq\Spo\leq1$, discontinuous limit.
\begin{align}
\label{6.2}
\lim\limits_{t\rightarrow\infty}f=
\begin{cases}
2(1+p_3)  &\text{along} \ \ \ r=0\\
4p_1  &\text{along} \ \ r\neq0.
\end{cases}
\end{align}
In Chapter 5, we had, in addition, the following discontinuous limits.
For $-1<\Spo<0$ and $\omega_{0}k>0$,
\begin{align}
\label{6.3}
\lim\limits_{t\rightarrow 0}f=
\begin{cases}
4p_1  &\text{along} \ \ \ r=\sqrt{-\frac{\omega_0}{k\Spo}}\\
0    &\text{along} \ \ r\neq \sqrt{-\frac{\omega_0}{k\Spo}}.
\end{cases}
\end{align}
For $0<\Spo<\frac12$ and $\omega_{0}k<0$,
\begin{align}
\label{6.4}
\lim\limits_{t\rightarrow 0}f=
\begin{cases}
4p_3  &\text{along} \ \ \ r=\sqrt{-\frac{\omega_0}{k\Spo}}\\
0    &\text{along} \ \ r\neq \sqrt{-\frac{\omega_0}{k\Spo}}.
\end{cases}
\end{align}
These correspond to an early-time permanent spike on the cylindrical shell $r=\sqrt{-\frac{\omega_0}{k\Spo}}$.

It is not necessary that we always have a discontinuous limits. If we see in the Chapter 4 we have a continuous limit at $r=0$ at early times,
\begin{align}
\lim\limits_{t\rightarrow 0}f=
\begin{cases}
0  &\text{along} \ \ \ r=0\\
0  &\text{along} \ \ r\neq0.
\end{cases}
\end{align}
and if we see in the Chapter 5, we also have the continuous limits
for $-1\leq\Spo\leq 0$ and $\omega_{0}k<0$,
\begin{align}
\lim\limits_{t\rightarrow \infty}f=
\begin{cases}
2(1+p_3)  &\text{along} \ \ \ r=0\\
2(1+p_3)   &\text{along} \ \ r\neq0.
\end{cases}
\end{align}

In (\ref{6.1}), the dominant term in the limit is contributed by \[\lambda=r^2t^{2p_1},\]
unless its coefficient $r^2$ is zero.

In (\ref{6.2}), the dominant term is again contributed by the $r^2t^{2p_1}$ term in $\lambda$, unless the coefficient $r^2$ is zero.

In (\ref{6.3}) and (\ref{6.4}) the dominant term is contributed by the time-independent term in $\omega$, unless the term is zero.

For another example, consider the OT $G_2$ spike solution (\cite{art:Lim2015},case $n_{10}=n_{20}=0$). These we have, for $|w|<1$, discontinuous limits
\[\lim\limits_{\tau \rightarrow\infty}f=
\begin{cases}
-|w|+1  &\text{along} \ \ \ z=0\\
0  &\text{along} \ \ z\neq0,
\end{cases}\]
which corresponds to an early-time permanent spike on the plane $z=0$.\footnote{$f=-(\ln F)_\tau$ in the time variable $\tau$.} The dominant term in the limit is contributed by the time-independent 
\[\omega=kz,\]
where $k$ is a nonzero constant, unless $z=0$.

In all the above examples, permanent spike occur because $f$ has a discontinuous limit, which in term is due to the fact that the dominant term has a spatially dependent coefficient that can become zero along certain worldlines.

This gives a heuristic to quickly determine whether a given seed solution leads to a generated solution with permanent spikes. The steps are: 
\begin{flushleft}
	1:	Compute $\lambda$, $\omega$ and $f$.
\end{flushleft} 
\begin{flushleft}
	2: Find the dominant term (for each asymptotic regime) and look at its coefficient.	
\end{flushleft}
\begin{flushleft}
	3: If the coefficient vanishes along a certain worldline, then expect a permanent spike to form along the worldline.
\end{flushleft}

This also explains why the rotational KVF can leads to permanent spike at the cylindrical axis -- its length vanishes at the cylindrical axis. Translational KVFs, whose length is non-vanishing  every where in space, do not have this mechanism.
\chapter{Transient spike and other inhomogeneous structures}
\label{Chapter7}
Figure \ref{fig5.12} and \ref{fig5.13} exhibit some interesting features of $\Sp$. $\Sp$ transitions from one equilibrium state to the next, at certain transition time that is spatially dependent. Each equlibrium state is coordinate independent. Recall from (\ref{GSigma})  that the dynamics of $\Sp$ is solely due to the dynamics of $f$. It is therefore important to take a closer look at $f$. In this chapter, we shall employ a method of analysis that has not been used to analyse spike solutions before. 
\section{Dynamical analysis of $f$}
\label{Section7.1}
Recall that $F=\lambda^2+\omega^2$ and $\lambda =r^2t^{2p_1}+k^2t^{2p_3}$ and $\omega=\frac{2k}{1+p_3}t^{1+p_3}+k\Spo r^2+\omega_{0}$. Observe that $\lambda$ and $\omega$ are sums of powers of $t$. There are four different powers, so we group them into 4 terms on the basis of the power of $t$:
\begin{equation}
\label{transient_division}
T_1= r^2t^{2p_1}, \ \ T_2=k^2t^{2p_3}, \ \ \ T_3= \frac{2k}{1+p_3}t^{1+p_3},\ \  \ T_4= k\Spo r^2+\omega_{0}.
\end{equation} 

\begin{figure}
	\begin{center}
		\includegraphics[width=10.5cm]{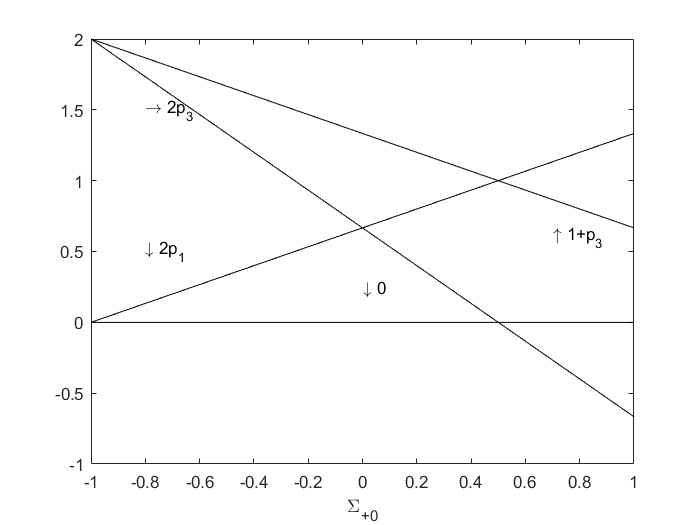}
	\end{center}
	\caption{Power of $t$ of the terms in (\ref{transient_division}) against $\Spo$.}
	\label{Combine4stage}
\end{figure}
Figure \ref{Combine4stage} plots the power of $t$ of each term in (\ref{transient_division}) against the parameter $\Spo$. In general, the four powers are distinct, except for 3 special values of $\Spo$. For $\Spo=-1$, there are two distinct powers; for $\Spo=0$, three distinct powers; and for $\Spo=\frac12$, two distinct powers.
The term with the largest power of $t$ dominates at late times; the term with smallest power of $t$ dominates at early times; and the terms with intermediate power of $t$ may or may not dominate for a finite time interval, depending on how big their coefficient is. 
Expressed in terms of $T_1$, $T_2$, $T_3$, $T_4$, 
\begin{align}
\label{fGeneralT}
f=\frac{2(T_1+T_2)(2p_1T_1+2p_3T_2)+2(T_3+T_4)(1+p_3)T_3}{(T_1+T_2)^2+(T_3+T_4)^2}.
\end{align}
Observe that
\begin{align}
f \approx \begin{cases}
4p_1 \ \ \ \ &\text{when } T_1 \text{ dominates}\\
4p_3 \ \ \ \ &\text{when } T_2 \text{ dominates}\\
2(1+p_3) \ \ \ \ &\text{when } T_3 \text{ dominates}\\
0 \ \ \ \ &\text{when } T_4 \text{ dominates}.
\end{cases}
\end{align}
That is, $f$ is approximately twice the value of the power of the dominant term. Furthermore the powers depend only on the parameter $\Spo$. Its independence of coordinates gives the graph of $f$ a cascading appearance.
An equilibrium state corresponds to a dominant term. Therefore, there are up to 4 distinct equilibrium states for general $\Spo$; 3 for $\Spo=0$ and 2 for $\Spo=-1$ and $\Spo=1/2$. 
The value of $f$ at successive equilibrium states is strictly increasing in time. Among the four values, $4p_3$ is negative for  $0.5<\Spo\leq 1$, with a minmum value of $-\frac{4}{3}$ at $\Spo=1$, which is still greater that $-2$, so the Hubble scalar $H$ is positive at each equilibrium state. But we will see later that $f$ can become less than $-2$ during so-called overshoot transitions. 

We define the transition time between two equilibrium states or dominant terms to be the time when both terms are equal in magnitude. For example, solving $T_1^2=T_2^2$ for $t$ yields the transition time 
\begin{equation}
t_{12}=\left(\frac{k^2}{r^2}\right)^\frac{1}{2\Spo}.
\end{equation}
Comparing the transition times will determine how many transitions an observer with fixed $r$  undergoes.
The coefficients of $T_1$ and $T_4$ have spatial dependence. They can even vanish for certain worldline ($r=0$ for $T_{1}$, and $r=\sqrt{\frac{-\omega_0}{k\Spo}}$ for $T_4$, provided that $\frac{\omega_{0}}{k \Spo}\leq0$), which create spikes along these worldlines. The spikes are called transient if the term dominates for an intermediate, finite time interval.
Some transition times have spatially dependence as a consequence of the spatially dependent coefficient. This means there are inhomogeneities in transition times except $t_{23}$.

The transition time between two dominant terms can be regarded as roughly the boundary between the two corresponding equilibrium states. We say ``roughly" because the transition is a smooth, continuous process, so there is no sharp boundary. If a transition time has spatial dependence, it also gives the spatial location of the boundary at a fixed time. The spacetime is partitioned into regions of equlibrium states, separated by transition times. When viewed at a fixed time, we can regard space as being partitioned into cells of equilibrium states, separated by walls (around which spatial gradient is large). If two walls are near each other, we see a narrow cell.
The neighbourhood of the narrow cell shall be called a spike if certain additional conditions are met. We will discuss these conditions later in Section \ref{Section7.2}.
\subsection{Case $\Spo=-1$}
For $\Spo=-1$, we have 
\[T_1= r^2, \ \ T_2=k^2t^{2}, \ \ \ T_3= kt^{2},\ \  \ T_4= -kr^2+\omega_{0}.\]
There are only two distinct powers of $t$, with $T_1$ and $T_4$ dominating at early times, and $T_2$ and $T_3$ dominating at late times. This conclusion can also be arrived at by inspecting  Figure \ref{Combine4stage}. We denote the sequence of dominant equilibrium states as 
\begin{equation}
T_{1} \ \ \ 
\&  \ \ T_{4} \ \ \longrightarrow\ \ \ T_{2}\ \ \ \&\ \ \ T_{3}.\ \ \
\end{equation}
Solving the equation
\[T_{1}^2+T_{4}^2 \  =\ T_{2}^2+ T_{3}^2\  \]
for $t$ yields the transition time
\begin{align}
t_{\left(1 \& 4\right)\left(2 \& 3\right)}= &\left(\frac{(\omega_{0}-kr^2)^2+r^4}{k^2(k^2+1)}\right)^{\frac{1}{4}}.
\end{align}

\begin{figure}
	\begin{center}
		\includegraphics[width=10.5cm]{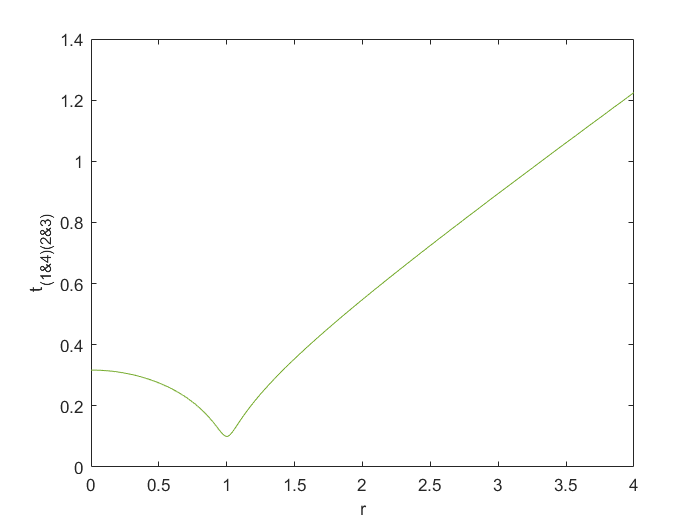}
	\end{center}
	\caption{Transition time $t_{\left(1 \& 4\right)\left(2 \& 3\right)}$ as a function of $r$ for $\Spo=-1$, $k=10$ and $\omega_{0}=10.1$. $t_{\left(1 \& 4\right)\left(2 \& 3\right)}$ has a global minimum at $r=\sqrt{\frac{\omega_0 k}{k^2+1}}= 1$.} 
	\label{fig_2stagem1tr}
\end{figure}
\begin{figure}
	\begin{center}
		\includegraphics[width=10.5cm]{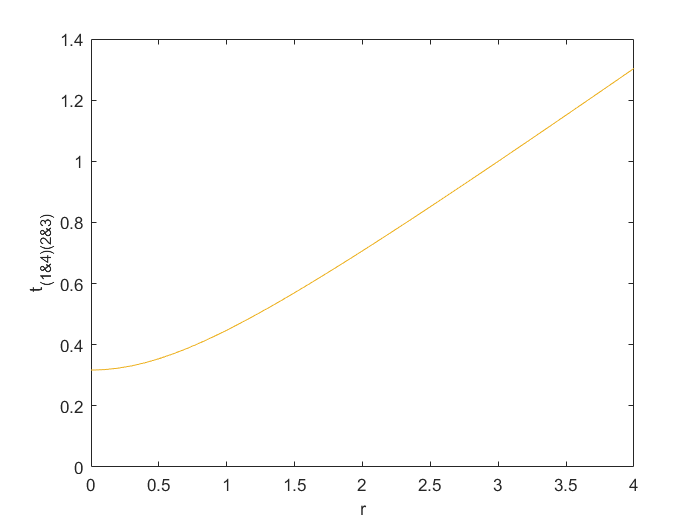}
	\end{center}
	\caption{Transition time $t_{\left(1 \& 4\right)\left(2 \& 3\right)}$ as a function of $r$ for  $\Spo=-1$, $k=10$ and $\omega_{0}=-10.1$. $t_{\left(1 \& 4\right)\left(2 \& 3\right)}$ has a global minimum at $r=0$.}
	\label{fig_2stagem1trn}
\end{figure}
\begin{figure}
	\begin{center}
		\includegraphics[width=10.5cm]{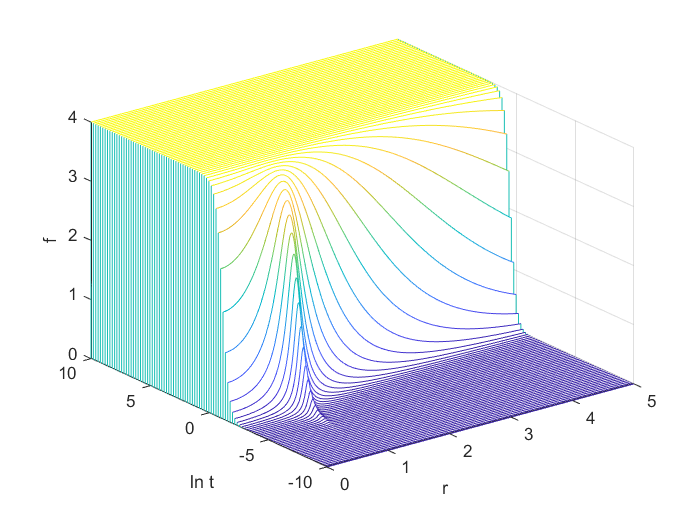}
	\end{center}
	\caption{$f$ against $\ln t$ and $r$ for $\Spo=-1$, $k=10$ and $\omega_{0}=10.1$. The transition time has a global minimum at $r=1$.}
	\label{fig_2stagem1}
\end{figure}
\begin{figure}
	\begin{center}
		\includegraphics[width=10.5cm]{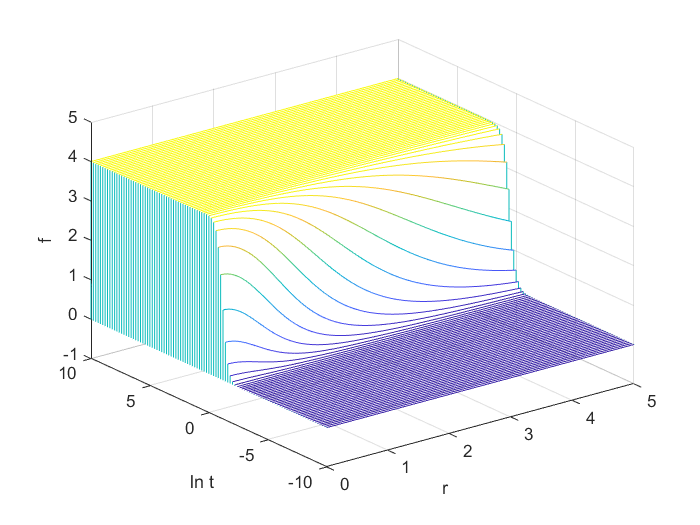}
	\end{center}
	\caption{$f$ against $\ln t$ and $r$ for $\Spo=-1$, $k=10$ and $\omega_{0}=-10.1$. The transition time has a global minimum at $r=0$.}
	\label{fig_2stagem1n}
\end{figure}
We now analyse the behaviour of $t_{\left(1 \& 4\right)\left(2 \& 3\right)}$ as a function of $r$. Observe that $\lim\limits_{r\rightarrow \infty}t_{\left(1 \& 4\right)\left(2 \& 3\right)}=\infty$. If $\omega_{0} k >0$, then $t_{\left(1 \& 4\right)\left(2 \& 3\right)}$ has a global minimum at $r=\sqrt{\frac{\omega_0 k}{k^2+1}}$, otherwise it has a global minimum at $r=0$. For example, if we have $\Spo=-1$, $k=10$ and $\omega_{0}=10.1$, then   $t_{\left(1 \& 4\right)\left(2 \& 3\right)}$ has a global minimum  at $r=\sqrt{\frac{\omega_0 k}{k^2+1}}= 1$ (see Figure \ref{fig_2stagem1tr}). But if we have $\Spo=-1$, $k=10$ and $\omega_{0}=-10.1$, then $t_{\left(1 \& 4\right)\left(2 \& 3\right)}$ has a global minimum at $r=0$ (see Figure \ref{fig_2stagem1trn}).
We plot $f$ against $\ln t$ and $r$ for these examples in Figures \ref{fig_2stagem1} and \ref{fig_2stagem1n}.

Note that, we do not have spikes in this case, even through $f$ looks spiky around $r=1$ during transition in Figure \ref{fig_2stagem1}. But if $\omega_{0}=0$ then we have permanent spike at early time (see Section \ref{Dynamics of the solution}).
\subsection{Case $-1<\Spo<0$}
\label{Section7.1.2}
\begin{figure}
	\begin{center}
		\includegraphics[width=12.5cm]{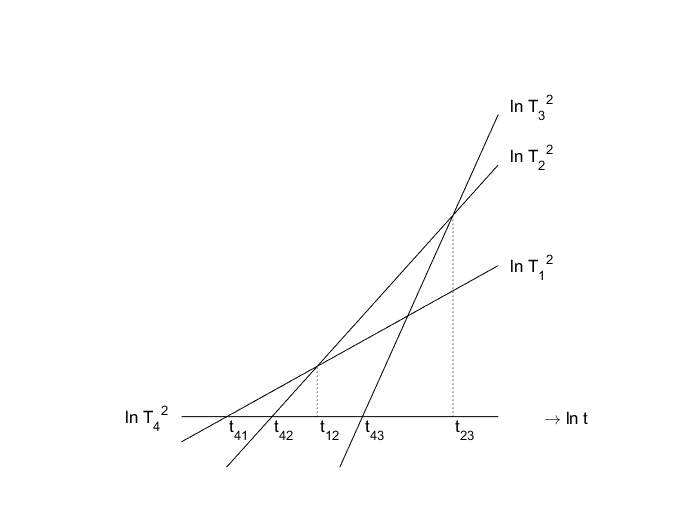}
	\end{center}
	\caption{Qualitative plot of the log of each term squared against $\ln t$, showing 4 dominant equilibrium states, for any value of $\Spo$ satisfying $-1<\Spo<0$.}
	\label{fig_4eqstatem05c}
\end{figure}

\begin{figure}
	\begin{subfigure}[t]{7cm}
		\includegraphics[width=6.5cm]{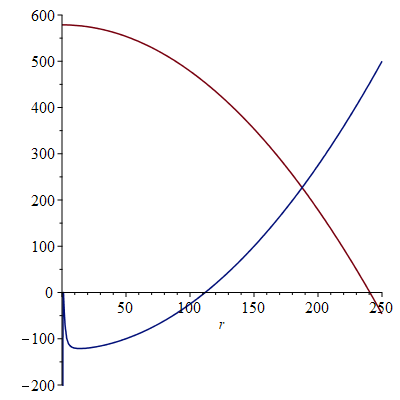}
		\caption{For $0\leq r\leq250$}
	\end{subfigure}
	\begin{subfigure}[t]{5.5cm}
		\includegraphics[width=5.5cm]{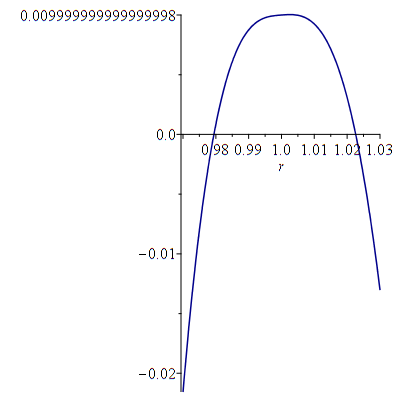}
		\caption{For $0.97\leq r\leq1.03$}
	\end{subfigure}
	\caption{Blue line is the plot of $t_{12}-t_{41}$  and red line is $ t_{23}-t_{12}$, when $\Spo=-0.5$, $k=10$, $\omega_{0}=5$. The blue line is positive for a small interval around $r=1$ and for $r>111.7900$. The red line is positive for $r<240.5626$.}
	\label{fig_4stagem05term}
\end{figure}

\begin{figure}
	\begin{center}
		\includegraphics[width=10.5cm]{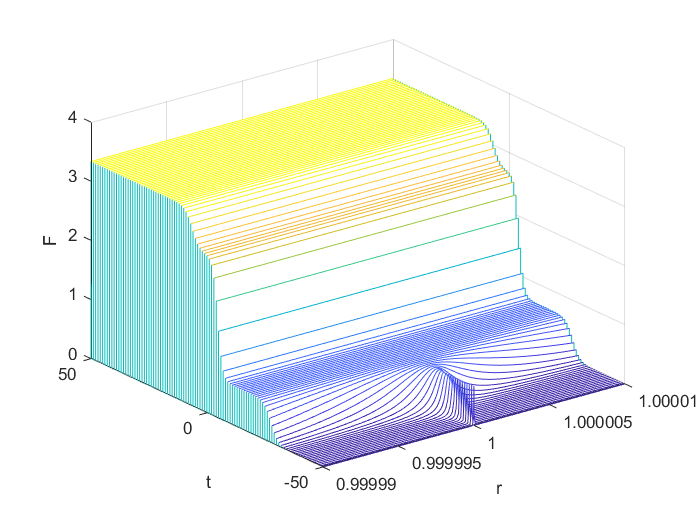}
	\end{center}
	\caption{$f$ against $\ln t$ and $r$ for $\Spo=-0.5$, $k=10$ and $\omega_{0}=5$ for the interval $0.99999<r<1.00001$, showing 4 distinct states along $r\neq1$.}
	\label{fig_4stagem05}
\end{figure}

\begin{figure}
	\begin{center}
		\includegraphics[width=10.5cm]{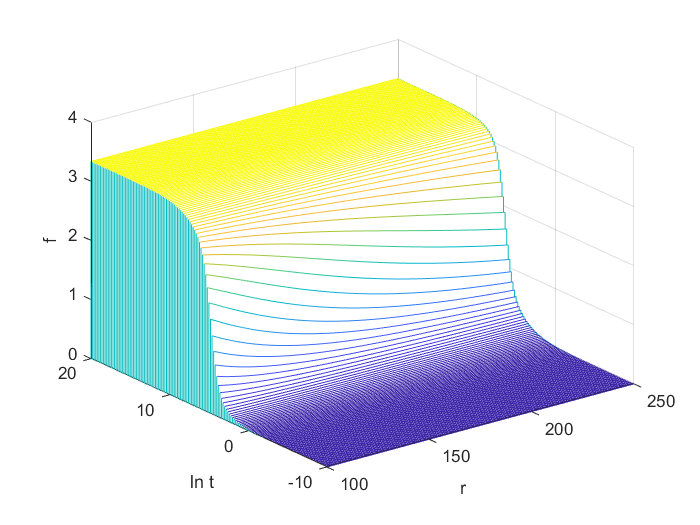}
	\end{center}
	\caption{$f$ against $\ln t$ and $r$ for $\Spo=-0.5$, $k=10$ and $\omega_{0}=5$ for the interval $100<r<250$, showing 2 visible distinct states because the transition times are too close together.}
	\label{fig_4stagem05interval2}
\end{figure}

\begin{figure}
	\begin{center}
		\includegraphics[width=8.5cm]{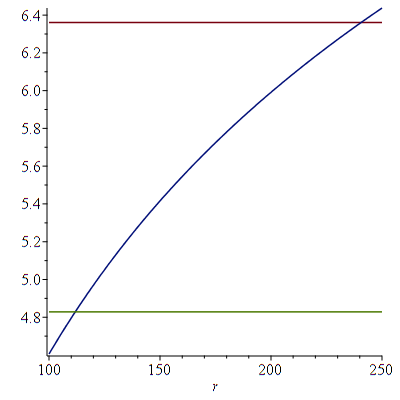}
	\end{center}
	\caption{The log of transition times $t_{41}$ (green), $t_{12}$ (blue) and $t_{23}$ (red)  against $r$ for $\Spo=-0.5$, $k=10$ and $\omega_{0}=5$ for the interval  $100<r<250$, showing that the transition times are close together.}
	\label{fig_4stagem05plot3t}
\end{figure}

For the case $-1<\Spo<0$, Figure \ref{Combine4stage} gives the ordering $T_4$,  $T_1$, $T_2$, $T_3$, in increasing power of $t$. We have up to 4 distinct equilibrium states, and along general worldlines there are 4 possible sequences of dominant equilibrium states, which we shall refer to as scenarios:
\begin{enumerate}
	\item  $T_{4}$ $\longrightarrow$ $T_{1}$ $\longrightarrow$ $T_{2}$ $\longrightarrow$ $T_{3}$
	\item $T_{4}$ $\longrightarrow$ $T_{2}$ $\longrightarrow$ $T_{3}$
	\item  $T_{4}$ $\longrightarrow$ $T_{1}$  $\longrightarrow$ $T_{3}$
	\item  $T_{4}$  $\longrightarrow$ $T_{3}$.
\end{enumerate}
There are two special worldlines. The first one is $r=0$, where $T_1$ vanishes. The possible scenarios along this worldline are:
\begin{enumerate}
	\item $T_{4}$ $\longrightarrow$ $T_{2}$ $\longrightarrow$ $T_{3}$
	\item  $T_{4}$  $\longrightarrow$ $T_{3}$,
\end{enumerate}
which are qualitatively the same as scenarios 2 and 4 above. The second special worldline is $r=\sqrt{\frac{-\omega_{0}}{k \Spo}}$, where $T_4$ vanishes, giving an early-time permanent spike.
The possible scenarios along this worldline are:
\begin{enumerate}
	\item $T_{1}$ $\longrightarrow$ $T_{2}$ $\longrightarrow$ $T_{3}$
	\item  $T_{1}$  $\longrightarrow$ $T_{3}$.
\end{enumerate}
The two special worldlines coincide if $\omega_{0}=0$. In this case the only possible scenarios along this worldline is
\[T_{2} \longrightarrow T_{3}.\]
We now introduce a useful diagram. From (\ref{transient_division}), we see that the logarithm of the square of each term is a linear function of $\ln t$. Figure \ref{fig_4eqstatem05c} shows a qualitative plot of the log of each term squared against $\ln t$, for the scenario 

\[T_{4} \longrightarrow  T_{1} \longrightarrow   T_{2}  \longrightarrow  T_{3}.\]
It is clear from the diagram that the transition times 
\begin{align}
t_{41}=\left(\frac{|k \Spo r^2 +\omega_{0}|}{r^2}\right)^{\frac{1}{2p_1}},\ \ \ &t_{12}=\left(\frac{k^2 }{r^2}\right)^{\frac{1}{2\Spo}}, \ \ \ \  t_{23}=\left(\frac{|k|(2- \Spo)}{3}\right)^{\frac{1}{2p_1}}
\end{align}
must satisfy the condition
\begin{align}
\label{4stageconditionm05}
t_{41}<t_{12}<t_{23}.
\end{align}
$t_{41}<t_{12}$ implies 
\begin{align}
\label{eq1hash}
|k \Spo r^2 +\omega_{0}| < \left(\frac{|k|^{p_1}}{r^{p_3}}\right)^\frac{2}{\Spo},
\end{align}
which gives one or more intervals of $r$. 
$t_{12}<t_{23}$ gives an upper bound on $r$:
\begin{equation}
\label{eq1hashr}
r<|k|\left(\frac{3}{|k|(2-\Spo)}\right)^{\frac{\Spo}{2 p_1}}.
\end{equation}
So the condition (\ref{4stageconditionm05}) restricts $r$ to one or more intervals. 
As a concrete  example, take $\Spo=-0.5$, $k=10$ and $\omega_{0}=5$.
(\ref{eq1hash}) can be solved numerically to give the intervals
\begin{align}
\label{eq2hash}
0.9794<r<1.0226 \ \ \ \ & \ \ \text{and} \ \ \ \ \ \ \ 111.7900<r.
\end{align}
Note that $r=1$ is the second special worldline, so it must be excluded from this scenario.
(\ref{eq1hashr}) gives $r<240.5626$. Together, the scenario occurs for the interval 
\begin{align}
0.9794<r<1,\ \ \  1<r&<1.0226 \ \text{ and} \ \ \ 111.7900<r<240.5626.
\end{align} 
See Figure \ref{fig_4stagem05term}. We plot $f$ against $\ln t$ and $r$ on a small interval around $r=1$ in Figure \ref{fig_4stagem05}, showing 4 distinct states along $r\neq1$. Along $r=1$, there is a permanent spike at early times.
We plot $f$ against $\ln t$ and $r$  for the interval $100<r<250$ in Figure \ref{fig_4stagem05interval2}, showing 2 visible distinct states because the transition times are too close together (see Figure \ref{fig_4stagem05plot3t}).
So if transition times are too close together, we see fewer visible distinct state than the actual number of states predicted by the scenario.

\begin{figure}
	\begin{center}
		\includegraphics[width=12.5cm]{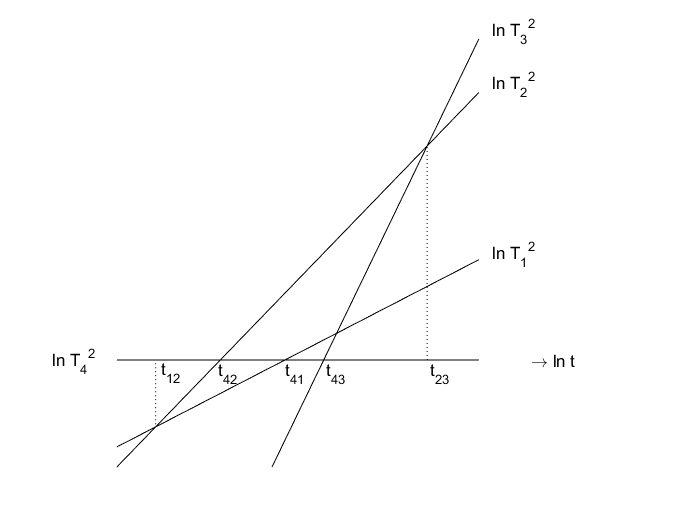}
	\end{center}
	\caption{Qualitative plot of the log of each term squared against $\ln t$, showing 3 dominant equilibrium states, for any value of $\Spo$ satisfying $-1<\Spo<0$.}
	\label{fig_3eqstate1m05c}
\end{figure}

What happens in other intervals of $r$?
From (\ref{eq2hash}), we know that $t_{41}$ becomes greater than $t_{12}$ for values of $r$ just beyond the boundaries. From the diagram in Figure \ref{fig_4eqstatem05c}, this happens if the graph of $\ln T_1^2$ becomes too low, as shown in Figure \ref{fig_3eqstate1m05c}. 
Now, the diagram in Figure \ref{fig_3eqstate1m05c} shows the scenario
\begin{align}
T_{4}  \longrightarrow T_{2} &  \longrightarrow  T_{3},
\end{align}
with transition times
\begin{align}
t_{42}=\left(\frac{|k \Spo r^2 +\omega_{0}|}{k^2}\right)^{\frac{1}{2p_3}}\ \ \, &\ \ \  t_{23}=\left(\frac{|k|(2- \Spo)}{3}\right)^{\frac{1}{2p_1}}.
\end{align}
They must satisfy the condition
\begin{align}
\label{3stage1conditionm05}
t_{12}<t_{42}<t_{23}.
\end{align}
The condition $t_{12}<t_{42}$ is equivalent to $t_{12}<t_{41}$, so it gives (\ref{eq1hash}) with the opposite inequality direction:
\begin{align}
\label{eq3hash}
|k \Spo r^2 +\omega_{0}| > \left(\frac{|k|^{p_1}}{r^{p_3}}\right)^\frac{2}{\Spo}.
\end{align}
$t_{42}<t_{23}$ implies
\begin{equation}
\label{eq4hash}
|k \Spo r^2 +\omega_{0}|<|k|^{\frac1{p_1}}\left(\frac{2-\Spo}{3}\right)^\frac{p_3}{p_1},
\end{equation}
which gives rise to one interval of $r$. Together, the condition restricts $r$ to one or more intervals. Continuing with the same example, (\ref{eq3hash}) gives the intervals
\begin{align}
r<0.9794 \ \ \ \ & \ \ \text{and}\ \ \ \ \ \ \ 1.0226<r<111.7900,
\end{align}
while
(\ref{eq4hash}) gives the interval $r<310.5666$. Together, the scenario occurs for the intervals
\begin{align}
0\leq r<0.9794 \ \ \ \ & \ \ \text{and}\ \ \ \ \ \ \ 1.0226<r<111.7900.
\end{align}

\begin{figure}
	\begin{subfigure}[t]{7.5cm}
		\includegraphics[width=6.5cm]{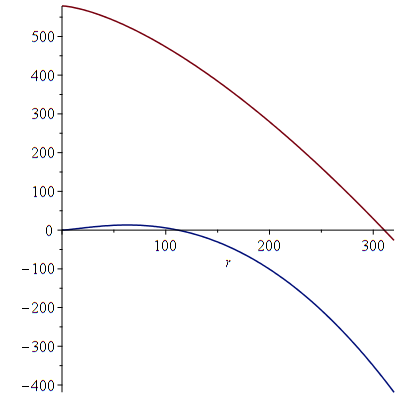}
		\caption{For $0\leq r\leq350$}
	\end{subfigure}
	\begin{subfigure}[t]{5.5cm}
		\includegraphics[width=5.5cm]{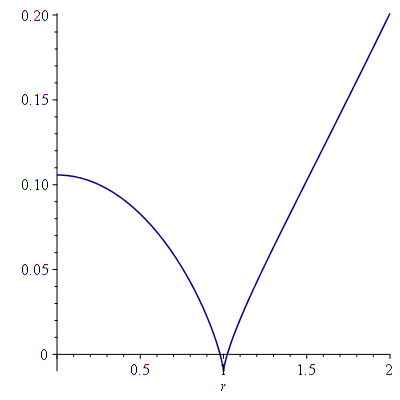}
		\caption{For $0\leq r\leq 2$}
	\end{subfigure}
	\caption{Blue line is the plot of $t_{42}-t_{12}$  and red line is $ t_{23}-t_{42}$, when $\Spo=-0.5$, $k=10$, $\omega_{0}=5$. The blue line is positive for a small interval $0\leq r<0.9794$ and for $1.0226<r<111.7900$. The red line is positive for $r<310.5666$.}
	\label{fig_3stage1m05term}
\end{figure}

\begin{figure}
	\begin{center}
		\includegraphics[width=10.5cm]{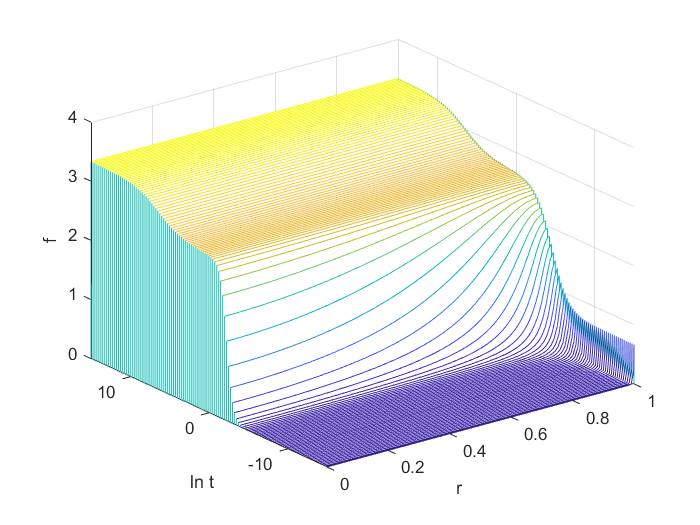}
	\end{center}
	\caption{$f$ against $\ln t$ and $r$ for $\Spo=-0.5$, $k=10$ and $\omega_{0}=5$ for the interval $0\leq r<0.9794$, showing 3 distinct states. From $r$ greater than $0.9794$, we have the 4-state scenario in Figure \ref{fig_4stagem05}.}
	\label{fig_3stage1m05}
\end{figure}

\begin{figure}
	\begin{center}
		\includegraphics[width=10.5cm]{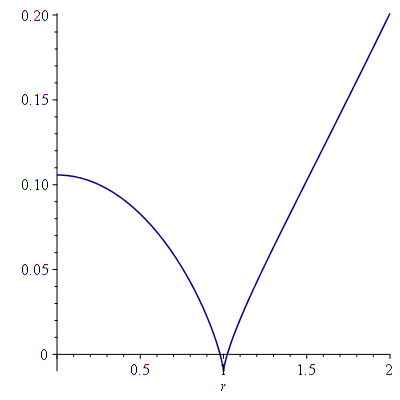}
	\end{center}
	\caption{$f$ against $\ln t$ and $r$ for $\Spo=-0.5$, $k=10$ and $\omega_{0}=5$ for the interval $1< r<120$, showing 3 distinct states for small $r$ which fade away to two visible states as the transition times become closer together as $r$ increases.}
	\label{fig_3stage1m05interval2}
\end{figure}

\begin{figure}
	\begin{center}
		\includegraphics[width=8.5cm]{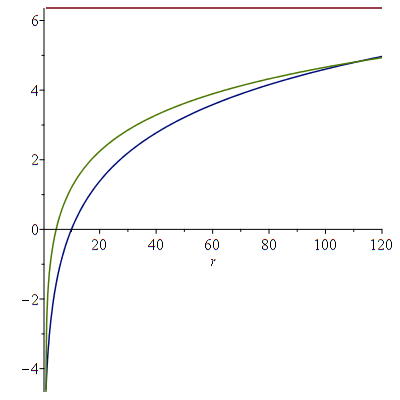}
	\end{center}
	\caption{The log of transition times  $t_{12}$ (blue), $t_{42}$ (green) and $t_{23}$ (red)  against $r$ for $\Spo=-0.5$, $k=10$ and $\omega_{0}=5$ for the interval  $1<r<120$, showing that the transition time $t_{42}$ becomes closer to $t_{23}$ as $r$ increases.}
	\label{fig_3stagem105plot3t}
\end{figure}

See Figure \ref{fig_3stage1m05term}. We plot $f$ against $\ln t$ and $r$  on these intervals showing 3 distinct states in Figures \ref{fig_3stage1m05} and \ref{fig_3stage1m05interval2}. Figure \ref{fig_3stage1m05} shows $f$ against $\ln t$ and $r$  for the interval $0\leq r<0.9794$, showing 3 distinct states. From $r$ greater than $0.9794$, we have the 4-state scenario in Figure \ref{fig_4stagem05}. 
Figure \ref{fig_3stage1m05interval2} shows $f$ against $\ln t$ and $r$ for the interval $1< r<120$, showing 3 distinct states for small $r$ which fade away to two visible states as the transition times become closer together as $r$ increases. Figure \ref{fig_3stagem105plot3t} shows the log of transition times  $t_{12}$ (blue), $t_{42}$ (green) and $t_{23}$ (red)  against $r$  for the interval  $1<r<120$, showing that the transition time $t_{42}$ becomes closer to $t_{23}$ as $r$ increases.

\begin{figure}
	\begin{center}
		\includegraphics[width=10.5cm]{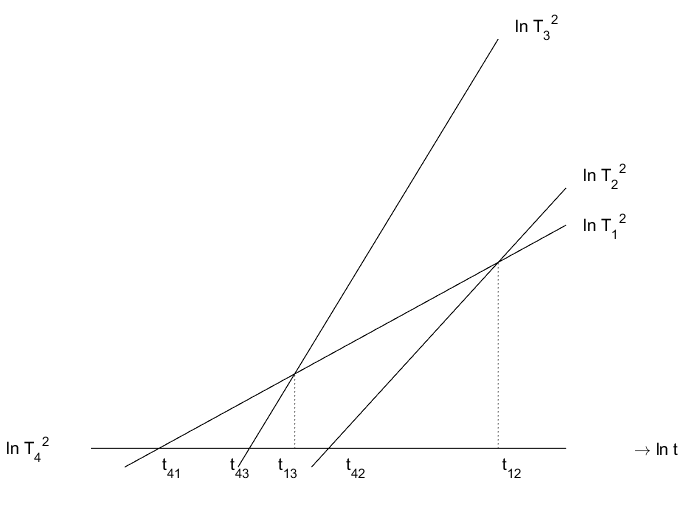}
	\end{center}
	\caption{Qualitative plot of the log of each term squared against $\ln t$, showing 3 dominant equilibrium states, for any value of $\Spo$ satisfying $-1<\Spo<0$.}
	\label{fig_3eqstate2m05c}
\end{figure}

\begin{figure}
	\begin{subfigure}[t]{5.5cm}
		\includegraphics[width=5.5cm]{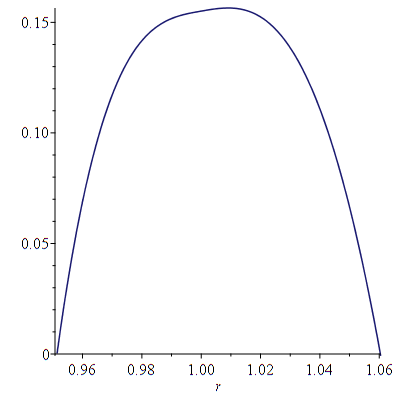}
		\caption{For $0.9514\leq r<1.0605$}
	\end{subfigure}
	\begin{subfigure}[t]{6.5cm}
		\includegraphics[width=6.5cm]{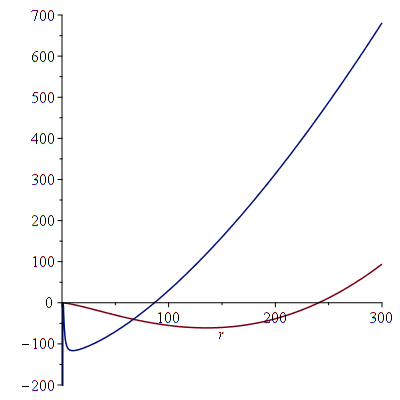}
		\caption{For $0\leq r\leq350$}
	\end{subfigure}
	\caption{Blue line is the plot of $t_{13}-t_{41}$  and red line is $ t_{12}-t_{13}$ when $\Spo=-0.5$, $k=10$, $\omega_{0}=5$. The blue line is positive for a small interval $0.9514\leq r<1.0605$ and for $r>86.5794$. The red line is positive for $r>240.5626$.}
	\label{fig_3stage2m05term}
\end{figure}

\begin{figure}
	\begin{center}
		\includegraphics[width=10.5cm]{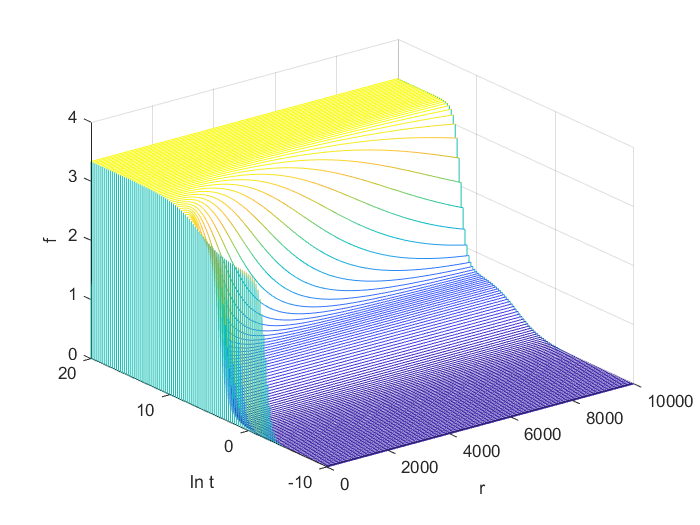}
	\end{center}
	\caption{$f$ against $\ln t$ and $r$ for $\Spo=-0.5$, $k=10$ and $\omega_{0}=5$ for the interval $0\leq r<5000$, showing 3 distinct states when $r>240.5626$. We have two different scenarios when $r<240.5626$. See Figures \ref{fig_4stagem05}, \ref{fig_3stage1m05}.}
	\label{fig_3stage2m05}
\end{figure}

To complete the example, we now look at what happens beyond $r=240.5626$, where $t_{12}$ becomes larger than $t_{23}$. From the diagram in Figure \ref{fig_4eqstatem05c}, this happens if the graph of $\ln T_2^2$ becomes too low, as shown in Figure \ref{fig_3eqstate2m05c}. Now, the diagram in Figure \ref{fig_3eqstate2m05c} shows the scenario 
\begin{align}
T_{4} \longrightarrow T_{1}  &\longrightarrow   T_{3},
\end{align}
with transition times
\begin{align}
t_{41}=\left(\frac{|k \Spo r^2 +\omega_{0}|}{r^2}\right)^{\frac{1}{2p_1}},\ \ \ &t_{13}=\left(\frac{r^2(2- \Spo)}{3|k|}\right)^{\frac{1}{2p_3}}.
\end{align}
They must satisfy the conditions
\begin{align}
\label{3stage2conditionm05}
t_{41}<t_{13}<t_{12}.
\end{align}
The condition $t_{13}<t_{12}$ is equivalent to $t_{23}<t_{12}$, so it gives (\ref{eq1hashr}) with the opposite inequality direction, a simple lower bound 
\begin{equation}
\label{eq7.15}
r>|k|\left(\frac{3}{|k|(2-\Spo)}\right)^{\frac{\Spo}{2 p_1}}.
\end{equation}
The condition $t_{41}<t_{13}$ implies  
\begin{align}
\label{eqstar}
|k \Spo r^2 +\omega_{0}|<r^\frac{1+p_3}{p_3}\left(\frac{2-\Spo}{3|k|}\right)^\frac{p_1}{p_3},
\end{align}
which restricts $r$ to one or more intervals.
Continuing with the same example, (\ref{eq7.15}) gives the interval $r>240.5626$, while (\ref{eqstar}) gives the intervals
\begin{align}
0.9514<r<1.0605 \ \ \ \ & \ \ \text{and}\ \ \ \ \ \ \ r>86.5794.
\end{align}
Together, the scenario occurs for $r>240.5626$.
See Figure \ref{fig_3stage2m05term}. We plot $f$ against $\ln t$ and $r$  showing 3 distinct states with a lower bound on $r$ in Figure \ref{fig_3stage2m05}.

\begin{figure}
	\begin{center}
		\begin{subfigure}{6.5cm}
			\includegraphics[width=7cm]{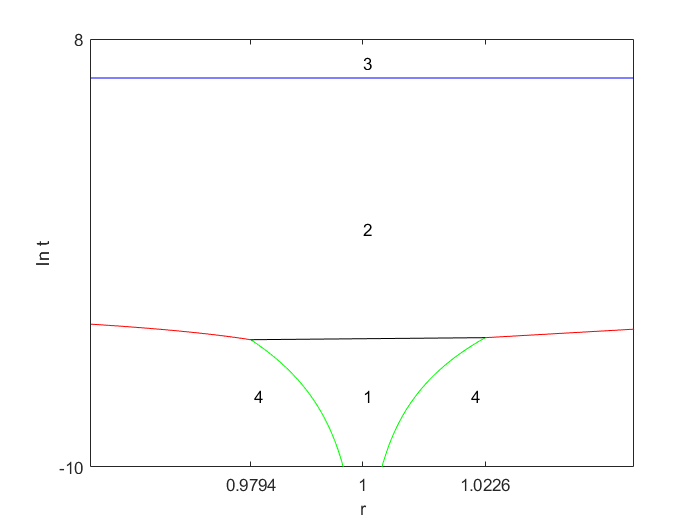}
		\end{subfigure}
		\begin{subfigure}{6.5cm}
			\includegraphics[width=7cm]{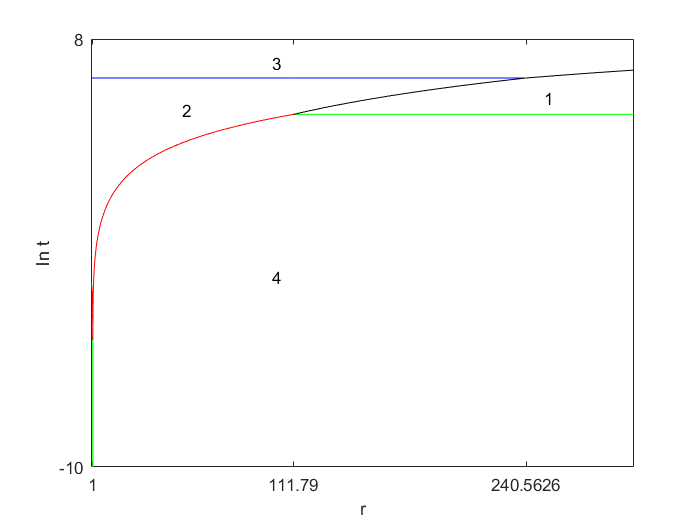}
		\end{subfigure}
	\end{center}
	\caption{Plot of the cells and transition times in the example  $\Spo=-0.5$, $k=10$ and $\omega_{0}=5$ showing the different scenarios along each fixed $r$. Each cell is labelled with the index of the dominant term.}
	\label{fig_4state3scenariosm05}
\end{figure}

This completes the example. We can summarise the different scenarios that occur in this example in another useful diagram, where we plot the transition times of each scenario, and label the dominant term in each cell. See Figure \ref{fig_4state3scenariosm05}.

\begin{figure}
	\begin{center}
		\includegraphics[width=10.5cm]{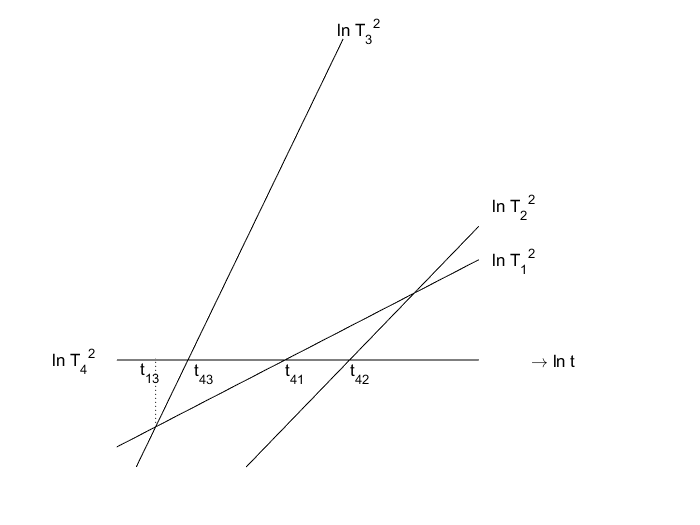}
	\end{center}
	\caption{Qualitative plot of the log of each term squared against $\ln t$, showing 2 dominant equilibrium states, for any value of $\Spo$ satisfying $-1<\Spo<0$.}
	\label{fig_2eqstatem05c}
\end{figure}

The fourth scenario is the 2-state sequence
\begin{align}
T_{4}   \longrightarrow T_{3} 
\end{align}
with transition time
\begin{align}
t_{43}= &\left(\frac{|k \Spo r^2 +\omega_{0}|(2- \Spo)}{3k}\right)^{\frac{1}{1+p_3}},
\end{align}
which is required to satisfy the conditions
\begin{align}
\label{2stageconditionm05}
t_{13}<t_{43}<t_{42}.
\end{align}
Figure \ref{fig_2eqstatem05c} shows a qualitative plot of the log of each term squared against $\ln t$, showing 2 dominant equilibrium states for value of $\Spo$ satisfying $-1<\Spo<0$. It is clear from the figure that, we have 2 distinct equilibrium states if and only if the transition times $t_{13}$, $t_{43}$ and $t_{42}$, satisfy the condition (\ref{2stageconditionm05}).
$t_{13}<t_{43}$  is equivalent to $t_{13}<t_{41}$, so it gives (\ref{eqstar}) with opposite inequality direction: 
\begin{align}
\label{eqstar2}
|k \Spo r^2 +\omega_{0}|>r^\frac{1+p_3}{p_3}\left(\frac{(2-\Spo)}{3|k|}\right)^\frac{p_1}{p_3}.
\end{align}
$t_{43}<t_{42}$ is equivalent to $t_{23}<t_{42}$, so it gives (\ref{eq4hash}) with the opposite inequality:
\begin{equation}
\label{eq4hash2}
|k \Spo r^2 +\omega_{0}|>|k|^{\frac1{p_1}}\left(\frac{2-\Spo}{3}\right)^\frac{p_3}{p_1}.
\end{equation}

\begin{figure}
	\begin{subfigure}[t]{6cm}
		\includegraphics[width=6cm]{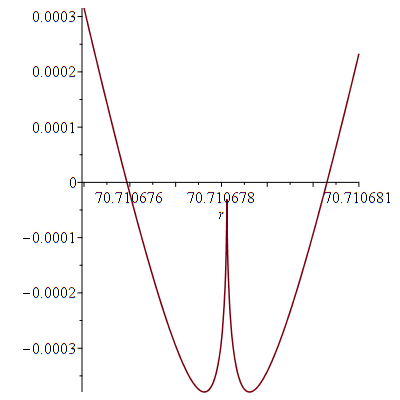}
		\caption{For $70.710675<r<70.710681$}
	\end{subfigure}
	\begin{subfigure}[t]{6cm}
		\includegraphics[width=6.5cm]{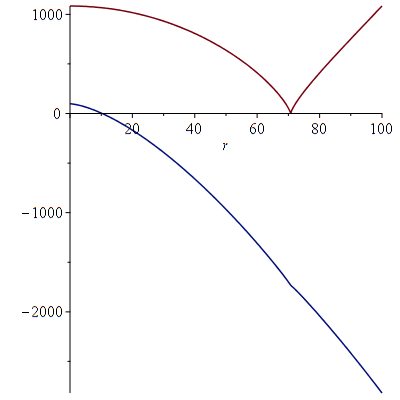}
		\caption{For $0\leq r\leq100$}
	\end{subfigure}
	\caption{Blue line is the plot of $t_{43}-t_{13}$  and red line is $ t_{42}-t_{43}$, when $\Spo=-0.5$, $k=0.2$ and $\omega_{0}=500$. The blue line is positive for a small interval $0\leq r<10.3244$. The red line is positive for for all $r$ except $70.7106759<r<70.7106803$.}
	\label{fig_2stagem05term}
\end{figure}
\begin{figure}
	\begin{center}
		\includegraphics[width=10.5cm]{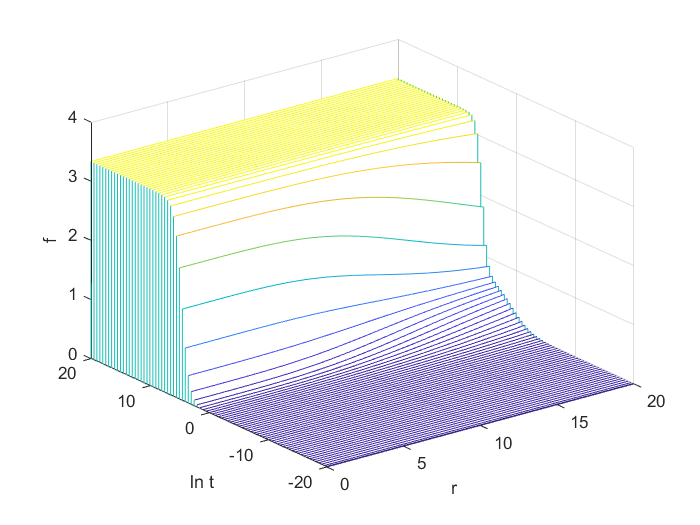}
	\end{center}
	\caption{$f$ against $\ln t$ and $r$ for $\Spo=-0.5$, $k=0.2$ and $\omega_{0}=500$, showing 2 distinct states for the interval $0\leq r<10$.}
	\label{fig_2stagem05f}
\end{figure}

In the previous example, the condition (\ref{2stageconditionm05}) is not satisfied anywhere. Consider a second example. Take  $\Spo=-0.5$, $k=0.2$ and $\omega_{0}=500$. (\ref{eqstar2}) gives $0\leq r<10.3244$, while (\ref{eq4hash2}) gives
\begin{align}
0\leq r<70.71067593 \ \ \ \text{and} \ \ \ r>70.71068030.
\end{align}
Together, they give the interval
\begin{align}
0\leq r<10.3244.
\end{align}
See Figure \ref{fig_2stagem05term}. We plot $f$ against $\ln t$ and $r$ for the values $\Spo=-0.5$, $k=0.2$ and $\omega_{0}=500$ on these intervals showing 2 distinct states in Figure \ref{fig_2stagem05f}. Beyond $r=10.3244$ the scenario is $T_{4}$ $\longrightarrow$ $T_{2}$ $\longrightarrow$ $T_{3}$.

\newpage
\subsection{Case $\Spo=0$}
For $\Spo=0$, we have 
\[T_1= r^2t^{\frac23}, \ \ T_2=k^2t^{\frac23}, \ \ \ T_3= \frac{3k}{2}t^{\frac43},\ \  \ T_4= \omega_{0}.\]
There are only three distinct powers of $t$, with $T_4$ dominating at early times, $T_3$ dominating at late times, and $T_1$ and $T_2$ possibly at intermediate times. The first scenario is the 3-state sequence.
\begin{align}
\label{Scenariozero1}
T_{4} \longrightarrow T_{1} \ &\& \ T_{2}   \longrightarrow T_{3}
\end{align}
with transition times
\begin{align}
t_{4\left(1\&2\right)}=\left(\frac{\omega_{0}^2}{r^4+k^4}\right)^{\frac{3}{4}},\ \ \ &t_{\left(1\&2\right)3}=\left(\frac{4(r^4+k^4)}{9k^2}\right)^{\frac{3}{4}},
\end{align}
which are required to satisfy the condition
\begin{align}
t_{4\left(1\&2\right)}<t_{\left(1\&2\right)3}.
\end{align}
The condition gives a lower bound on $r$
\begin{equation}
r>\left(\frac32|k\omega_{0}|-k^4\right)^\frac14.
\end{equation}

\begin{figure}
	\begin{center}
		\includegraphics[width=8.5cm]{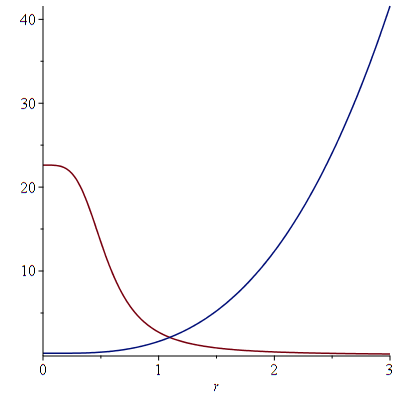}
	\end{center}
	\caption{Blue line is the plot of $t_{\left(1\&2\right)3}$  and red line is $ t_{4\left(1\&2\right)}$, when $k=0.5$ and $\omega_{0}=2$. Figure shows that for $r>1.0950$, $t_{4\left(1\&2\right)}<t_{\left(1\&2\right)3}$ and for $r<1.0950$, $t_{4\left(1\&2\right)}>t_{\left(1\&2\right)3}$.}
	\label{fig_3stage0term}
\end{figure}

\begin{figure}
	\begin{center}
		\includegraphics[width=10.5cm]{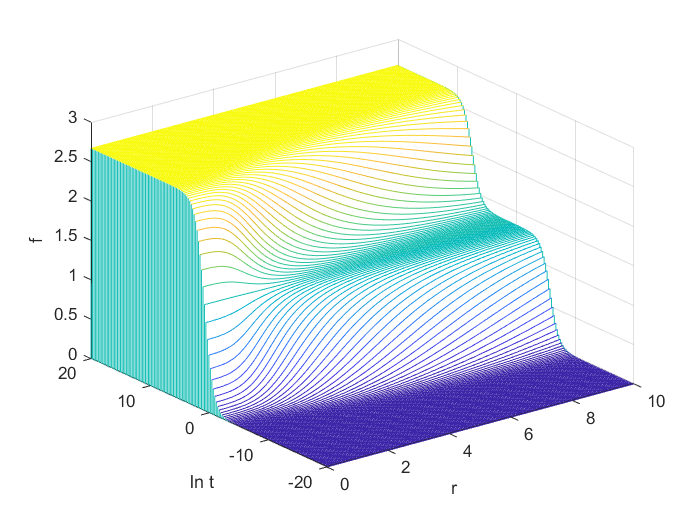}
	\end{center}
	\caption{$f$ against $\ln t$ and $r$ for $k=0.5$ and $\omega_{0}=2$ for the interval $0\leq r<10$, showing scenario (\ref{Scenariozero1}) for $r>1.0950$ and scenario (\ref{Scenariozero2}) for $r<1.0950$.}
	\label{fig_3stage0}
\end{figure}

If the lower bound is positive, then for $r$ less than this we have the second scenario, the 2-state sequence
\begin{align}
\label{Scenariozero2}
T_{4}  \longrightarrow T_{3}  
\end{align}
with transition time
\begin{align}
t_{43}= &\left|\frac{2\omega_{0}}{3k}\right|^{\frac{3}{4}}.
\end{align}
For example, given $k=0.5$ and $\omega_{0}=2$, for $r>1.0950$ we have the scenario (\ref{Scenariozero1}) and for $r<1.0950$ we have the scenario (\ref{Scenariozero2}). See Figure \ref{fig_3stage0term}. We plot $f$ against $\ln t$ and $r$ on the interval $0\leq r \leq 10$ showing both scenarios in Figure \ref{fig_3stage0}. 




\newpage
\subsection{Case $0<\Spo<0.5$}
\label{Section7.1.4}
\begin{figure}
	\begin{center}
		\includegraphics[width=12.5cm]{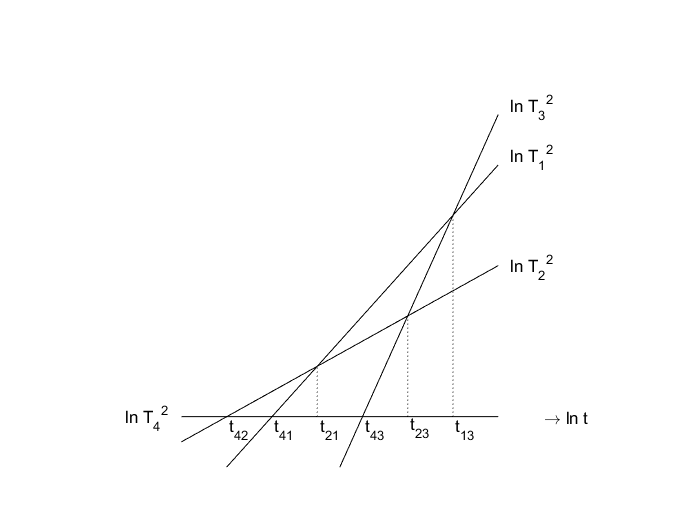}
	\end{center}
	\caption{Qualitative plot of the log of each term squared against $\ln t$, showing 4 dominant equilibrium states, for any value of $\Spo$ satisfying $0<\Spo<0.5$.}
	\label{fig_4eqstate025c}
\end{figure}

\begin{figure}
	\begin{subfigure}[t]{7cm}
		\includegraphics[width=6.5cm]{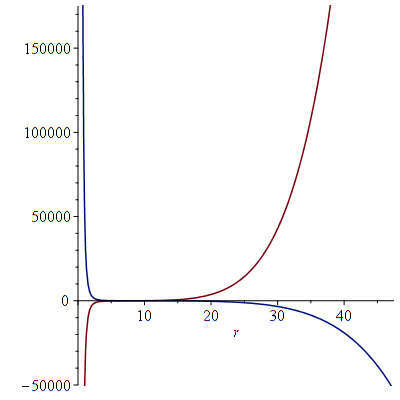}
		\caption{For $0\leq r\leq50$}
	\end{subfigure}
	\begin{subfigure}[t]{5.5cm}
		\includegraphics[width=5.5cm]{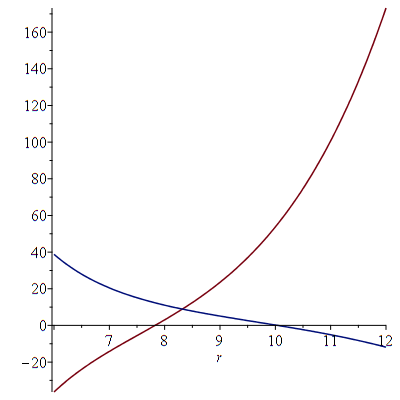}
		\caption{For $6\leq r\leq12$}
	\end{subfigure}
	\caption{Blue line is the plot of $t_{21}-t_{42}$  and red line is $ t_{13}-t_{21}$, when $\Spo=0.25$, $k=15$, $\omega_{0}=6$. The blue line is positive for a interval $r<10.1374$. The red line is positive for $r>7.8251$. Together they give the interval $7.8251<r<10.1374$.}
	\label{fig_4stage025term}
\end{figure}

\begin{figure}
	\begin{center}
		\includegraphics[width=10.5cm]{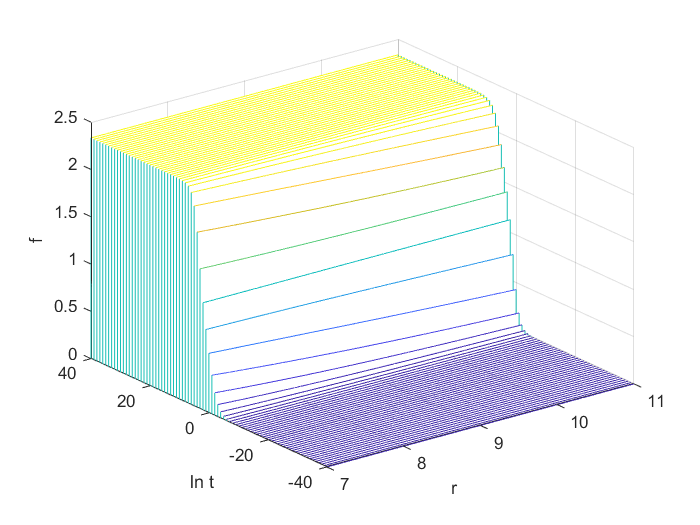}
	\end{center}
	\caption{$f$ against $\ln t$ and $r$ for $\Spo=0.25$, $k=15$ and $\omega_{0}=6$ for the interval $7<r<11$, showing 2 visible distinct states instead of 4, because the transition times are too close together.}
	\label{fig_4stage025}
\end{figure}

\begin{figure}
	\begin{center}
		\includegraphics[width=8.5cm]{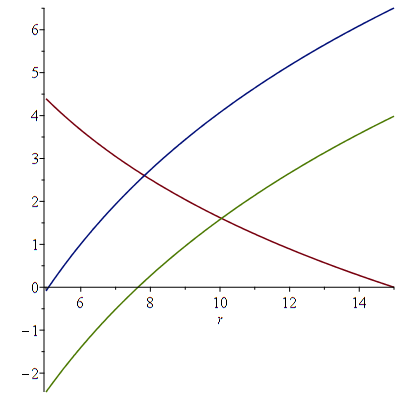}
	\end{center}
	\caption{The log of transition times $t_{42}$ (green), $t_{21}$ (red) and $t_{13}$ (blue)  against $r$ for $\Spo=0.25$, $k=15$, $\omega_{0}=6$ for the interval  $5<r<15$, showing that the transition times are close together.}
	\label{fig_4stage025plot3t}
\end{figure}

The case $0<\Spo<0.5$ is qualitatively similar to the case $-1<\Spo<0$ in Section \ref{Section7.1.2}. Figure \ref{Combine4stage} gives the ordering $T_4$,  $T_2$, $T_1$, $T_3$, in increasing power of $t$. The possible scenarios along general worldlines are
\begin{enumerate}
	\item  $T_{4}$ $\longrightarrow$ $T_{2}$ $\longrightarrow$ $T_{1}$ $\longrightarrow$ $T_{3}$
	\item $T_{4}$ $\longrightarrow$ $T_{2}$ $\longrightarrow$ $T_{3}$
	\item  $T_{4}$ $\longrightarrow$ $T_{1}$  $\longrightarrow$ $T_{3}$
	\item  $T_{4}$  $\longrightarrow$ $T_{3}$.
\end{enumerate}
There are two special worldlines. The first one is $r=0$, where $T_1$ vanishes. The possible scenarios along this worldline are the scenarios 2 and 4 above. The second special worldline is $r=\sqrt{\frac{-\omega_{0}}{k \Spo}}$, where $T_4$ vanishes, giving an early-time permanent spike.
The possible scenarios along this worldline are:
\begin{enumerate}
	\item $T_{2}$ $\longrightarrow$ $T_{1}$ $\longrightarrow$ $T_{3}$
	\item  $T_{2}$  $\longrightarrow$ $T_{3}$.
\end{enumerate}
The two special worldlines coincide if $\omega_{0}=0$. In this case the only possible scenario along this worldline is
\[T_{2} \longrightarrow T_{3}.\]
Figure \ref{fig_4eqstate025c} shows a qualitative plot of the $\log$ of each term squared against $\ln t$, for the scenario
\begin{align}
T_{4} \longrightarrow  &  T_{2}  \longrightarrow  T_{1}  \longrightarrow  T_{3}.
\end{align}
The transition times
\begin{align}
t_{42}=\left(\frac{|k \Spo r^2 +\omega_{0}|}{k^2}\right)^{\frac{1}{2p_3}},\ \ \ &t_{21}=\left(\frac{k^2 }{r^2}\right)^\frac{1}{2\Spo} ,\ \ \ \  t_{13}=\left( \frac{r^2(2- \Spo)}{3|k|}\right)^\frac{1}{2p_3}
\end{align}
must satisfy the condition
\begin{align}
\label{4stagecondition025}
t_{42}<t_{21}<t_{13}.
\end{align}
$t_{42}<t_{21}$ implies (\ref{eq1hash}):
\begin{align}
\label{eqstar1}
|k \Spo r^2 +\omega_{0}| < \left(\frac{|k|^{p_1}}{r^{p_3}}\right)^\frac{2}{\Spo},
\end{align}
while $t_{21}<t_{13}$ implies (\ref{eq7.15}):
\begin{equation}
\label{eqstar2b}
r>|k|\left(\frac{3}{|k|(2-\Spo)}\right)^{\frac{\Spo}{2 p_1}}.
\end{equation}
As a concrete example, take $\Spo=0.25$, $k=15$ and $\omega_{0}=6$. (\ref{eqstar1}) gives 
\begin{align}
\label{eqstar3}
r<10.1374
\end{align}
while (\ref{eqstar2}) gives 
\begin{align}
\label{eqstar4}
r>7.8251.
\end{align}
Together they give the interval 
\begin{align}
7.8251<r<10.1374.
\end{align}
See Figure \ref{fig_4stage025term}.
We plot $f$ against $\ln t$ and $r$ for the interval $7<r<11$ in Figure \ref{fig_4stage025}, showing 2 visible distinct states because the transition times are too close together (see Figure \ref{fig_4stage025plot3t}).


\begin{figure}
	\begin{center}
		\includegraphics[width=12.5cm]{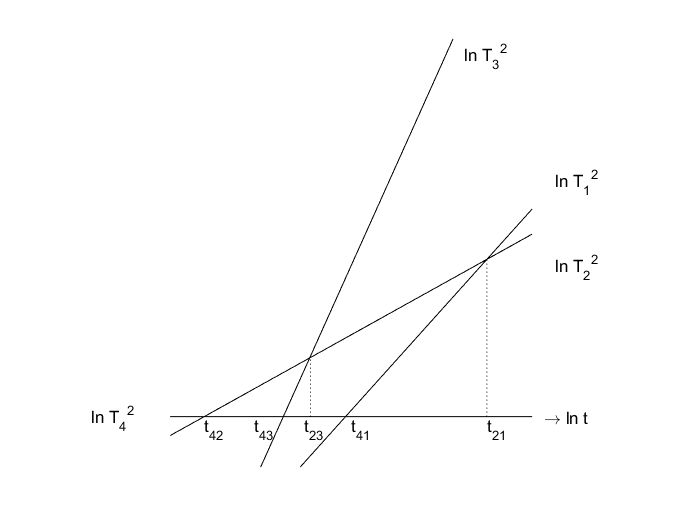}
	\end{center}
	\caption{Qualitative plot of the log of each term squared against $\ln t$, showing 3 dominant equilibrium states, for any value of $\Spo$ satisfying $-1<\Spo<0$.}
	\label{fig_3eqstate1025c}
\end{figure}

\begin{figure}
	\begin{center}
		\includegraphics[width=6.5cm]{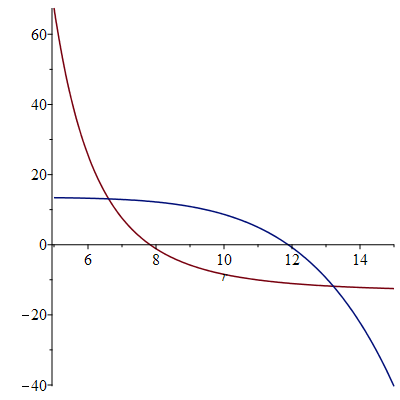}
	\end{center}
	\caption{Blue line is the plot of $t_{23}-t_{42}$  and red line is $ t_{21}-t_{23}$, when $\Spo=0.25$, $k=15$ and $\omega_{0}=6$. The blue line is positive for a small interval $r<11.8859$. The red line is positive for $r<7.8251$. Together they give the interval $0\leq r<7.8251$.}
	\label{fig_3stage1025term}
\end{figure}

\begin{figure}
	\begin{center}
		\includegraphics[width=10.5cm]{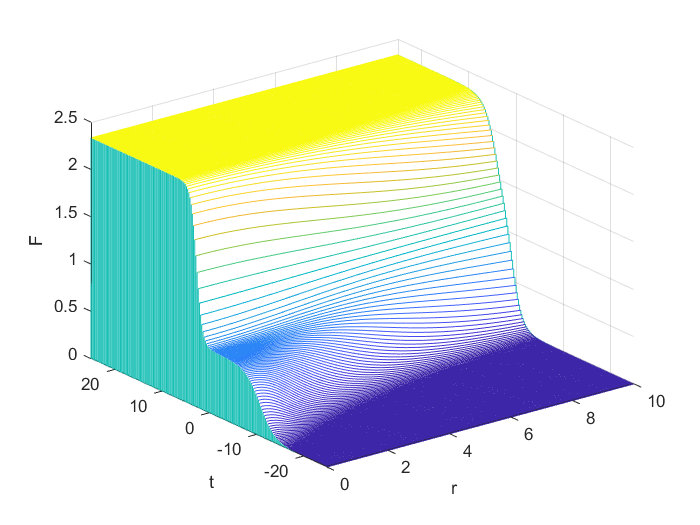}
	\end{center}
	\caption{$f$ against $\ln t$ and $r$ for $\Spo=0.25$, $k=15$ and $\omega_{0}=6$ for the interval $0\leq r<7.8251$, showing 3 distinct states only for $r<4$.}
	\label{fig_3stage1025}
\end{figure}

From (\ref{eqstar4}), we know that $t_{21}$ becomes greater than $t_{13}$ for values of $r$ just below $7.8251$. From the diagram in Figure $\ref{fig_4eqstate025c}$, this happens if the graph of $\ln T_1^2$ becomes too low, as shown in Figure \ref{fig_3eqstate1025c}. This gives the scenario
\begin{align}
T_{4}\longrightarrow  T_{2} &  \longrightarrow  T_{3},
\end{align}
with transition times
\begin{align}
t_{42}=\left(\frac{|k \Spo r^2 +\omega_{0}|}{k^2}\right)^{\frac{1}{2p_3}},  \ \ &  t_{23}=\left(\frac{|k|(2- \Spo)}{3}\right)^\frac{1}{2p_1}.
\end{align}
They must satisfy the condition
\begin{align}
\label{3stage1condition1025}
t_{42}<t_{23}<t_{21}.
\end{align}
$t_{42}<t_{23}$ implies (\ref{eq4hash}):
\begin{align}
\label{eqstar5}
|k \Spo r^2 +\omega_{0}|<|k|^{\frac1{p_1}}\left(\frac{2-\Spo}{3}\right)^\frac{p_3}{p_1},
\end{align}
while $t_{23}<t_{21}$ implies (\ref{eq1hashr}):
\begin{equation}
\label{eqstar6}
r<|k|\left(\frac{3}{|k|(2-\Spo)}\right)^{\frac{\Spo}{2 p_1}}.
\end{equation}
Continuing with the same example, (\ref{eqstar5}) gives 
\begin{align}
r<11.8859
\end{align}
while (\ref{eqstar6}) gives 
\begin{align}
r<7.8251.
\end{align}
Together they give the interval 
\begin{align}
0\leq r < 7.8251.
\end{align}
See Figures \ref{fig_3stage1025term}. We plot $f$ against $\ln t$ and $r$  on the interval $0\leq r<10$ showing 3 distinct states in Figure \ref{fig_3stage1025}. 

\begin{figure}
	\begin{center}
		\includegraphics[width=10.5cm]{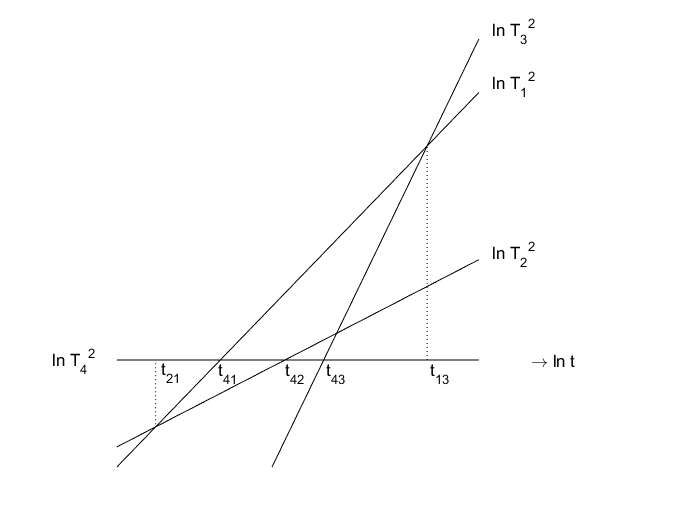}
	\end{center}
	\caption{Qualitative plot of the log of each term squared against $\ln t$, showing 3 dominant equilibrium states, for any value of $\Spo$ satisfying $0<\Spo<0.5$.}
	\label{fig_3eqstate2025c}
\end{figure}

\begin{figure}
	\begin{center}
		\includegraphics[width=6.5cm]{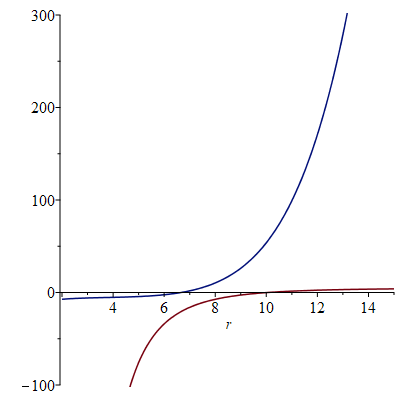}
	\end{center}
	\caption{Blue line is the plot of $t_{41}<t_{13}$  and red line is $ t_{12}<t_{41}$ when $\Spo=0.25$, $k=15$ and $\omega_{0}=6$. The blue line for a small interval $r>6.6524$. The red line is positive for $r>10.1374$. Together they give the interval $r>10.1374$.}
	\label{fig_3stage2025term}
\end{figure}

\begin{figure}
	\begin{center}
		\includegraphics[width=10.5cm]{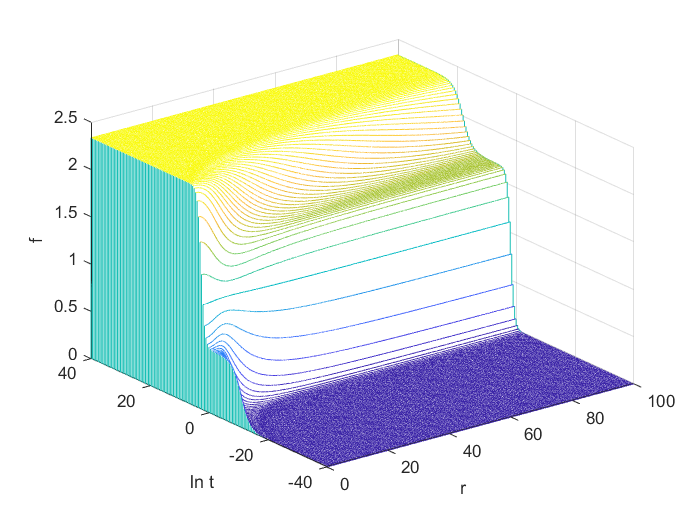}
	\end{center}
	\caption{$f$ against $\ln t$ and $r$ for $\Spo=0.25$, $k=15$ and $\omega_{0}=6$ , showing 3 distinct states for $r>10.1374$. We have two different scenarios for $r<10.1374$. See Figures \ref{fig_4stage025}, \ref{fig_3stage1025}.}
	\label{fig_3stage2025}
\end{figure}

\begin{figure}
	\begin{center}
		\includegraphics[width=10cm]{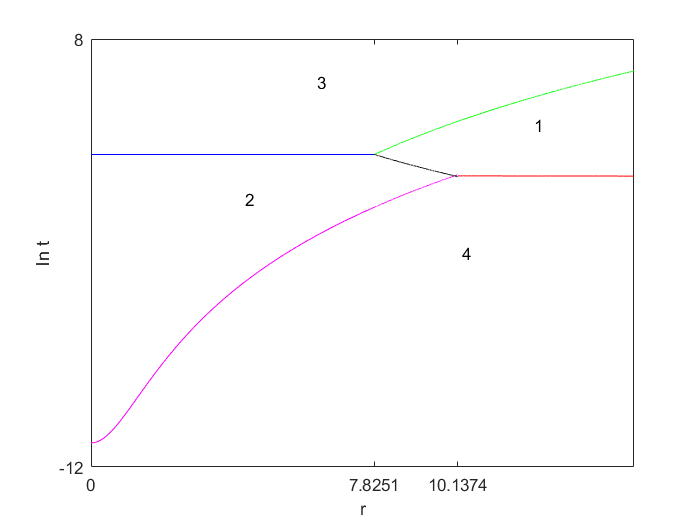}
	\end{center}
	\caption{Plot of the cells and transition times in the example  $\Spo=0.25$, $k=15$ and $\omega_{0}=6$ showing the different scenarios along each fixed $r$. Each cell is labelled with the index of the dominant term.}
	\label{fig_4state3scenarios025}
\end{figure}

To complete the example, from (\ref{eqstar3}) we know that $t_{42}$ becomes greater than $t_{21}$ for values of $r$ just above $10.1374$. From the diagram in Figure $\ref{fig_4eqstate025c}$, this happens if the graph of $\ln T_2^2$ becomes too low, as shown in Figure $\ref{fig_3eqstate2025c}$. This gives the scenario
\begin{align}
T_{4} \longrightarrow\  T_{1}  &\longrightarrow  T_{3},
\end{align}
with transition times
\begin{align}
t_{41}=\left(\frac{|k \Spo r^2 +\omega_{0}|}{r^2}\right)^{\frac{1}{2p_1}},\ \ \ &t_{13}=\left(\frac{r^2(2- \Spo)}{3|k|}\right)^{\frac{1}{2p_3}}.
\end{align}
They must satisfy the condition
\begin{align}
\label{3stage2condition025}
t_{21}<t_{41}<t_{13}.
\end{align}
$t_{21}<t_{41}$ implies (\ref{eq3hash}):
\begin{align}
\label{eqstar7}
|k \Spo r^2 +\omega_{0}| > \left(\frac{|k|^{p_1}}{r^{p_3}}\right)^\frac{2}{\Spo}
\end{align}
while $t_{41}<t_{13}$ implies:
\begin{equation}
\label{eqstar8}
|k \Spo r^2 +\omega_{0}|<r^\frac{1+p_3}{p_3}\left(\frac{(2-\Spo)}{3|k|}\right)^\frac{p_1}{p_3}.
\end{equation}
Continuing with the same example, (\ref{eqstar7}) gives 
\begin{align}
r>10.1374,
\end{align}
while (\ref{eqstar8}) gives 
\begin{align}
r>6.6524.
\end{align}
Together they give the interval 
\begin{align}
r>10.1374.
\end{align}
See Figures \ref{fig_3stage2025term}. We plot $f$ against $\ln t$ and $r$ showing 3 distinct states with a lower bound on $r$ in Figure \ref{fig_3stage2025}. We summarise the scenarios in Figure \ref{fig_4state3scenarios025}.


\begin{figure}
	\begin{center}
		\includegraphics[width=10.5cm]{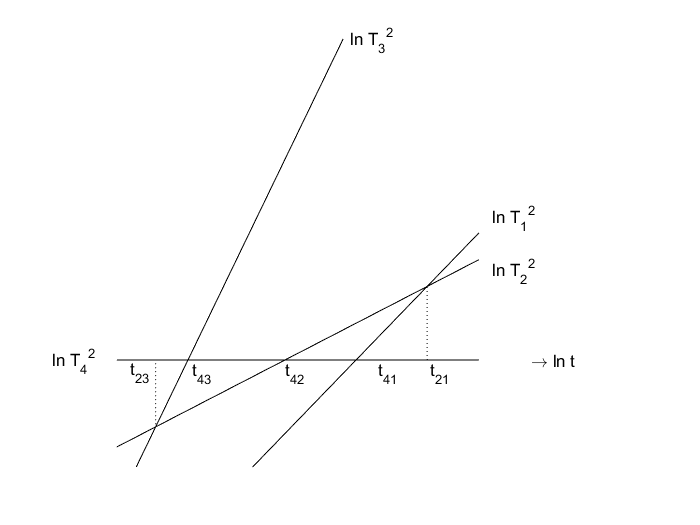}
	\end{center}
	\caption{Qualitative plot of the log of each term squared against $\ln t$, showing 3 dominant equilibrium states, for any value of $\Spo$ satisfying $0<\Spo<0.5$.}
	\label{fig_2eqstate025c}
\end{figure}

\begin{figure}
	\begin{center}
		\includegraphics[width=6.5cm]{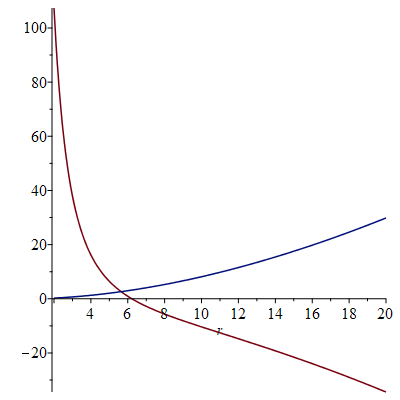}
	\end{center}
	\caption{Blue line is the plot of $t_{43}-t_{23}$  and red line is $ t_{41}-t_{43}$, when $\Spo=0.25$, $k=10$ and $\omega_{0}=200$. The blue line is positive for a  interval $r>0.9945$. The red line is positive for $r<6.2223$. Together they give the interval $0.9945<r<6.2223$.}
	\label{fig_2stage025term}
\end{figure}

\begin{figure}
	\begin{center}
		\includegraphics[width=10.5cm]{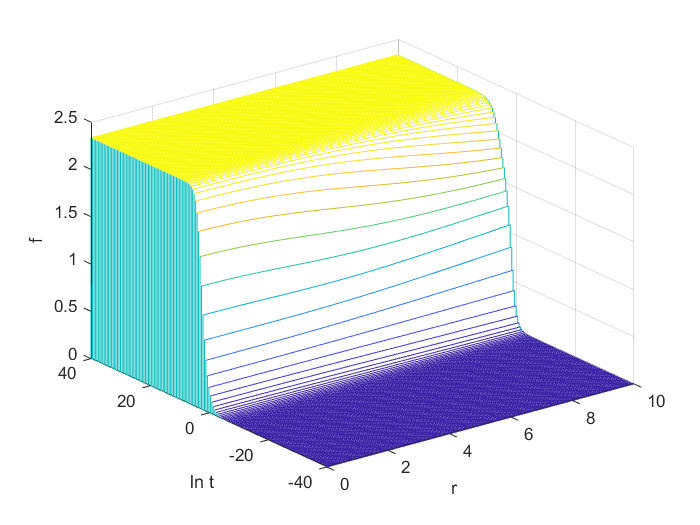}
	\end{center}
	\caption{$f$ against $\ln t$ and $r$ for $\Spo=0.25$, $k=10$ and $\omega_{0}=200$ for the interval $0\leq r<10$, showing 2 distinct states.}
	\label{fig_2stage025}
\end{figure}

\newpage
The fourth scenario is the 2-state sequence
\begin{align}
T_{4} \longrightarrow T_{3}  
\end{align}
with transition times
\begin{align}
t_{43}= &\left(\frac{|k \Spo r^2 +\omega_{0}|(2- \Spo)}{3|k|}\right)^{\frac{1}{1+p_3}}
\end{align}
which are required to satisfy the condition
\begin{align}
\label{2stagecondition025}
t_{23}<t_{43}<t_{41}.
\end{align}
Figure \ref{fig_2eqstate025c} shows a qualitative plot of the log of each term squared against $\ln t$, showing 2 dominant equilibrium states for value of $\Spo$ satisfying $0<\Spo<0.5$. It is clear from the figure that, we have 2 distinct equilibrium states if and only if the transition times $t_{23}$, $t_{43}$ and $t_{41}$, satisfy the condition (\ref{2stagecondition025}).
$t_{23}<t_{43}$ implies  (\ref{eq4hash2}):
\begin{align}
\label{eqstar9}
|k \Spo r^2 +\omega_{0}|>|k|^{\frac{1}{p_1}}\left(\frac{2-\Spo}{3}\right)^\frac{p_3}{p_1}
\end{align} 
while $t_{43}<t_{41}$ implies (\ref{eqstar2}):
\begin{align}
\label{eqstar10}
|k \Spo r^2 +\omega_{0}|>r^{\frac{1+p_3}{p_3}}\left(\frac{2-\Spo}{3|k|}\right)^\frac{p_1}{p_3}.
\end{align}
Consider a second example. Take $\Spo=0.25$, $k=10$ and $\omega_{0}=200$,  (\ref{eqstar9}) implies $ r>0.9945$, while (\ref{eqstar10}) implies $r<6.2223$. Together they give the interval $0.9945<r<6.2223$. See Figure \ref{fig_2stage025term}. We plot $f$ against $\ln t$ and $r$ showing 2 distinct states with an upper bound on $r$ in Figure \ref{fig_2stage025}.
Beyond $r=6.2223$ the scenario is $T_{4}$ $\longrightarrow$ $T_{1}$ $\longrightarrow$ $T_{3}$. For $0\leq r<0.9945$ the scenario is $T_{4}$ $\longrightarrow$ $T_{2}$ $\longrightarrow$ $T_{3}$.

\newpage
\subsection{Case $\Spo=0.5$}
For $\Spo=\frac12$, we have 
\[T_1= r^2 t, \ \ T_2=k^2, \ \ \ T_3= 2kt,\ \  \ T_4= \frac{kr^2}{2}+\omega_{0}.\]
There are only two distinct powers of $t$, with $T_2$ and $T_4$ dominating at early times, and $T_1$ and $T_3$ dominating at late times. The scenario is
\begin{equation}
T_{2} \ \ \ 
\&  \ \ T_{4} \ \ \longrightarrow\ \ \ T_{1}\ \ \ \&\ \ \ T_{3}.\ \ \
\end{equation}
Solving the equation
\[T_{2}^2+T_{4}^2 \  =\ T_{1}^2+ T_{3}^2\  \]
for $t$ yields the transition time
\begin{align}
t_{\left(2 \& 4\right)\left(1 \& 3\right)}= & \left(\frac{k^2 r^4+4 k^4+4 k \omega_0 r^2+4 \omega_0^2}{4(r^4+4k^2)}\right)^{\frac{1}{2}}.
\end{align}

\begin{figure}
	\begin{center}
		\includegraphics[width=10.5cm]{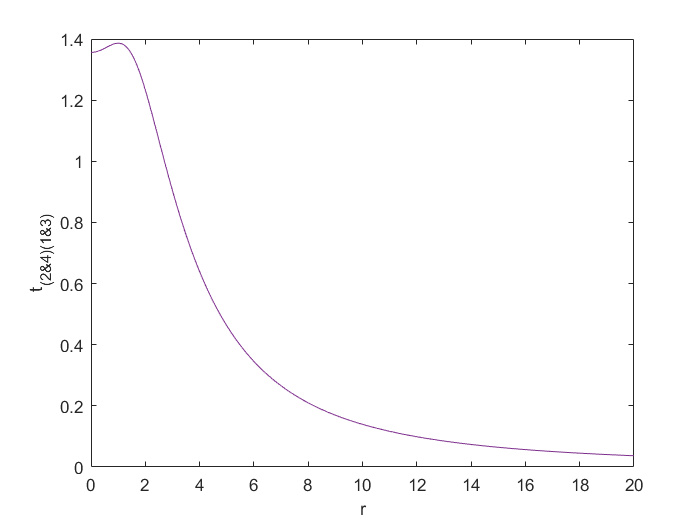}
	\end{center}
	\caption{Transition time $t_{\left(2 \& 4\right)\left(1 \& 3\right)}$ as a function of $r$ for $\Spo=0.5$, $k=2$ and $\omega_{0}=15$.}
	\label{fig_2stagep5tr}
\end{figure}
\begin{figure}
	\begin{center}
		\includegraphics[width=10.5cm]{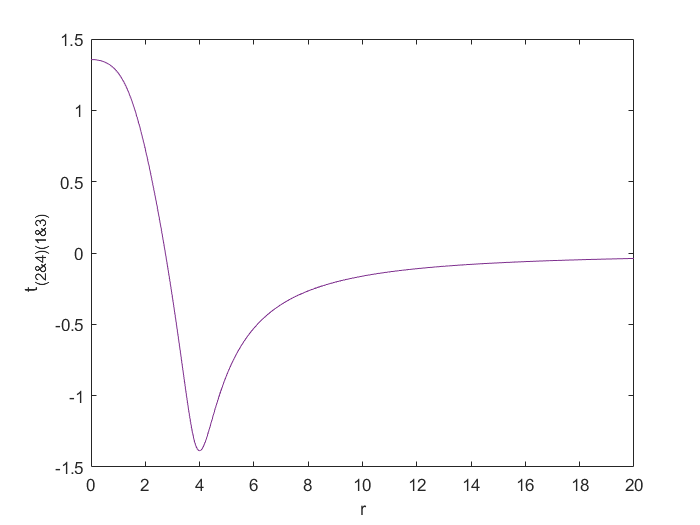}
	\end{center}
	\caption{Transition time $t_{\left(2 \& 4\right)\left(1 \& 3\right)}$ as a function of $r$ for  $\Spo=0.5$, $k=-2$ and $\omega_{0}=15$.}
	\label{fig_2stagep5trn}
\end{figure}
\begin{figure}
	\begin{center}
		\includegraphics[width=10.5cm]{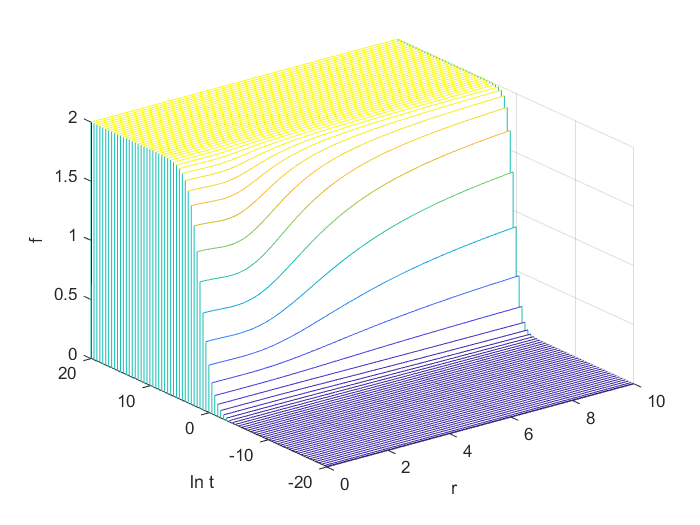}
	\end{center}
	\caption{$f$ against $\ln t$ and $r$ for $\Spo=0.5$, $k=2$ and $\omega_{0}=15$. The transition time has a local maximum at $r=2$.}
	\label{fig_2stagep5}
\end{figure}
\begin{figure}
	\begin{center}
		\includegraphics[width=10.5cm]{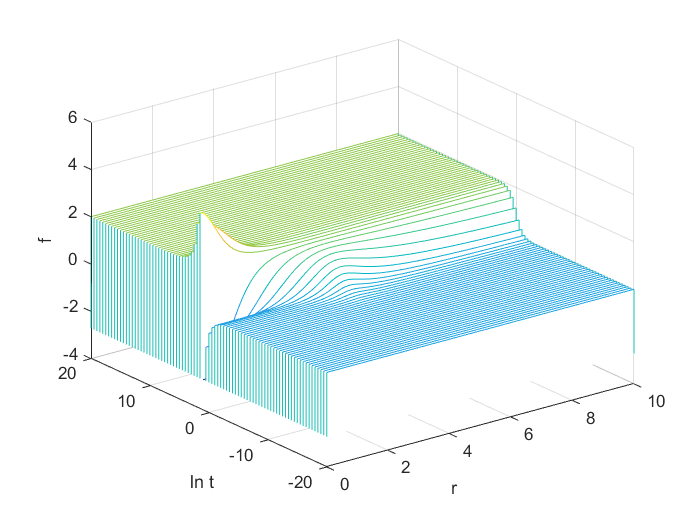}
	\end{center}
	\caption{$f$ against $\ln t$ and $r$ for $\Spo=0.5$, $k=-2$ and $\omega_{0}=15$. The transition time has a local minimum at $r=4$.}
	\label{fig_2stagep5n}
\end{figure}
We now analyse the behaviour of $t_{\left(2 \& 4\right)\left(1 \& 3\right)}$ as a function of $r$. Observe that $\lim\limits_{r\rightarrow \infty}t_{\left(2 \& 4\right)\left(1 \& 3\right)}=\frac{|k|}{2}$. $t_{\left(2 \& 4\right)\left(1 \& 3\right)}$ has one non-trivial critical point at 
\begin{equation}
r=
\begin{cases}
\frac{\sqrt{k(-\omega_0+\sqrt{4k^4+\omega_0^2})}}{k},& \text{if $k>0$} \\
-\frac{\sqrt{-k(\omega_0+\sqrt{4k^4+\omega_0^2})}}{k},& \text{if $k<0$}.
\end{cases}
\end{equation}
If $\omega_{0} k >0$, then $t_{\left(2 \& 4\right)\left(1 \& 3\right)}$ has a local maximum at the critical point; and if $\omega_{0} k <0$, then $t_{\left(2 \& 4\right)\left(1 \& 3\right)}$ has a local minimum at the critical point. For example, if we have $\Spo=0.5$, $k=2$ and $\omega_{0}=15$, then at $r=\frac{\sqrt{k(-\omega_0+\sqrt{4k^4+\omega_0^2})}}{k}= 1$, we have a local maximum (see Figure \ref{fig_2stagep5tr}). But if we have $\Spo=0.5$, $k=-2$ and $\omega_{0}=15.1$, then at $r=\frac{\sqrt{-k(\omega_0+\sqrt{4k^4+\omega_0^2})}}{k}= 4$ we have a local minimum (see Figure \ref{fig_2stagep5trn}).
We plot $f$ against $\ln t$ and $r$ for these examples in Figures \ref{fig_2stagep5} and \ref{fig_2stagep5n}. Figure \ref{fig_2stagep5n} shows an overshoot transition, which we will discuss in Section ??

Note that, we do not have spikes in this case.
\subsection{Case $0.5<\Spo\leq1$}
\label{Section7.1.6}
\begin{figure}
	\begin{center}
		\includegraphics[width=12.5cm]{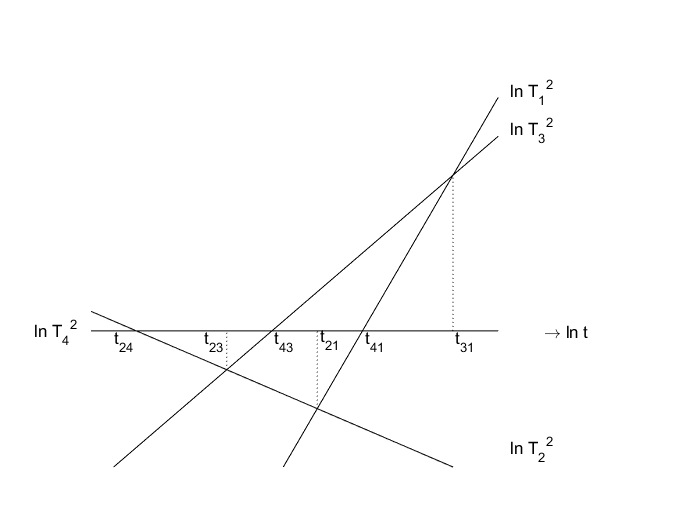}
	\end{center}
	\caption{Qualitative plot of the log of each term squared against $\ln t$, showing 4 dominant equilibrium states, for any value of $\Spo$ satisfying $0.5<\Spo\leq1$.}
	\label{fig_4eqstate075c}
\end{figure}

\begin{figure}
	\begin{center}
		\includegraphics[width=6.5cm]{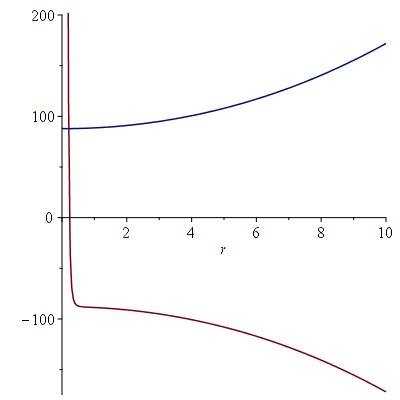}
	\end{center}
	\caption{Blue line is the plot of $t_{43}-t_{24}$  and red line is $ t_{43}-t_{31}$, for $\Spo=0.75$, $k=0.1$ and $\omega_{0}=10$. The blue line is positive for all the values of $r$. The red line is positive for $r<0.2323$.}
	\label{fig_4stage075term}
\end{figure}

\begin{figure}
	\begin{subfigure}[t]{7cm}
		\includegraphics[width=7.25cm]{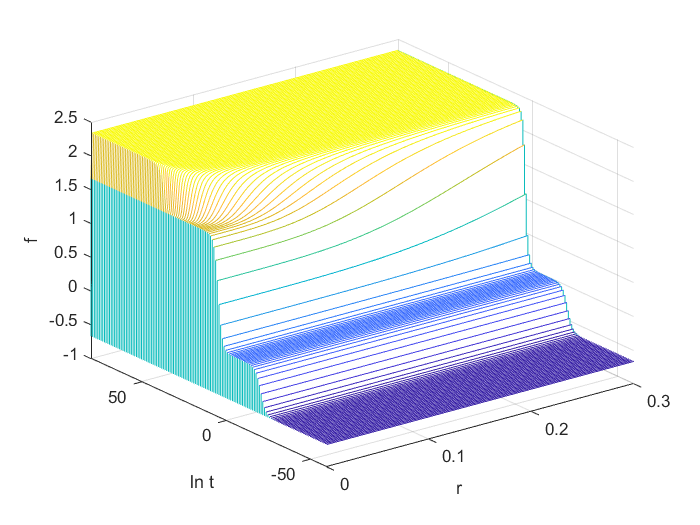}
		\caption{For $0\leq r\leq0.3$}
	\end{subfigure}
	\begin{subfigure}[t]{7cm}
		\includegraphics[width=7.25cm]{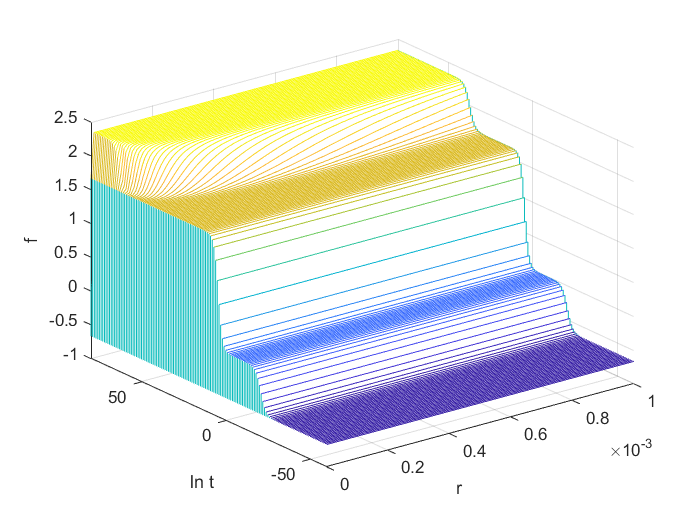}
		\caption{For $0\leq r\leq0.001$}
	\end{subfigure}
	\caption{$f$ against $\ln t$ and $r$ for $\Spo=0.75$, $k=0.1$ and $\omega_{0}=10$ for the interval $0<r<0.3$, showing 4 distinct states. At $r=0$, we have a permanent spike at late times.}
	\label{fig_4stage075}
\end{figure}

The case $0.5<\Spo \leq 1$ is qualitatively similar to the case $-1<\Spo<0$ in Section \ref{Section7.1.2} and the case $0<\Spo<0.5$ in Section \ref{Section7.1.4}. But it also has some differences, which we will point out. Figure \ref{Combine4stage} gives the ordering $T_2$,  $T_4$, $T_3$, $T_1$, in increasing power of $t$. The possible scenarios along general worldlines are
\begin{enumerate}
	\item  $T_{2}$ $\longrightarrow$ $T_{4}$ $\longrightarrow$ $T_{3}$ $\longrightarrow$ $T_{1}$
	\item $T_{2}$ $\longrightarrow$ $T_{4}$ $\longrightarrow$ $T_{1}$
	\item  $T_{2}$ $\longrightarrow$ $T_{3}$  $\longrightarrow$ $T_{1}$
	\item  $T_{2}$  $\longrightarrow$ $T_{1}$.
\end{enumerate}
There are two special worldlines. The first one is $r=0$, where $T_1$ vanishes, giving a late-time permanent spike. This is different from the other cases, as $T_1$ has the highest power only in this case. The possible scenarios along this worldline are:
\begin{enumerate}
	\item  $T_{2}$ $\longrightarrow$ $T_{4}$ $\longrightarrow$ $T_{3}$ 
	\item  $T_{2}$  $\longrightarrow$ $T_{3}$
\end{enumerate}
The second special worldline is $r=\sqrt{\frac{-\omega_{0}}{k \Spo}}$, where $T_4$ vanishes.
The possible scenario along this worldline are the scenarios 3 and 4 above.
The two special worldlines coincide if $\omega_{0}=0$. In this case the only possible scenario along this worldline is
\[T_{2} \longrightarrow T_{3}.\]
Figure \ref{fig_4eqstate075c} shows a qualitative plot of the $\log$ of each term squared against $\ln t$, for the scenario
\begin{align}
T_{2} \longrightarrow  &  T_{4}  \longrightarrow  T_{3}  \longrightarrow  T_{1}.
\end{align}
Note the negative slope for $\ln T_2^2$.
The transition times
\begin{align}
t_{24}=\left(\frac{|k \Spo r^2 +\omega_{0}|}{k^2}\right)^{\frac{1}{2p_3}},  t_{43}=\left(\frac{|k \Spo r^2 +\omega_{0}|(2- \Spo)}{3|k|}\right)^{\frac{1}{1+p_3}},& \  t_{31}=\left( \frac{r^2(2- \Spo)}{3|k|}\right)^{\frac{1}{2p_3}}
\end{align}
must satisfy the condition
\begin{align}
\label{4stagecondition075}
t_{24}<t_{43}<t_{31}.
\end{align}
$t_{24}<t_{43}$ implies (\ref{eq4hash}):
\begin{equation}
\label{eq1star}
|k \Spo r^2 +\omega_{0}|>k^{\frac1{p_1}}\left(\frac{2-\Spo}{3}\right)^\frac{p_3}{p_1}.
\end{equation}
while $t_{43}<t_{31}$ implies (\ref{eqstar8}):
\begin{equation}
\label{eq2star}
|k \Spo r^2 +\omega_{0}|<r^\frac{1+p_3}{p_3}\left(\frac{2-\Spo}{3|k|}\right)^\frac{p_1}{p_3}.
\end{equation} As a concrete example, take $\Spo=0.75$, $k=0.1$ and $\omega_{0}=10$ (\ref{eq1star}) gives $r>0$ \footnote{The special worldline $r=0$ is excluded.} while (\ref{eq2star}) gives $r<0.2323$. Together they give
\begin{align}
0<r<0.2323.
\end{align}
See Figures \ref{fig_4stage075term}.
We plot $f$ against $\ln t$ and $r$  for the interval $0<r<0.3$ in Figure \ref{fig_4stage075}, showing 4 visible distinct states for small values of $r$, and 3 distincts states for larger values of $r$. Along $r=0$, we have a late-time permanent spike.


\begin{figure}
	\begin{center}
		\includegraphics[width=10.5cm]{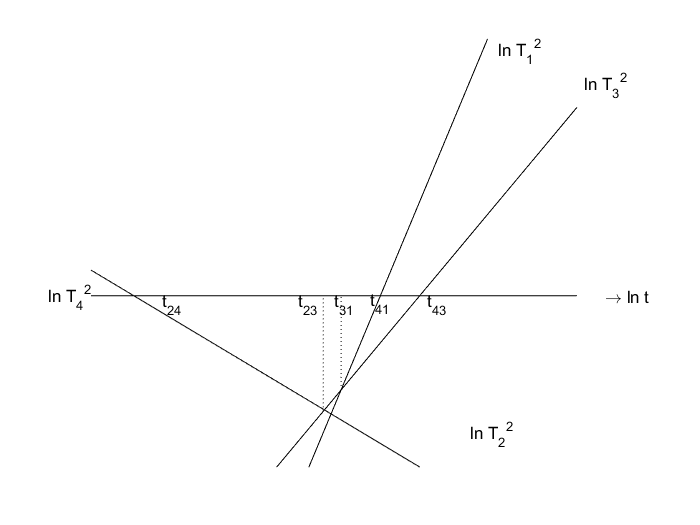}
	\end{center}
	\caption{Qualitative plot of the log of each term squared against $\ln t$, showing 3 dominant equilibrium states, for any value of $\Spo$ satisfying $0.5<\Spo\leq1$.}
	\label{fig_3eqstate1075c}
\end{figure}
\begin{figure}
	\begin{center}
		\includegraphics[width=6.5cm]{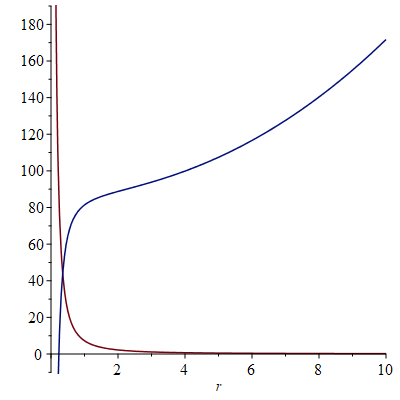}
	\end{center}
	\caption{Red line is the plot of $t_{24}-t_{41}$  and blue line is $ t_{43}-t_{41}$, for $\Spo=0.75$, $k=0.1$ and $\omega_{0}=10$. The blue line is negative for a small interval $r<0.2323$. The red line is positive for all values of $r$. Together they give the interval $r>0.2323$.}
	\label{fig_3stage1075term}
\end{figure}
\begin{figure}
	\begin{center}
		\includegraphics[width=10.5cm]{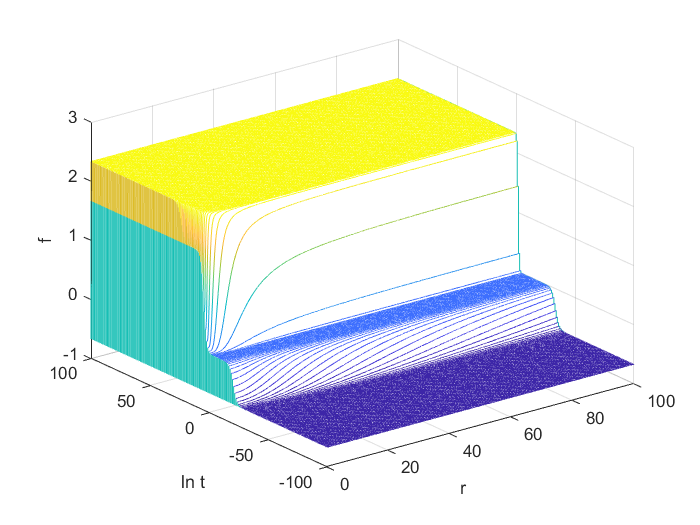}
	\end{center}
	\caption{$f$ against $\ln t$ and $r$ for $\Spo=0.75$, $k=0.1$ and $\omega_{0}=10$ for the interval $0\leq r<100$, showing 3 distinct states only for $r>0.2323$.}
	\label{fig_3stage1075}
\end{figure}

For values of $r$ just above $0.2323$, $t_{43}$ becomes greater  than $t_{31}$. From Figure \ref{fig_4eqstate075c}, this happens if the graph of $\ln T_3^2$ becomes too low, as shown in Figure \ref{fig_3eqstate1075c}. This gives the scenario
\begin{align}
T_{2} \longrightarrow T_{4} &\longrightarrow T_{1},
\end{align}
with transition times
\begin{align}
t_{24}=\left(\frac{|k \Spo r^2 +\omega_{0}|}{k^2}\right)^{\frac{1}{2p_3}}\ \ \ , &\ \ \  t_{41}=\left(\frac{|k \Spo r^2 +\omega_{0}|}{r^2}\right)^{1/2p_1}.
\end{align}
They must satisfy the condition
\begin{align}
\label{3stage1condition1075}
t_{24}<t_{41}<t_{43}.
\end{align}
$t_{24}<t_{41}$ implies (\ref{eq1hash}): 
\begin{align}
\label{eq3star}
|k \Spo r^2 +\omega_{0}| < \left(\frac{|k|^{p_1}}{r^{p_3}}\right)^\frac{2}{\Spo},
\end{align}
while $t_{43}<t_{31}$ implies (\ref{eqstar}):
\begin{align}
\label{eq4star}
|k \Spo r^2 +\omega_{0}|<r^\frac{1+p_3}{p_3}\left(\frac{2-\Spo}{3|k|}\right)^\frac{p_1}{p_3},
\end{align}
Continuing with the same example, (\ref{eq3star}) gives $r>0$ while (\ref{eq4star}) gives $r>0.2323$. Together they give
\begin{align}
r>0.2323.
\end{align}
See Figure \ref{fig_3stage1075term}. We plot $f$ against $\ln t$ and $r$  on the interval $0\leq r<100$ showing 3 distinct states in Figure \ref{fig_3stage1075}.
This completes the example.


\begin{figure}
	\begin{center}
		\includegraphics[width=10.5cm]{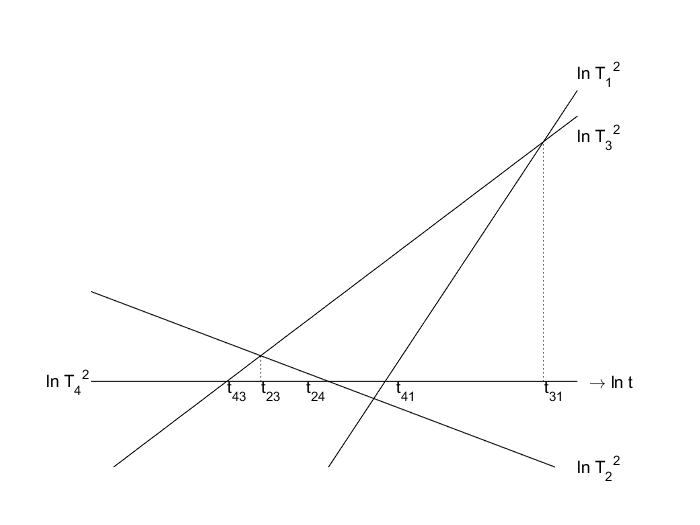}
	\end{center}
	\caption{Qualitative plot of the log of each term squared against $\ln t$, showing 3 dominant equilibrium states, for any value of $\Spo$ satisfying $0<\Spo\leq1$.}
	\label{fig_3eqstate2075c}
\end{figure}

\begin{figure}
	\begin{center}
		\includegraphics[width=6.5cm]{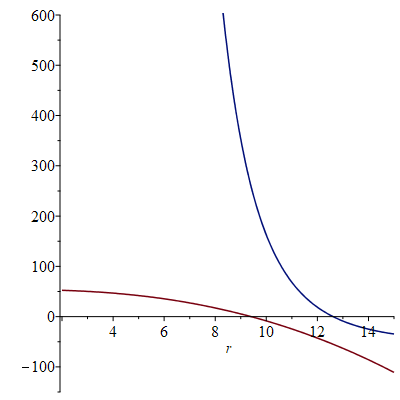}
	\end{center}
	\caption{Blue line is the plot of $t_{31}-t_{23}$  and red line is $ t_{23}-t_{43}$, for $\Spo=0.75$, $k=250$ and $\omega_{0}=0.1$. The red line is positive for a interval $0\leq r<9.4014$ . The blue line is positive for a interval $0\leq r<12.6133$. Together we have the interval $0\leq r<9.4014$.}
	\label{fig_3stage2075term}
\end{figure}

\begin{figure}
	\begin{center}
		\includegraphics[width=10.5cm]{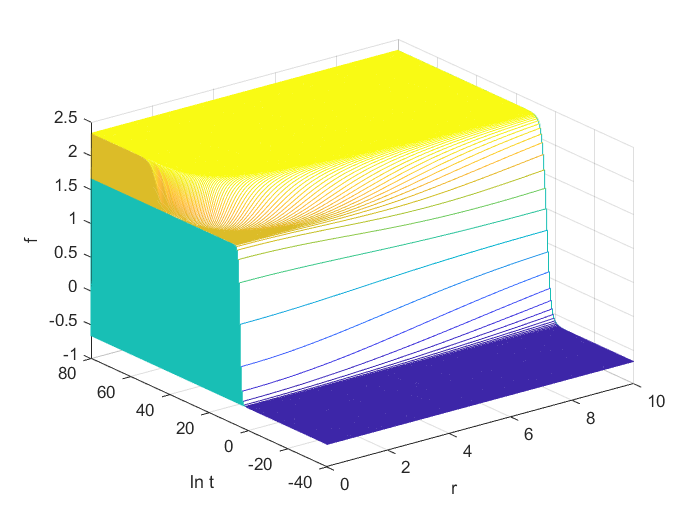}
	\end{center}
	\caption{$f$ against $\ln t$ and $r$ for $\Spo=0.75$, $k=250$ and $\omega_{0}=0.1$ for the interval $0\leq r<10$, showing 3 distinct states. From $r$ greater than $9.4014$, we have the 4-state scenario.}
	\label{fig_3stage2075}
\end{figure}

\begin{figure}
	\begin{center}
		\includegraphics[width=10.5cm]{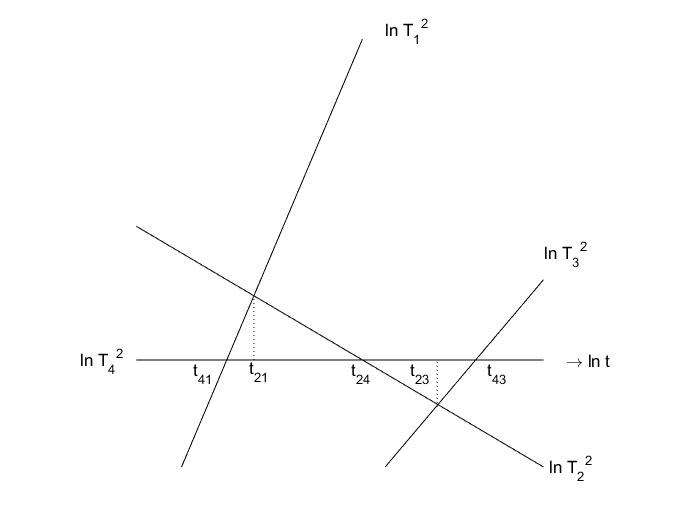}
	\end{center}
	\caption{Qualitative plot of the log of each term squared against $\ln t$, showing 2 dominant equilibrium states, for any value of $\Spo$ satisfying $0.5<\Spo\leq1$.}
	\label{fig_2eqstate075c}
\end{figure}

\begin{figure}
	\begin{subfigure}[t]{7.5cm}
		\includegraphics[width=6.5cm]{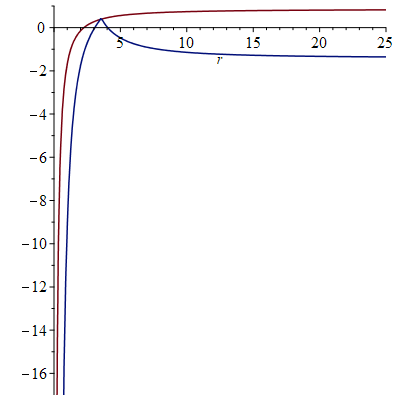}
		\caption{For $0< r<25$}
	\end{subfigure}
	\begin{subfigure}[t]{5.5cm}
		\includegraphics[width=6.5cm]{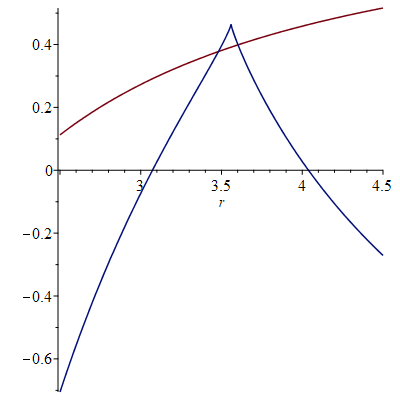}
		\caption{For $2.5< r<4.5$}
	\end{subfigure}
	\caption{Red line is the plot of $t_{23}-t_{21}$  and blue line is $ t_{21}-t_{41}$, for $\Spo=0.75$, $k=2$ and $\omega_{0}=-19$. The blue line is positive for small interval  $3.07236<r<4.0387$. The red line is positive for $r>2.2487$.  Together they give the interval $3.07236<r<4.0387$.}
	\label{fig_2stage075term}
\end{figure}

\begin{figure}
	\begin{center}
		\includegraphics[width=10.5cm]{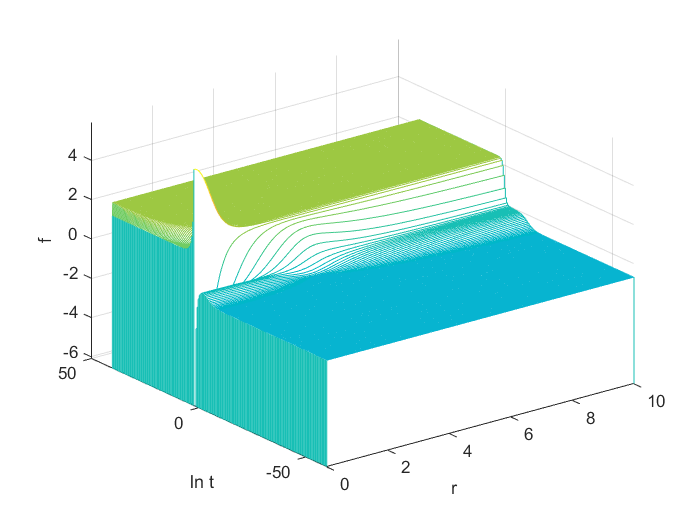}
	\end{center}
	\caption{$f$ against $\ln t$ and $r$ for $\Spo=0.75$, $k=2$ and $\omega_{0}=-19$ for the interval $0\leq r<10$, showing 2 distinct states  for $3.07236<r<4.0387$.}
	\label{fig_2stage075}
\end{figure}

\newpage
The third scenario is the 3-state sequence
\begin{align}
T_{2}  \longrightarrow  T_{3} &  \longrightarrow  T_{1}
\end{align}
with the transition times
\begin{align}
t_{23}=\left(\frac{|k|(2- \Spo)}{3}\right)^{\frac{1}{2p_1}}\ \ \ , &\ \ \  t_{31}=\left( \frac{r^2(2- \Spo)}{3|k|}\right)^\frac{1}{2p_3},
\end{align}
which are required to satisfy the condition
\begin{align}
\label{3stage1condition2075}
t_{43}<t_{23}<t_{31}.
\end{align}
Figure \ref{fig_3eqstate2075c} shows a qualitative plot of the log of each term squared against $\ln t$, showing 3 dominant equilibrium states.
$t_{43}<t_{23}$ implies (\ref{eq4hash}):
\begin{align}
|k \Spo r^2 +\omega_{0}|<|k|^{\frac{1}{p_1}}\left(\frac{2-\Spo}{3}\right)^\frac{p_3}{p_1}
\end{align}
while $t_{23}<t_{31}$ implies (\ref{eq1hashr}):
\begin{equation}
r<|k|\left(\frac{3}{|k|(2-\Spo)}\right)^{\frac{\Spo}{2 p_1}}.
\end{equation}
For example, given $\Spo=0.75$, $k=250$ and $\omega_{0}=0.1$, the first condition $t_{43}<t_{23}$ implies $r<9.4014$, while the second condition $t_{23}<t_{31}$ implies $r<12.6133$. Together they give the interval $0<r<9.4014$. See Figure \ref{fig_3stage2075term}. We plot $f$ against $\ln t$ and $r$  showing 3 distinct states in Figure \ref{fig_3stage2075}.
\newpage
The fourth scenario is the 2-state sequence
\begin{align} 
T_{2} &  \longrightarrow  T_{1}
\end{align}
with the transition time
\begin{align}
t_{21}= &\left(\frac{k^2 }{r^2}\right)^{\frac{1}{2\Spo}},
\end{align}
which is required to satisfy the condition
\begin{align}
\label{2stagecondition075}
t_{41}<t_{21}<t_{23}.
\end{align}
Figure \ref{fig_2eqstate075c} shows a qualitative plot of the log of each term squared against $\ln t$, showing 2 distinct equilibrium states. 
$t_{41}<t_{21}$ implies (\ref{eq3hash}):
\begin{align}
|k \Spo r^2 +\omega_{0}| > \left(\frac{|k|^{p_1}}{r^{p_3}}\right)^\frac{2}{\Spo},
\end{align}  
while $t_{21}<t_{23}$ implies (\ref{eq7.15}): 
\begin{equation}
r>|k|\left(\frac{3}{|k|(2-\Spo)}\right)^{\frac{\Spo}{2 p_1}},
\end{equation} 
For example, given $\Spo=0.75$, $k=2$ and $\omega_{0}=-19$, the first condition $t_{41}<t_{21}$ implies $3.07236<r<4.0387$, while the second condition $t_{21}<t_{23}$ implies $r>2.2487$. Together they give the interval $3.07236<r<4.0387$. See Figure \ref{fig_2stage075term}. We plot $f$ against $\ln t$ and $r$   showing 2 distinct states in Figure \ref{fig_2stage075}.

\newpage
\section{Transient spikes}
\label{Section7.2}
We now take a closer look at transient spikes. In section \ref{Section7.1}, we saw a number of examples with transient and narrow inhomogeneity.  See Figures \ref{fig_2stagem1}, \ref{fig_3stage0}, \ref{fig_3stage1025}, \ref{fig_2stagep5n} and \ref{fig_2stage075}. Do we want to call all these features transient spikes? What is the definition for transient spikes?

To explore how to define transient spikes, we begin by looking at the definition of permanent spikes. A permanent spike is a feature in a region of spacetime characterized by a discontinuous limit of the type
\begin{align}
f \rightarrow
\begin{cases}
L_1 \ \ \ \ &\text{if} \ \ \ r=r_0 \\
L_2 \ \ \ \ &\text{if} \ \ \ r\neq r_0
\end{cases}
\end{align}
as $t\rightarrow \infty$ (late-time) or as $t\rightarrow0$ (early-time). In other words, the limit is different at $r=r_0$, which we call the spike worldline. The examples in Section \ref{Section7.1.6} all have a late-time permanent spike along $r=0$. Moreover, the final transition time ($t_{31}$ or $t_{41}$) tends to infinity as $r$ tends to zero. In other words, the cell that contains $r=0$ is increasingly narrow as $t\rightarrow \infty$. At what time  should a permanent spike begin to be called as such? It is not clear. In the early stage of its formation, the inhomogeneous structure is still rather wide. As it becomes narrower at later times, we become more likely to call the inhomogeneous structure a permanent spike. At late enough time, everyone would agree to call the structure a permanent spike. Therefore, while there is vagueness about when it starts, a permanent spike is easily identified by the asymptotic narrowing of the inhomogeneous structure.

By analogy, we can attempt to characterise transient spikes as a feature in a region of spacetime where an inhomogeneous structure becomes narrow temporarily. There is the vagueness about how narrow is considered narrow. The second issue is that this definition is too broad. It would include the features in Figures \ref{fig_2stagem1},  \ref{fig_2stagep5n} and \ref{fig_2stage075} as transient spikes.

In the original context where transient spikes were first named, the worldlines in a small neighbourhood undergo a scenario that is different from those undergone by worldlines further away. Adding this criterion rules out the feature in Figure \ref{fig_2stagem1}.

The features in  Figures \ref{fig_2stagep5n} and \ref{fig_2stage075} are overshoot transitions, which should be distinguished from transient spikes. We can add a criterion that transient spike is not a single transition, but something that lasts longer. In the next section, we will discuss overshoot transition in details. An overshoot transition can occur inside a transient spike. 
\section{Overshoot transition}
\label{Section7.3}
We noted earlier in Section \ref{Section7.1} that $f$ has a cascading appearance. Despite this, Figures \ref{fig_2stagep5n} and \ref{fig_2stage075} show that $f$ can fluctuate wildly when it makes a transition between equilibrium states.

Under what condition does this happen? If we examine $f$ from (\ref{fGeneralT}):
\begin{align}
\label{fGeneralT2}
f=\frac{2(T_1+T_2)(2p_1T_1+2p_3T_2)+2(T_3+T_4)(1+p_3)T_3}{(T_1+T_2)^2+(T_3+T_4)^2},
\end{align}
we see that the magnitude of $f$ becomes large if the denominator becomes small due to cancellation. Among $T_1$, $T_2$, $T_3$, $T_4$, only $T_4$ can become negative, so cancellation is only possible if $T_4$ is negative. Recall from (\ref{Combine4stage}) that 
\begin{align}
T_4=k \Spo r^2 +\omega_{0},
\end{align}
So $T_4$ is negative if and only if
\begin{align}
\label{OS1}
r<\sqrt{\frac{-\omega_{0}}{k \Spo}}, \quad \frac{\omega_{0}}{k \Spo}<0.
\end{align}
Cancellation happens when
\begin{align}
T_3 +T_4 \approx 0.
\end{align}
Its effect is most prominent when cancellation occurs during the 
\begin{align}
\label{ost1}
T_4 \rightarrow T_3
\end{align}
transition in a scenario.
Heuristically, when $T_3 +T_4= \epsilon$, where $|\epsilon|$ is small, and suppose $T_1$ and $T_2$ are $o(\epsilon)$, then \eqref{fGeneralT2} implies
\begin{align}
\label{ost3}
f\approx \frac{2(1+p_3)}{\epsilon}T_3.
\end{align}
Then $f$ becomes negative in the first stage of the transition (when $\epsilon<0$), then positive in the second stage (when $\epsilon>0$). This produces overshoots, whose amplitude can be large if $T_1$ and $T_2$ are much smaller than $T_3$ and $T_4$ when this happens. We therefore call such a transition an overshoot transition.

The overshoot transition occurs on the interval 
\begin{align}
0\leq r < \sqrt{\frac{-\omega_{0}}{k \Spo}}.
\end{align}
In the exceptional case $\Spo=0$, $T_4$ is negative if and only if $\omega_{0}$ is negative.
Additional restriction on the interval is provided by the condition that the transition $T_4\rightarrow T_3$ occurs. This condition is broken if $T_1$ and $T_2$ becomes large enough that the scenario changes to $T_4\rightarrow T_1 \rightarrow T_3$, $T_4\rightarrow T_2 \rightarrow T_3$, or $T_4\rightarrow T_1 \ \& \ T_2 \rightarrow T_3$.

\begin{figure}
	\begin{center}
		\includegraphics[width=10.5cm]{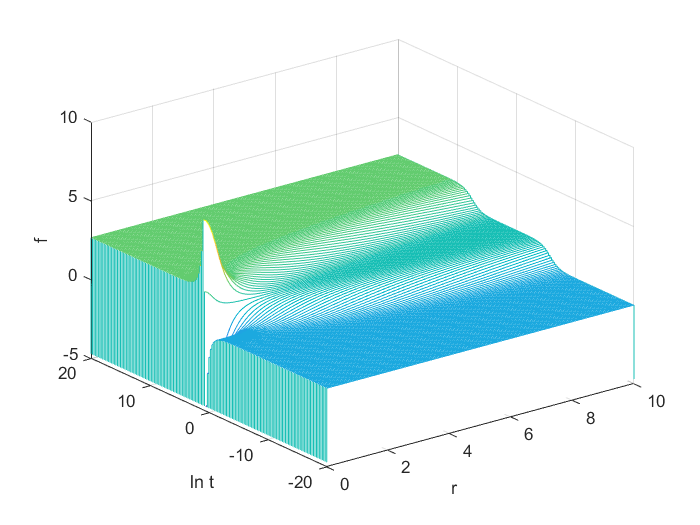}
	\end{center}
	\caption{Plot of $f$ against $\ln t$ and $r$ for $\Spo=0$, $k=0.5$ and $\omega_{0}=-2$. An overshoot transition occurs on $r\lesssim1$, around $\ln t \approx 0.7356$. This occurs within a transient spike, which occurs on $r  \lesssim 2$, $-2  \lesssim  \ln t \lesssim 3$.}
	\label{fig_r0Sp0zerosir}
\end{figure}

An overshoot transition can occur during a transient spike. For example, take $\Spo=0$, $k=0.5$ and $\omega_{0}=-2$. See Figures \ref{fig_r0Sp0zerosir}, \ref{fig_r0Sp0zerosirrf} and \ref{fig_r0Sp0zerosirtf}. An overshoot transition occurs on $r\lesssim1$, around $\ln t \approx 0.7356$. This occurs within a transient spike, which occurs on $r  \lesssim 2$, $-2  \lesssim  \ln t \lesssim 3$.

\begin{figure}
	\begin{center}
		\includegraphics[width=10.5cm]{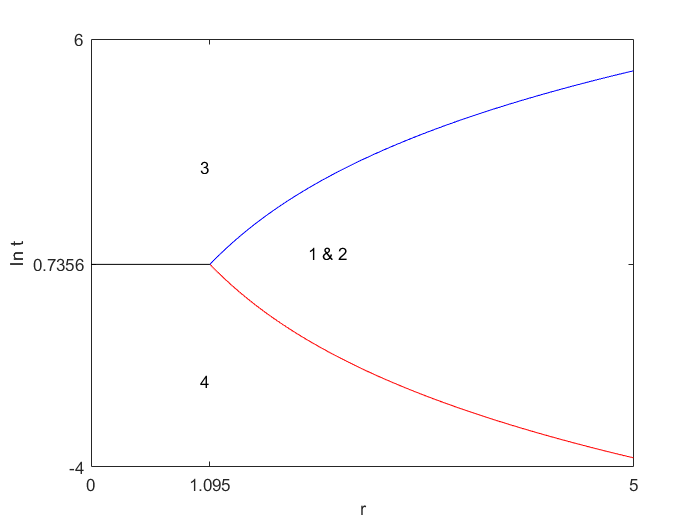}
	\end{center}
	\caption{Plot of the cells and transition times in the example  $\Spo=0$, $k=0.5$ and $\omega_{0}=-2$ showing the different scenarios along each fixed $r$. Each cell is labelled with the index of the dominant term. An overshoot transition occurs on $r\lesssim 1$, around $\ln t \approx 0.7356$. This occurs within a transient spike, which occurs on $r  \lesssim 2$, $-2  \lesssim  \ln t \lesssim 3$.}
	\label{fig_4state2scenarios0}
\end{figure}
\begin{figure}
	\begin{center}
		\includegraphics[width=10.5cm]{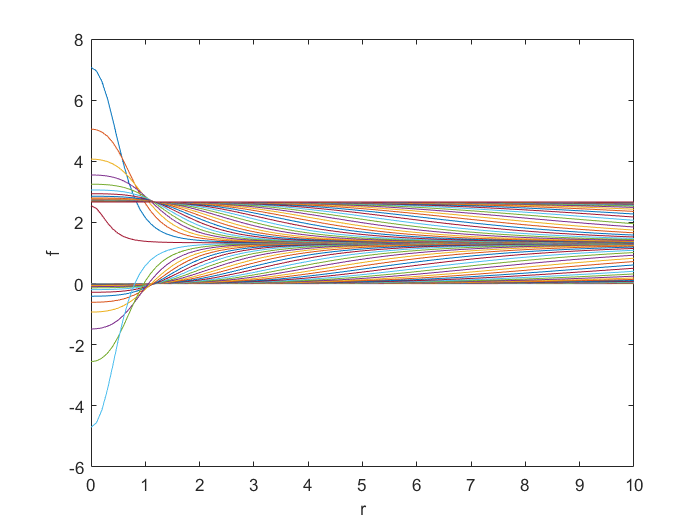}
	\end{center}
	\caption{$f$ against $r$ for $\Spo=0$, $k=0.5$ and $\omega_{0}=-2$, showing the overshoots occurring on $r\lesssim 1$.}
	\label{fig_r0Sp0zerosirrf}
\end{figure}
\begin{figure}
	\begin{center}
		\includegraphics[width=10.5cm]{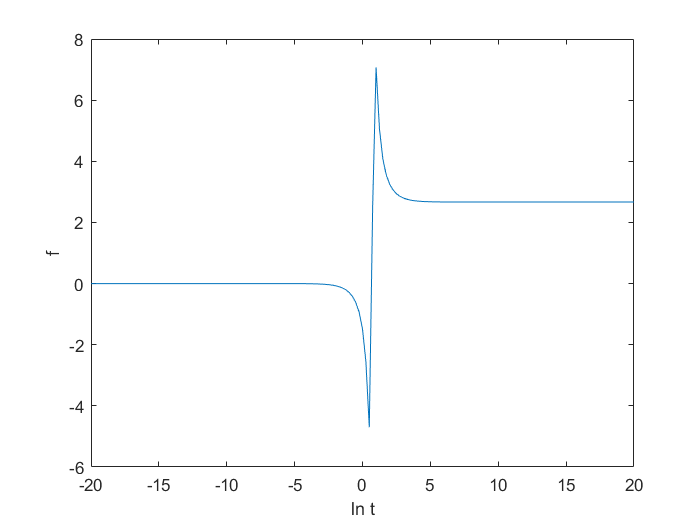}
	\end{center}
	\caption{$f$ against $\ln t$ along $r=0$ for $\Spo=0$, $k=0.5$ and $\omega_{0}=-2$, showing the overshoots.}
	\label{fig_r0Sp0zerosirtf}
\end{figure}
\newpage
\section{Summary}
The main aim of this chapter is to explore and describe the transient dynamics of $f$. For this, we introduced a new technique. We grouped $F$ into four terms $T_1$, $T_2$, $T_3$, $T_4$ based on the power of $t$. We plotted their power against the parameter $\Spo$ in Figure \ref{Combine4stage}, and we get three specific values and three intervals of $\Spo$ that are analysed separately in subsections of Section \ref{Section7.1}. The term with largest and smallest power dominates at late times and early times respectively, and a permanent spike occurs where its coefficient vanishes. The term with intermediate power may dominate for a finite time, depending on the size of their coefficient, and give rise to transient structures.

When a term dominates, $f$ is in an equilibrium state. The transition time between two equilibrium states is considered a boundary between the two states.  Viewed along a worldline, the observer undergoes a sequence of  2-4 equilibrium states, which we call a scenario. Viewed at a fixed time, space is divided into cells of equilibrium states, separated by walls of (spatially dependent) transition times. Viewed as a whole, the spacetime is divided into cells separated by transition times (see Figures \ref{fig_4state3scenariosm05}, \ref{fig_4state3scenarios025},  \ref{fig_4state2scenarios0}). Such a picture reveals the extent of all transient and permanent structures.

In Section \ref{Section7.2}, we revised the description of transient spikes. Our new analysis shows that transient spikes occur on a spatial region rather than along a single worldline. For an inhomogeneous structure to be called a transient spike, it must meet the following criteria:
\begin{itemize}
	\item Its cell is narrow,
	\item Its worldlines undergo a transition different from neighbouring worldlines,
	\item It lasts longer than a single transition.
\end{itemize}
The last criterion distinguishes a transient spike from a new phenomenon called overshoot transition, during which $f$ overshoots when it transitions between equlibrium states. The overshoots occur because the dominant terms involved have opposite signs. An overshoot transition can occur inside a transient spike (see Figure \ref{fig_r0Sp0zerosir}).

\chapter{Conclusion}
The thesis is about the spiky solution with a major focus on finding late-time permanent spike and analysing transient spike. The journey of the exact spike solution started in 2008, when the OT $G_2$ spike solution was discovered. Its non-OT $G_2$ generalised solution was discovered in 2015 by using the Geroch transformation and its stiff fluid generalised  solution was found in 2016 by using Stephani transformation, which is also used in this thesis. 

The above solutions produce spikes  at early times and our first aim is to find a solution that produces a late-time permanent spike. To achieve this, we applied the Stephani transformation using the rotational KVF from the LRS Jacobs solution. It generated a new cylindrically symmetric OT $G_2$ solution that produces the desired late-time permanent spike along the axis of rotation. This is the first non-silent solution with a late-time permanent spike. It is also the first instance of a spike along a line (previous spikes found occur along a plane). Matter density is higher at the spike. The physical radius of the spike turns out to be constant.

Our second aim was to explore and analyse the $k\neq0$ solutions, which feature a rich variety of structures that include a second spike along the cylindrical shell $r=\sqrt{\frac{-\omega_{0}}{k \Spo}}$, transient spikes and  the newly discovered overshoot transitions. To achieve this, we introduced a new technique to analyse the dynamics of a key function, $f$. The analysis helped us revise the description of transient spikes and describe the overshoot transition.

To summarise, in this thesis we have
\begin{itemize}
	\item found the first non-silent solution with a late-time permanent spike.
	\item found the first spike along a line. 
	\item introduced a new technique to analyse a key function, $f$.
	\item revised the description of transient spikes.
	\item discovered and described overshoot transitions.
\end{itemize}
We conclude this thesis by commenting on future research. Firstly, the family of exact solutions we found in this thesis make up only a set of measure zero in the class of cylindrically symmetric solutions. How does a typical cylindirically symmetric solution evolve? To answer this question, it is necessary to conduct a numerical study of the class of cylindrically symmetric solutions, like the numerical study done for the class of non-OT $G_2$ vacuum solutions \cite{art:Limetal2009}.

Secondly, we have used the rotational KVF of the LRS Jacobs solution. Exact solutions that admit a rotational KVF include the LRS Taub solution,  the NUT (LRS Bianchi type VIII) solution, and the Taub-NUT (LRS Bianchi type IX) solution \cite[page 198]{book:WainwrightEllis1997}. It would be interesting to see what spiky solutions are generated from these solutions.

Thirdly, our exact solutions are OT $G_2$ solutions. In principle, non-OT $G_2$ solutions and $G_1$ solutions can be generated from a rotational KVF. Are there simple enough seed solutions that generate spiky solutions with such isometries? 

\begin{appendices}
\chapter{Kinematic variables}
\label{App:formulas}
Let $\mathbf{u}$ be unit timelike vector field. The covariant derivative $u_{a;b}$ can be decompose into irreducible parts according to \cite[Section 1.1.3]{book:WE}
\begin{equation}
u_{a;b}=\sigma_{ab}+\omega_{ab}+\frac13 \Theta h_{ab}-\dot{u}_au_b,
\end{equation}
where $\sigma_{ab}$ is the rate of shear tensor and is symmetric and trace-free, $\omega_{ab}$ is the rate of vorticity vector and is antisymmetric. Also  $u^a\sigma_{ab}=0=u^a\omega_{ab}$. $\dot{u}_a$ the acceleration vector, and the scalar $\Theta$ is the rate of expansion scalar. It follows that
\begin{align}
\sigma_{ab}&=u_{(a;b)}-\frac13 \Theta h_{ab}+\dot{u}_{(a}u_{b)},\\
\omega_{ab}&=u_{[a;b]}+\dot{u}_{[a}u_{b]},\\
\dot{u}_a&=u_{a;b}u^b,\\
\Theta&=u^a{}_{;a}.
\end{align}
The expansion tensor $\Theta_{ab}=\sigma_{ab}+\frac13 \Theta h_{ab}$. In a cosmological context we shall replace $\Theta$ by the Hubble Sclar $H$ defined as $H=\frac13 \Theta h_{ab}$.
Below are the kinematics variables in Iwasawa frame. For more details see  \cite[Appendix A]{art:HeinzleUgglaRohr2009}.
\begin{align}
H &= - \frac13 \frac{1}{N} \partial_0 (b^1+b^2+b^3)
\\
\Theta_{11} &= - \frac{1}{N} \partial_0 b^1
\\
\Theta_{22} &= - \frac{1}{N} \partial_0 b^2
\\
\Theta_{33} &= - \frac{1}{N} \partial_0 b^3
\\
\label{sigma12_n1}
\sigma_{12} &= \frac12 \frac{1}{N} e^{b^2-b^1} \partial_0 n_1
\\
\label{sigma23_n3}
\sigma_{23} &= \frac12 \frac{1}{N} e^{b^3-b^2} \partial_0 n_3
\\
\label{sigma31_n1n2}
\sigma_{31} &= \frac12 \frac{1}{N} e^{b^3-b^1} (-n_3 \partial_0 n_1 + \partial_0 n_2 )\\
\dot{u}_1 &= - e^{b^1} \partial_1 \ln |N|
\\
\dot{u}_2 &= - e^{b^2} [ - n_1 \partial_1 \ln |N| + \partial_2 \ln |N| ]
\\
\dot{u}_3 &= - e^{b^3} [ (n_1 n_3 - n_2) \partial_1 \ln |N| - n_3 \partial_2 \ln |N| + \partial_3 \ln |N| ]
\end{align}
\begin{align}
n_{11} &= e^{b^2+b^3-b^1} \left[ n_1 \partial_1 n_2 - n_2 \partial_1 n_1 - \partial_2 n_2 + \partial_3 n_1 \right]
\\
n_{22} &= e^{b^3+b^1-b^2} \partial_1 n_3
\\
n_{33} &= 0
\\
n_{12} &= \tfrac12 e^{b^3} [ (n_1 n_3 - n_2) \partial_1(b^1-b^2) + n_1 \partial_1 n_3 - n_3 \partial_1 n_1 + \partial_1 n_2 
\notag\\
&\qquad\qquad	- \partial_2 n_3 - n_3 \partial_2(b^1-b^2) + \partial_3(b^1-b^2) ]
\\
n_{23} &= \tfrac12 e^{b^1} \partial_1(b^2-b^3)
\\
n_{31} &=  \tfrac12 e^{b^2} \left[ - \partial_1 n_1 - n_1 \partial_1(b^3-b^1) + \partial_2(b^3-b^1) \right]
\\
a_1 &= \tfrac12 e^{b^1} \partial_1(b^2+b^3)
\\
a_2 &= \tfrac12 e^{b^2} \left[ \partial_1 n_1 - n_1 \partial_1(b^3+b^1) + \partial_2(b^3+b^1) \right]
\\
a_3 &= \tfrac12 e^{b^3} [ (n_1 n_3 - n_2) \partial_1(b^1+b^2) - \partial_1(n_1 n_3-n_2)
\notag\\
&\qquad\qquad + \partial_2 n_3 - n_3 \partial_2(b^1+b^2) + \partial_3(b^1+b^2) ]
\end{align}
\chapter{Killing vector fields and their group actions}
\label{App:KVFsgroup_actions}
A vector field $\xi ^a$ is a KVF of a given metric $g_{ab}$ if it satisfies the Killing equations 
\begin{align}
\xi_{a;b}+\xi_{a;b}=0,
\end{align}
where semicolon denotes covariant derivative.
Two vectors $\xi^a$ and $\eta ^a$ commute if they satisfy
\begin{align}
\xi^b \eta_{a;b}-\eta^b \xi_{a;b}=0.
\end{align}
It is standard result that the set of all isometries of a given manifold $(\mathcal{M}, \mathbf{g})$ forms a Lie group $G_r$ of dimension $r$ called isometry group of $(\mathcal{M}, \mathbf{g})$. Each one-dimensional subgroup of $G_r$ defines a family of curves whose tangent fields is a KVF. In this way the Lie Group $G_r$ generates the Lie algebra of KVFs. If two KVFs commute, then they form an Abelian $G_2$ group. Two KVFs $\xi^a$ and $\eta ^a$ in an Abelian $G_2$ group act orthogonally transitively if they satisfy
\begin{align}
\xi_{[a;b} \xi_{c}\eta _{d]}=0, \ \ &\ \ \eta_{[a;b} \eta_{c}\xi _{d]}=0.
\end{align}

A locally rotationally symmetric (LRS) model admits at least 3 KVFs that form a $G_3$ group whose group orbits are two dimensional. In other words, the 3 KVFs span only a two dimensional surface.

\chapter{Weyl curvature  invariants}
\label{App:Weyl_S}
The four Weyl scalar invariants are 
\begin{align}
CC 
&= C_{abcd}C^{abcd}
\\
CCs
&= C_{abcd}{}^*C^{abcd}
\\
CCC
&= C_{ab}{}^{cd}C_{cd}{}^{ef}C_{ef}{}^{ab}
\\
CCCs
&= C_{ab}{}^{cd}C_{cd}{}^{ef} {}^*C_{ef}{}^{ab},
\end{align}
where ${}^*C_{abcd}=\tfrac12\varepsilon_{ab}{}^{ef}C_{efcd}$,
and $\varepsilon^{abcd}$ is the totally antisymmetric permutation tensor, with 
$\varepsilon^{0123}=\frac{1}{\sqrt{-g}}$.
\end{appendices}
\bibliographystyle{unsrt}
\bibliography{cites1}
\end{document}